\documentclass[a4paper,10pt]{article}
\RequirePackage{pdf14}
\pdfoutput=1
\usepackage[x11names]{xcolor}
\usepackage{jheppub,parskip,comment,bm,amsmath,amsfonts,amssymb,mathtools,tikz,float,graphicx,ytableau}
\hypersetup{colorlinks=true,allcolors=blue}

\newcommand{\nn}{\nonumber}
\newcommand*{\Surjrightarrow}{%
    \mathrel{\ooalign{$\longrightarrow$\cr$\mkern 8.5mu\rightarrow$}}%
}
\newcommand{\ADAIF}[2]{D_{#1}\left(\mathfrak{sl}_{#2}\right)}
\newcommand{\ADAF}[3]{D_{#1}^{#2}\left(\mathfrak{sl}_{#3}\right)}
\newcommand{\ADAI}[3]{D_{#1}\left(\mathfrak{sl}_{#2},#3\right)}
\newcommand{\ADAN}[4]{D_{#1}^{#2}\left(\mathfrak{sl}_{#3},#4\right)}
\newcommand{\ADA}[4]{\left(A_{#1}^{#2}\left[#3\right],#4\right)}
\newcommand*{\thinbraceleft}[1]{%
  \vcenter{\hbox{%
    \begin{tikzpicture}[
      decoration=brace,
      inner sep=0pt,
    ]
      \node (M) {$\displaystyle #1$};
      \draw[decorate] (M.south west) -- (M.north west);
    \end{tikzpicture}%
  }}%
}

\usetikzlibrary{arrows,arrows.meta,shapes,shapes.misc,patterns,decorations.pathmorphing,decorations.pathreplacing,positioning,chains,fit}
\pgfdeclarepatternformonly{coarsedots}{\pgfqpoint{-1pt}{-1pt}}{\pgfqpoint{5pt}{5pt}}{\pgfqpoint{12pt}{12pt}}{
    \pgfpathcircle{\pgfqpoint{.5pt}{.5pt}}{.5pt}
    \pgfusepath{fill}}
\tikzset{gauge/.style={rounded rectangle, draw=black!100, thick, minimum size=5mm},  gaugeD/.style={rounded rectangle, draw=black!100,double,thick,minimum size=5mm},  empty/.style={rounded rectangle, draw=white!100, thick, minimum size=5mm}, flavor/.style={rectangle, draw=black!100, thick, minimum size=5mm},flavorD/.style={rectangle, draw=black!100, double,thick, minimum size=5mm}}

\title{Simplifying the Type $A$ Argyres--Douglas Landscape}

\author[a]{Christopher Beem,\!}
\author[b]{Mario Martone,\!}
\author[a]{Matteo Sacchi,\!}
\author[a]{Palash Singh,\!}
\author[b]{and Jake Stedman}

\affiliation[a]{Mathematical Institute, University of Oxford, Woodstock Road, Oxford, OX2 6GG, UK}
\affiliation[b]{Department of Mathematics, King's College London, The Strand, London, WC2R 2LS, UK}

\emailAdd{christopher.beem@maths.ox.ac.uk}
\emailAdd{mario.martone@kcl.ac.uk}
\emailAdd{matteo.sacchi@maths.ox.ac.uk}
\emailAdd{palash.singh@maths.ox.ac.uk}
\emailAdd{jake.williams@kcl.ac.uk}

\abstract{A well-established organisational principle for Argyres--Douglas-type $\mathcal{N}=2$ superconformal field theories in four dimensions is to characterise such theories by the data defining a(n irregular) Hitchin system on $\mathbb{CP}^1$. The dictionary between Hitchin system data and various features of the corresponding SCFT has been studied extensively, but the overall structure of the resulting space of SCFTs still appears quite complicated. In this work, we systematically delineate a variety of simplifications that arise within this class of constructions due to several large classes of isomorphisms between SCFTs associated with inequivalent Hitchin system data (and their exactly marginal gaugings). We restrict to the most studied class of theories, namely the type $A$ theories without outer automorphism twists.}

\begin{document}
\maketitle
\flushbottom

\section{\label{sec:intro}Introduction}

Four-dimensional superconformal field theories (SCFTs) with eight supercharges provide a rich playground for studying strongly-coupled physics and non-trivial dualities. Over the years, a large number of such theories have been constructed as non-Lagrangian SCFTs. A special class of such theories is class $\mathcal{S}$ \cite{Gaiotto:2009we,Gaiotto:2009hg}. These are 4d $\mathcal{N}=2$ SCFTs obtained as the infrared (IR) limit of the compactification of 6d $\mathcal{N}=(2,0)$ SCFTs on a Riemann surface with marked points carrying co-dimension two defect operators. The IR physics has close connections to the two-dimensional Hitchin integrable system on the same Riemann surface where the Higgs field has simple poles of a specified form at the marked points. Such marked points are normally referred to as regular punctures in theories of class $\mathcal{S}$.

A related, but largely distinct, class (though see, \emph{e.g.}, \cite{Beem:2020pry}) of strongly-coupled SCFTs are \emph{Argyres--Douglas theories} \cite{Argyres:1995jj,Argyres:1995xn}. A defining feature of these theories is that the Coulomb branch operators have fractional scaling dimensions, which obstructs an $\mathcal{N}=2$ Lagrangian description (though some of these theories are known to be realised as the low energy limit of $\mathcal{N}=1$ Lagrangian theories \cite{Maruyoshi:2016tqk,Maruyoshi:2016aim,Agarwal:2016pjo,Agarwal:2017roi,Benvenuti:2017bpg,Maruyoshi:2018nod,Garozzo:2020pmz}, which gives access to a wealth of useful information about these theories such as their full superconformal indices). Originally discovered as special points in the Coulomb branch/parameter space of gauge theories, stringy realisations of these SCFTs have been much studied subsequently. Popular realisations are either as certain scaling limits of the compactifications of 6d $\mathcal{N}=(2,0)$ SCFTs \cite{Gaiotto:2009we,Gaiotto:2009hg} or via geometric engineering in type IIB string theory on singular Calabi--Yau three-folds \cite{Shapere:1999xr,Cecotti:2010fi}. The former approach is the subject of this paper.

The six-dimensional perspective on Argyres--Douglas theories has been initiated in \cite{Bonelli:2011aa} and developed extensively in a long series of papers by D.~Xie and collaborators \cite{Nanopoulos:2010zb,Xie:2012hs,Xie:2013jc,Wang:2015mra,Xie:2016uqq,Xie:2016evu,Xie:2017vaf,Song:2017oew,Xie:2017aqx,Wang:2018gvb,Xie:2019yds,Xie:2019zlb,Xie:2019vzr,Xie:2021ewm,Li:2022njl,Shan:2023xtw}; in these works, four-dimensional theories are associated directly with certain irregular/wild Hitchin systems that are expected to arise by taking certain scaling limits of conventional (regular) class $\mathcal{S}$ theories. In order to describe a superconformal field theory in four dimensions, the corresponding limiting Hitchin systems are defined on the Riemann sphere with at most two marked points, with the Higgs field developing a simple pole/tame singularity at one of the marked point and a higher order pole/irregular singularity at the other.

In this work, we focus in particular on irregular singularities for $\mathfrak{sl}_N$ Hitchin systems that are labelled by two positive integers, $p$ and $b$. We postpone the detailed characterisation of these irregular singularities to Section \ref{subsec:hitchin_review}. We denote these theories by\footnote{See Appendix \ref{app:notations} for a dictionary between our notation and other naming conventions appearing in the literature.}
\begin{equation}
    \ADAN{p}{b}{N}{[Y]}~,
\end{equation}
where $[Y]$ is an integer partition of $N$ that labels the regular puncture and the integers $p$ and $b$ characterise the irregular singularity. The special cases $b=N$ and $b=N-1$ correspond to what are called type I and type II Argyres--Douglas theories, respectively, in the literature. (For $b=N$ we will suppress the $b$ label entirely). For regular punctures that are either empty ($[Y]=[N]$) or full ($[Y]=[1^N]$), the corresponding type I and type II theories admit well-known geometric engineering realisations in type IIB string theory \cite{Xie:2012hs,Cecotti:2012jx,Cecotti:2013lda,Wang:2015mra}.

Inequivalent microscopic compactifications of (often distinct) 6d $\mathcal N=(2,0)$ theories are known to occasionally give rise to the same four-dimensional SCFT, both in the regular and irregular setting. A principal aim of this paper is to produce an economical account of such isomorphisms in the setting of the aforementioned class of (untwisted) type $A$ Argyres--Douglas SCFTs.

We initially restrict to the case of type I theories. For these theories, we identify two fundamental classes of isomorphisms,
\begin{eqnarray}
    \ADAI{p}{N}{[N^{l_N},\dots,1^{l_1}]} &\cong& \ADAI{p}{pL-N}{[(p-1)^{l_1},\dots,(p-N)^{l_N}]}~,\qquad L=\sum_{i=1}^N l_i~,\label{eq:TypeIConjIntro}\\
    \ADAI{p}{N}{[Y]} &\cong& \ADAI{p}{pl+N}{[p^l,Y]}~,\qquad l=1,2,\ldots~.\label{eq:TypeIColIntro}
\end{eqnarray}
A special case of \eqref{eq:TypeIConjIntro} and the $\gcd(p,N)=1$ case of \eqref{eq:TypeIColIntro} were first identified in \cite{Xie:2019yds}; we extend these to the general case as well as provide additional evidence by identifying and comparing circle compactifications of these theories to three dimensions. For the coprime case, these identifications are particularly elegant when interpreted at the level of the associated vertex operator algebras for these theories. There the above can be understood entirely in terms of the \emph{collapsing level} condition for boundary admissible level affine Kac--Moody vertex algebras \cite{ADAMOVIC2018117,Xie:2019yds,Arakawa:2021ogm} along with the realisation of those same current algebras at the intersection of truncation curves for the matrix-extended $\mathcal{W}_{1+\infty}$ algebra \cite{Eberhardt:2019xmf}.

The picture is further simplified for type I theories for which $p$ and $N$ are not coprime. In particular, this includes all type I theories with $\mathcal{N}=2$ supersymmetric exactly marginal couplings. Such theories are generally expected to arise by conformally gauging together other Argyres--Douglas theories that do not have any exactly marginal couplings\footnote{Such theories without any exactly marginal couplings and non abelian flavour symmetry have been dubbed Argyres--Douglas matter theories \cite{Xie:2017vaf,Xie:2017aqx}.} along with free hypermultiplets. It was proposed in \cite{Xie:2017aqx,Xie:2017vaf} to organise these (generalised) gauge theories by associating auxiliary Riemann surfaces to such theories and identifying the $\mathcal{N}=2$ conformal manifold of the SCFT with the complex structure moduli space \emph{\`a la} regular class $\mathcal{S}$. Three-punctured spheres are then identified with Argyres--Douglas matter theories. A variety of examples were worked out in this framework, though it appears to the authors that in general the strategy requires a degree of case-by-case analysis to implement.

In the interest of producing a general and simple picture of this space of theories, we determine a single relation \eqref{eq:typeIwEMD} that expresses any type I Argyres--Douglas theory with $\gcd(p,N)\neq1$ as a conformal gauging of some number of coprime type I Argyres--Douglas theories (which, in particular, have no exactly marginal couplings) along with a definite number of hypermultiplets. Though restricted to a single duality frame, this presentation has the benefit of involving only the basic, coprime type I theories as building blocks. We verify the relation by comparing Coulomb branch spectra as well as 3d reductions \cite{Giacomelli:2020ryy,Closset:2020afy}.

Looking beyond the $b=N$ case, we observe that all theories with $b\neq N$ can be identified with type I theories up to the addition of an appropriate number of free hypermultiplets:
\begin{equation}\label{eq:ResLanConjIntro}
        \ADAN{p}{b}{N}{[Y]} \cong\; \ADAI{p}{(N-b)p+b}{[(p-1)^{N-b},Y]} \otimes H_{\text{free}}\text{ free hypermultiplets}~,    
\end{equation}
where
\begin{equation}\label{eq:ResLanConjFreeHyperIntro}
    H_{\text{free}}=(N-b)\sum_{i=p}^N (i-p+1)l_i~,\qquad[Y]=[N^{l_N},\dots,1^{l_1}]~.
\end{equation}
This extends a relation proposed in \cite{Xie:2019yds} for the case $\gcd(p,N)=1$ with maximal regular puncture. We support this relation by identifying the corresponding three-dimensional mirrors \cite{Giacomelli:2020ryy,Closset:2020afy,Xie:2021ewm}, as well as by matching Coulomb branch spectra and various central charges (though general formulae for $a$ and $c$ central charges of the theories on the left for $b\neq N,N-1$ do not appear in the literature). A benefit of this reduction to type I is that for type I theories, the non-abelian part of the global flavour symmetry algebra is determined entirely by the regular puncture contribution, with the irregular singularity contributing $\gcd(p,N)-1$ $\mathfrak{u}(1)$ factors. Thus it is simple to detect symmetry enhancements by mapping to the corresponding type I theory.

Thus, all the type $A$ Argyres--Douglas theories discussed in this paper reduce to the type I theories, and furthermore the most general such theory is obtained by exactly marginal gauging of coprime type I theories and free hypermultiplets. These coprime type I theories are further related by \eqref{eq:TypeIConjIntro} and \eqref{eq:TypeIColIntro}, which ultimately yields a surprisingly economical accounting of this entire class of theories.

The remainder of this paper is organised as follows. In Section \ref{sec:ADreview} we review relevant aspects of type $A$ Argyres--Douglas theories, emphasising the associated Hitchin systems. We identify the class of irregular singularities to be considered and further discuss what is known of the flavour symmetries, central charges, Coulomb branch spectra, three-dimensional mirrors, circle compactifications, associated vertex operator algebras, and Higgs branches of these theories. In Section \ref{sec:typeI} we highlight some previously proposed relations between type I theories (for coprime $N$ and $p$) and argue for additional or extended versions. We provide a variety of checks for these equivalences. Here we comment on the vertex algebra perspective on these relations, which is illuminated by the realisation of boundary admissible level affine current algebras as truncations of matrix-extended $\mathcal W_{1+\infty}$-algebras. We then present a realisation of general type I theories with $\gcd(p,N)\neq1$ as conformal gaugings of coprime type I theories. This allows us to extend our first set of type I equivalences to the non-coprime case and identify some exceptional equivalences. In Section \ref{sec:iso} we present the third class of isomorphisms whereby all theories under consideration can be related back to a type I theory. We provide a general argument in terms of a match of 3d mirrors and give details in a couple of examples. We conclude with a summary and a short discussion of future directions. In two appendices we provide details regarding three dimensional unitary linear quiver gauge theories as well as a dictionary relating various established notations for the Argyres--Douglas theories considered here.

\section{\label{sec:ADreview}Argyres--Douglas theories of type \texorpdfstring{$A$}{A}}

\subsection{\label{subsec:hitchin_review}Irregular Hitchin systems and notation}

The theories that we consider in this paper are realised by compactification of 6d $\mathcal N=(2,0)$ theories of type $A_{N-1}$ on the sphere ($\mathbb{CP}^1$) with two marked points carrying co-dimension two defects \cite{Gukov:2006jk,Gaiotto:2009hg,Xie:2012hs}. The corresponding Hitchin system is the space of Higgs bundles on $\mathbb{CP}^1$ with a regular singularity at one marked point and a specified irregular singularity at the other.\footnote{Setups with more than one irregular singularity typically give rise to theories consisting of a non-conformal gauging of non-Lagrangian conformal Argyres--Douglas theories. For the analysis of such theories arising from a Hitchin system with multiple irregular singularities in the $A_1$ case, see \cite{Bonelli:2011aa}; for higher rank cases see \cite{Nanopoulos:2010zb}.} The regular singularity is characterised by the presence of a simple pole for the Higgs field with residue lying in a nilpotent co-adjoint orbit of $\mathfrak{sl}_N$. The nilpotent orbit can be identified with an integer partition $[Y]$ of $N$. The general strategy of characterising these theories in terms of the corresponding Hitchin system, and in particular in terms of the choice of regular and irregular singularity, has been pursued extensively by Xie and collaborators starting with \cite{Xie:2012hs}.

Following \cite{witten2008gauge,Xie:2012hs} and many others in the mathematics literature, we adopt a local description of the Higgs field behaviour near the irregular singularity which involves potentially fractional powers of the local coordinate (\emph{i.e.}, passing to a ramified cover), but in exchange the coefficient matrices of the singular terms lie in a Cartan subalgebra so can in fact be simultaneously diagonalised. This requires the presence of a branch cut across which there is a compensating gauge transformation. The characterisation of irregular singularities using only the local behaviour of the Higgs field omits some details of the structure of the underlying Higgs bundle. A more careful treatment can be given in the setting of \emph{ramifiedly good filtered Higgs bundles} as in \cite{Fredrickson:2017yka,Fredrickson:2017jcf}. Fortunately, it seems that this level of detail is not required for the purpose of characterising SCFTs; the restriction to the simpler data of the Higgs field singularity is in line with the approach taken in the pioneering series of works by Xie and collaborators.

For $y$ a local coordinate centred at the irregular singularity, a general local expression for the Higgs field is then taken to have the form
\begin{equation}\label{eq:Higgsexpori}
	\Phi(y) = \frac{u_k}{y^{2+\frac{k}{n}}}\mathrm{d}y + \sum_{l=-n}^{k-1} \frac{u_l}{y^{2+\frac{l}{n}}} \mathrm{d}y + \cdots~,
\end{equation}
where $N\geqslant n\in\mathbb Z_{>0}$ and $-n<k\in\mathbb Z$ with $k$ and $n$ coprime.\footnote{The restriction to coprime $k$ and $n$ is not really restrictive, as the non-coprime case turns out to be equivalent to a specialisation of the coprime case. A possible exception would be in a case where the automorphism $\sigma$ in \eqref{eq:auto} has an order which is greater than (but a multiple of) $n$; we exclude this possibility here, as appears to have been done in previous work on the topic.} When $n>1$, the Higgs field is not single-valued as written; the singular terms in the expansion \eqref{eq:Higgsexpori} pick up $n$-th root of unity phases around the irregular singularity,
\begin{equation}
	\Phi(e^{2\pi i}y) = \frac{\omega^{-k}u_k}{y^{2+\frac{k}{n}}}\mathrm{d}y + \sum_{l=-n}^{k-1} \frac{\omega^{-l}u_l}{y^{2+\frac{l}{n}}} \mathrm{d}y + \cdots~,\qquad\omega\coloneqq e^{\frac{2\pi i}{n}}~.
\end{equation}
Single-valuedness is restored by a compensating $\mathfrak{sl}_N$ gauge transformation $\sigma\in \mathrm{Aut}(\mathfrak{sl}_N)^\circ$ such that
\begin{equation}\label{eq:auto}
	\Phi(y) = \sigma\left(\Phi(e^{2\pi i} y)\right) ~~~\Longleftrightarrow~~~ \sigma\left(u_j\right) = \omega^j\,u_j\quad\forall j~.
\end{equation}
Any such automorphism imparts a $\mathbb{Z}/n\mathbb{Z}$ grading on $\mathfrak{sl}_N$,
\begin{equation}
	\mathfrak{sl}_N = \bigoplus_{j\in\mathbb{Z}/n\mathbb{Z}} \mathfrak{sl}_N(j)~,
\end{equation}
where an element of $\mathfrak{sl}_N(j)$ has eigenvalue $\omega^j$ under the action of $\sigma$. Single-valuedness then requires that the coefficient matrix $u_j$ be taken from the graded subspace $\mathfrak{sl}_N(j\;\text{mod}\;n)$. The requirement that $u_k$ be semi-simple then requires that the grading be such that $\mathfrak{sl}_N(k\;\text{mod}\;n)$ contains a semisimple element. These are the so-called \emph{positive rank} gradings of $\mathfrak{sl}_N$, \emph{cf.} \cite{reeder2012gradings,levy2008kw,levy2009vinberg} for their classification (as well as the corresponding classification for general $\mathfrak{g}$).

A general order-$n$ inner automorphism $\sigma$ that induces a positive rank grading is characterised by a partition of $N$ of the form $[n^m,1^{N-n\cdot m}]$. Restricted to a Cartan subalgebra, the corresponding automorphism acts by permutations with $m$ cycles of length $n$ and the remaining cycles of length one. Thus the explicit general form for the leading matrix $u_k$ is parameterised by $m$ complex coefficients $a_i^{(k)}$ and takes the form
\begin{equation}\label{eq:ukmatrix}
	u_k = 
	\begin{pmatrix}
		a_1^{(k)} \Sigma_n & & & & \\
		& a_2^{(k)} \Sigma_n & & & \\
		& & \ddots & & \\
		& & & a_m^{(k)} \Sigma_n & \\
		& & & & 0_{N-mn} \\
	\end{pmatrix}\,,\qquad\Sigma_n \coloneqq
	\begin{pmatrix}
		1 & & & \\
		& \omega & & \\
		& & \ddots & \\
		& & & \omega^{n-1} \\
	\end{pmatrix} ,
\end{equation}
For $n=1$ there are only integer powers of $z$; this is the unramified case. With our conventions, $m$ then indicates the number of non-zero eigenvalues of $u_k$, and the leading matrix for the clumsily named partition $[1^m,1^{N-m}]$ takes the form,
\begin{equation}
    u_k =
	\begin{pmatrix}
		a_1^{(k)} & & & & \\
		& a_2^{(k)} & & & \\
		& & \ddots & & \\
		& & & a_m^{(k)} & \\
		& & & & 0_{N-m} \\
	\end{pmatrix}~,
\end{equation}
where now $\sum_{i=1}^{m}a_i^{(k)}=0$ by tracelessness.

The intersection of the graded space $\mathfrak{sl}_N(j)$ with the Cartan subalgebra can be determined explicitly. Sub-leading coefficients are then restricted to take the following form,
\begin{equation}\label{eq:subleadulform}
	u_l = 
	\begin{pmatrix}
		a_1^{(l)} \Xi_n^{(l)} & & & & \\
		& a_2^{(l)} \Xi_n^{(l)} & & & \\
		& & \ddots & & \\
		& & & a_m^{(l)} \Xi_n^{(l)} & \\
		& & & & 0_{N-mn} \\
	\end{pmatrix}~,\qquad\Xi_n^{(l)} \coloneqq 
	\begin{pmatrix}
		1 & & & \\
		& \omega^r & & \\
		& & \ddots & \\
		& & & \omega^{(n-1)r} \\
	\end{pmatrix}~,
\end{equation}
where $r\in\mathbb Z_n$ is determined by the requirement $r k \equiv l(\text{mod}\;n)$.

The diagonal matrices in \eqref{eq:Higgsexpori} may in principle have repeated eigenvalues, with the precise structure of degeneracy captured by a sequence of Levi subalgebra inclusions. In the unramified case this situation was rigorously analysed in the seminal work of Biquard and Boalch \cite{biquard_boalch_2004} (see also \cite{witten2008gauge}), and in the general case this leads to what have been dubbed type III Argyres--Douglas theories \cite{Xie:2012dw,Xie:2017vaf,Xie:2017aqx,Xie:2021ewm}. In the present work, we will exclude type III theories and henceforth assume that all non-zero numerical coefficients are generic. 

The coefficient matrices $u_l$ for $l=0(\text{mod}\;n)$ require some special attention. These terms are not affected by the monodromy around the irregular singularity and $\Xi_n^{0(\text{mod}\;n)}$ is just the identity matrix $\mathbb I_n$. Since $\sigma$ leaves the $N-n\cdot m$ block at the lower-right hand corner untouched, these coefficients could take the following general form,
\begin{equation}\label{eq:intordpolecoeff}
    u_{l} =
	\begin{pmatrix}
		a_1^{(l)} \mathbb I_n & & & & \\
		& a_2^{(l)} \mathbb I_n & & & \\
		& & \ddots & & \\
		& & & a_m^{(l)} \mathbb I_n & \\
		& & & & d_{N-n\cdot m} \\
	\end{pmatrix}~,
\end{equation}
where now $d_{N-n\cdot m}$ is a \emph{general} diagonal matrix. These diagonal matrices for successive integer poles are constrained to give rise to nested Levi subalgebras as in the above discussion of type III. Amongst such choices, it was argued in \cite{Xie:2017vaf} that consistent SCFTs can arise only if all of these blocks are identically zero with the exception of that appearing in the coefficient of the simple pole $u_{-n}$, which may be general. We restrict to the case where the corresponding $d_{N-n\cdot m}$ has all distinct eigenvalues, so is maximally non-degenerate.

The class of general Argyres--Douglas theories corresponding to the specified class of $\mathfrak{sl}_N$ Hitchin systems can thus be characterised by integers $n$, $m$, and $k$, as well as the integer partition of $N$, $[Y]$. To ease notation, we define the following parameters,
\begin{equation}\label{eq:bmnpq}
    b = mn~,\qquad p=m(n+k)~,\qquad q=n+k~,
\end{equation}
and use $N$, $b$, $p$, and $[Y]$ to label the Argyres--Douglas theories under consideration. These theories will be then denoted as
\begin{equation}
    \ADAN{p}{b}{N}{[Y]}~.
\end{equation}
We depict how the various parameters that can be read off from the block diagonal form of $u_j$, as in \eqref{eq:subleadulform}, in the following reference diagram; empty squares contain all zeroes.

\resizebox{0.624\hsize}{!}{
\begin{tikzpicture}
    \draw[very thin] (-1,3)--(-1,-2);
    \draw[very thin] (0,3)--(0,-2);
    \draw[very thin] (1,3)--(1,-2);
    \draw[very thin] (2,3)--(2,-2);
    \draw[very thin] (-2,-1)--(3,-1);
    \draw[very thin] (-2,0)--(3,0);
    \draw[very thin] (-2,1)--(3,1);
    \draw[very thin] (-2,2)--(3,2);
    \node at (-1.5,2.5) {{\footnotesize $a_1^{(j)}\Xi_n^{(j)}$}};
    \node at (-0.5,1.5) {{\footnotesize $a_2^{(j)}\Xi_n^{(j)}$}};
    \node at (0.5,0.5) {$\ddots$};
    \node at (1.5,-0.5) {{\footnotesize $a_m^{(j)}\Xi_n^{(j)}$}};
    \node at (2.6,-1.5) {{\footnotesize $0_{N-nm}$}};
    \raisebox{0.4cm}{\hspace*{-2.5cm}$\left(\vphantom{\begin{array}{c}~\\[4.25cm] ~\end{array}}\right.$}
    \raisebox{0.4cm}{\hspace*{5.2cm}$\left.\vphantom{\begin{array}{c}~\\[4.25cm] ~\end{array}}\right)$}
    \node at (-6.5,0.5) {$N\left\{\vphantom{\begin{array}{c}~\\[4.25cm] ~\end{array}}\right.$};
    \node at (-5,-2.5) {\hspace{-0.1cm}$\underbrace{\hspace{1cm}}_{\displaystyle n}$};
    \node at (-0.75,-2.5) {\hspace{-0.1cm}$\underbrace{\hspace{1cm}}_{\displaystyle N-b}$};
    \node at (-3.5,3.5) {\hspace{-0.1cm}$\overbrace{\hspace{4cm}}^{\displaystyle b=nm}$};
\end{tikzpicture}}

The cases $b=N$ and $b=N-1$ are singled out as those for which the coefficients of all the singular terms have generically distinct eigenvalues; thus they are regular semisimple elements of $\mathfrak{sl}_N$. These have been given the names \textit{type I} and \textit{type II} Argyres--Douglas theories, respectively. This subset of theories with no regular puncture ($[Y]=[N]$) or maximal regular puncture ($[Y]=[1^N]$) can also be realised via geometric engineering in type IIB. In particular, for $[Y]=[N]$ these theories are engineered by the following (quasi-homogeneous, compound du Val) hypersurface singularities in $\mathbb C^4$ \cite{Shapere:1999xr,Cecotti:2010fi,Wang:2015mra}:
\begin{equation}\label{eq:hypersurfN}
    \begin{split}
        \ADAN{p}{N}{N}{[N]} \;&:\quad W(w,t,x,z)=tw+x^N+z^{p-N}=0~,\\
        \ADAN{p}{N-1}{N}{[N]} \;&:\quad W(w,t,x,z)=tw+x^N+xz^{p-N}=0~,
    \end{split}
\end{equation}
where the holomorphic three-form is
\begin{equation}
    \Omega_3=\frac{\mathrm{d}w\,\mathrm{d}t\,\mathrm{d}x\,\mathrm{d}z}{\mathrm{d}W}~.
\end{equation}
For $[Y]=[1^N]$ the following hypersurfaces in $\mathbb C^3\times\mathbb C^*$ are used instead \cite{Cecotti:2012jx,Cecotti:2013lda}:
\begin{equation}
    \begin{split}
        \ADAN{p}{N}{N}{[1^N]} \;&:\quad W(w,t,x,z)=tw+x^N+z^p=0~, \\
        \ADAN{p}{N-1}{N}{[1^N]} \;&:\quad W(w,t,x,z)=tw+x^N+xz^p=0~,
    \end{split}
\end{equation}
where now the holomorphic three-form takes the form
\begin{equation}
    \Omega_3=\frac{\mathrm{d}w\,\mathrm{d}t\,\mathrm{d}x\,\mathrm{d}z}{z\mathrm{d}W}~.
\end{equation}
A convenient dictionary relating the different naming conventions for Argyres--Douglas theories appearing in the literature is supplied in Appendix \ref{app:notations}.

To lighten the notation, we will adopt simpler expressions for type I theories or theories with maximal regular puncture:
\begin{eqnarray}
    &\ADAN{p}{N}{N}{[Y]} &\eqqcolon \ADAI{p}{N}{[Y]}~, \\
    &\ADAN{p}{b}{N}{[1^N]} &\eqqcolon \ADAF{p}{b}{N}~, \\
    &\ADAN{p}{N}{N}{[1^N]} &\eqqcolon \ADAIF{p}{N}~.
\end{eqnarray}

\subsection{\label{subsec:CBspecCM}Coulomb branch spectrum}

The Seiberg--Witten curve of a 4d $\mathcal N=2$ SCFT captures many properties of the effective field theory on its Coulomb branch \cite{Seiberg:1994rs,Seiberg:1994aj}. The relevant data that we discuss here is the spectrum of Coulomb branch operators (\emph{i.e.}, the set of scaling dimensions/$\mathfrak{u}(1)_r$ charges of Coulomb branch operators), the rank of the flavour symmetry, and the number of $\mathcal{N}=2$ supersymmetric exactly marginal deformations (\emph{i.e.}, the dimension of the $\mathcal N=2$ conformal manifold). In the class $\mathcal S$ setting, the Seiberg--Witten curve is identified with the spectral curve of the Hitchin system \cite{Gaiotto:2009hg}. To match the notation most commonly used in the class $\mathcal S$ setting, we change coordinates from $y$ to $z=\frac{1}{y}$ so the irregular singularity is at $z\rightarrow\infty$. The spectral curve of the Hitchin system is determined by the Higgs field according to
\begin{equation}
    T^\ast\mathbb{CP}^1\supset\Sigma\;:\;\mathrm{det}(\lambda-\Phi(z))=0~,
\end{equation}
where $\lambda=x\,\mathrm{d}z$ is identified with the Seiberg--Witten differential and $x$ parameterises the fibre direction.

For $\mathfrak{g}=\mathfrak{sl}_N$, the spectral curve takes the following general form,
\begin{equation}\label{spectral}
    x^N + \sum_{i=2}^N \phi_i(z)x^{N-i}=0~,
\end{equation}
where the $\phi_i(z)$ are Laurent polynomials in $z$ and $\phi_i(z)(dz)^i \in H^0(\mathbb{CP}^1,K_{\mathbb{CP}^1}^{\otimes i})$. The integral of the Seiberg--Witten differential around closed cycles in $\Sigma$ gives the mass of BPS particles, and thus one assigns scaling dimensions
\begin{equation}\label{eq:xzscaling}
    \Delta[\lambda]=1 ~~\Longrightarrow~~ \Delta[x]+\Delta[z]=1~,
\end{equation}
For a monomial in the equation for the spectral curve of the form $v_{i,j}x^iz^j$, the dimension of the coefficient $v_{i,j}$ can be computed by demanding that each term in the spectral curve has the same scaling dimension overall. Given the leading term in \eqref{spectral}, this requires
\begin{equation}\label{eq:vijscalingbasic}
    \Delta[v_{i,j}] = (N-i)\Delta[x]-j\Delta[z]~.
\end{equation}
These parameters are assigned different physical interpretations depending on their scaling dimensions:
\begin{itemize}
    \item $\Delta[v_{i,j}]=0$~~$\Longrightarrow$~~ $\mathcal N=2$ exactly marginal deformations (EMD);
    \item $0<\Delta[v_{i,j}]<1$~~$\Longrightarrow$~~ relevant chiral couplings;
    \item $\Delta[v_{i,j}]=1$~~$\Longrightarrow$~~ mass deformations;
    \item $\Delta[v_{i,j}]>1$~~$\Longrightarrow$~~ Coulomb branch operator expectation values.
\end{itemize}
In any SCFT (without decoupled free vector multiplets), all parameters of dimension $0\leqslant\Delta_1<1$ correspond to sources for Coulomb branch operators and are paired with vevs of those operators, which carry dimension $1<\Delta_2\leqslant2$ such that $\Delta_1+\Delta_2=2$ \cite{Argyres:1995xn}. In particular, associated with EMDs there will be vevs of exactly marginal operators with $\Delta[v_{i,j}]=2$. We will be primarily interested in mass deformations, exactly marginal deformations, and the Coulomb branch spectrum of these theories.

For the $\ADAN{p}{b}{N}{[Y]}$ theories considered here, the Seiberg--Witten curve should in principle be computed as a spectral curve of the corresponding Hitchin system by identifying the global Higgs fields compatible with the prescribed regular and irregular singularities. However, absent a simple characterisation of these global solutions, a procedure for indirectly determining the Coulomb branch spectrum---the ``Newton polygon method''---has been proposed in \cite{Xie:2012hs}.\footnote{See, \emph{e.g.}, \cite{Xie:2015rpa,Xie:2017vaf,Xie:2017aqx} for the extension to more general Argyres--Douglas theories.}

In this approach, the differentials in \eqref{spectral} are taken to have the general form,
\begin{equation}\label{eq:phizpower}
    \phi_i(z) = v_{i,q_i}\,z^{q_i} + v_{i,q_i-1}\,z^{q_i-1} + \cdots + v_{i,-p_i}\,z^{-p_i}~.
\end{equation}
Here $p_i$ are the \emph{pole structures} associated to the partition $[Y]$; these are determined in a straightforward manner by $[Y]$ as explained in \cite{Gaiotto:2009we}. The $q_i$ are determined by finding the leading terms arising from the local expansion of the Higgs field \eqref{eq:Higgsexpori}.

The most na\"ive assumption---that all $v_{i,j}$ present in this expression are non-zero and can be varied independently---is incorrect as can be checked in various simple examples. However, it appears that \emph{for all monomials whose coefficients scale as $\Delta\geqslant1$} this na\"ive expectation does indeed hold true; the assumption that this is generally the case amounts to an important input into the techniques used to compute the Coulomb branch spectrum in \cite{Xie:2012hs,Xie:2017vaf,Xie:2017aqx}, and we will take this conjecture as given in the relevant calculations performed in the present paper. Allowed parameters with $\Delta<1$ can then be inferred (up to ambiguities in cases of degeneracy) according to the required pairing with parameters with scaling $\Delta>1$.

Summarising, the Seiberg--Witten curve for $\ADAN{p}{b}{N}{[Y]}$ can be computed by writing down all monomials that appear in \eqref{eq:phizpower} after computing the corresponding $p_i$ and $q_i$. This na\"ive expression contains some extra terms $v_{i,j}x^iz^j$ corresponding to those $\Delta[v_{i,j}]<1$ that do not pair with any other monomial such that the sum of their scaling dimensions is 2. These extra terms must be then removed from this expression to obtain the correct Seiberg--Witten curve.

A necessary self-consistency condition for this procedure is that the number of (na\"ive) terms with scaling dimension $0\leqslant \Delta < 1$ must be greater than or equal to the number of terms with scaling dimension $2-\Delta$. In \cite{Xie:2017vaf} it is proposed that the special case of this condition for $\Delta=0$ is in fact sufficient to guarantee a physically sensible interpretation of the corresponding Hitchin system as arising from a four-dimensional SCFT. Namely, if and only if there are more dimension-two terms than (na\"ive) dimension-zero terms then the construction is declared inconsistent. This condition, also referred to as the \emph{conformality constraint} in \cite{Xie:2017aqx}, leads to the following condition on the commutant (Levi) sub-algebras associated to the coefficients $u_j$,
\begin{equation}
    \mathfrak l_k = \mathfrak l_{k-1} = \cdots = \mathfrak l_{-n+1} \supset \mathfrak l_{-n}~,
\end{equation}
where $\mathfrak l_j$ is the Levi sub-algebra corresponding to the Cartan element $u_j$ in \eqref{eq:Higgsexpori}. This restriction was already used around \eqref{eq:intordpolecoeff} to restrict the behaviour of the integer order pole coefficients in the constructions under consideration.

\subsection{\label{subsec:ADflavourcc}Flavour symmetries and central charges}

The flavour symmetry algebras of the $\ADAI{p}{N}{[Y]}$ theories generically receive contributions from both the regular and the irregular singularities,
\begin{equation}
    \mathfrak g_F^{\text{irreg}}\oplus\mathfrak g_Y \subseteq \mathfrak g_F~.
\end{equation}
There can be accidental symmetry enhancements such that this is a proper inclusion. In practice, these symmetry patterns can be usually understood from the quiver description of the 3d mirror theory reviewed in the next subsection. Above, $\mathfrak g_Y$ is the (standard) symmetry associated to a regular puncture labelled by the partition $[Y]=[N^{l_N},\dots,1^{l_1}]$, which is given by \cite{Gaiotto:2009we}
\begin{equation}\label{eq:regpuncsymm}
    \mathfrak g_Y = \left(\bigoplus_{i=1}^N \mathfrak u(l_i)\right)/\mathfrak u(1)~.
\end{equation}
For the maximal puncture $[Y]=[1^N]$ one has an associated symmetry algebra of $\mathfrak g_{[1^N]}=\mathfrak{su}(N)$, while for the empty puncture $[Y]=[N]$ there is no contribution to the flavour symmetry---$\mathfrak g_{[N]}=\emptyset$.

The contribution from the irregular singularity has been argued to take the form \cite{Xie:2017vaf,Xie:2017aqx},
\begin{equation}
    \mathfrak g_F^{\text{irreg}} = \mathfrak u(N-b)\oplus\mathfrak u(1)^{m-1}~,
\end{equation}
where $m=\gcd(b,p)$. In particular, the rank is given by
\begin{equation}
    \text{rk}\left(\mathfrak g_F^{\text{irreg}}\right) = N-b+\gcd (b,p)-1~.
\end{equation}
This simplifies considerably for type I ($b=N$) and type II ($b=N-1)$ theories \cite{Giacomelli:2017ckh}.

The flavour central charge associated to some symmetry $\mathfrak g$ is given by the following 't Hooft anomaly \cite{Tachikawa:2011dz,Chacaltana:2012zy}:
\begin{equation}
    k_{\mathfrak g} = -2\text{Tr} R\,T^a\,T^b \eqqcolon -2\text{Tr} R\,\mathfrak{g}^2~,
\end{equation}
where $T^a$ are the generators of $\mathfrak g$. The $a$ and $c$ central charges are defined similarly using the conformal/Weyl anomaly. The problem of computing these central charges for $\ADAN{p}{b}{N}{[Y]}$ can be simplified to the problem of computing these for the cases with maximal regular puncture $\ADAN{p}{b}{N}{[1^N]}$, which we shall refer to as the UV SCFT for the remainder of this section. This is because the former can be thought of as the IR limit of the RG flow triggered by giving a nilpotent vev of the form $[Y]$ to the $\mathfrak{su}(N)$ moment map of the UV SCFT. Thus the central charges for the IR theory can be computed in terms of the UV central charges by matching anomalies along the RG flow.

Furthermore, a linear combination of $a$ and $c$ can be determined in terms of the Coulomb branch spectrum via the Shapere--Tachikawa relation \cite{Shapere:2008zf}\footnote{It is possible to directly access the values of $a$ and $c$ from purely Coulomb branch data, but it requires more detailed information about the stratification of the singular locus \cite{Martone:2020nsy,Argyres:2020wmq} which is not readily available in our case.}
\begin{equation}\label{eq:ShapereTachikawa}
    2a-c = \frac14\sum_{i=1}^r (2\Delta_i-1)~,
\end{equation}
where $r$ is the rank of the 4d SCFT, defined as the complex dimension of the Coulomb branch, and $\Delta_i$ are the scaling dimensions of the Coulomb branch operators.

Note that the flavour central charge of the $\mathfrak{su}(N-b)\subset\mathfrak g_F^{\text{irreg}}$ symmetry does not change under the nilpotent Higgsing discussed above and has been conjectured to be \cite{Xie:2017aqx,Xie:2017vaf}
\begin{equation}
    k_{\mathfrak{su}(N-b)} = 2\left(N-b+\frac{n}{n+k}\right)~.
\end{equation}
Meanwhile the flavour central charge of the $\mathfrak{su}(N)$ symmetry associated to the maximal puncture of the UV SCFT is conjectured to be \cite{Xie:2016evu}
\begin{equation}\label{eq:ksuNfullpunc}
    k_{\mathfrak{su}(N)} = 2\left(N-\frac{n}{n+k}\right)~.
\end{equation}
The $c$ central charge of the UV SCFT is not known in complete generality. However for $b=N,N-1$, the $c$ central charge is known from the geometric engineering construction to be \cite{Shapere:2008zf,Xie:2015rpa,Giacomelli:2017ckh}
\begin{equation}\label{eq:cfullpunc}
    c = \Big(2r+\text{rk}(\mathfrak g_F)\Big)\frac{\Delta_{\text{max}}}{12}+\frac{r}{6}~,
\end{equation}
where $\Delta_{\text{max}}$ is the scaling dimension of the Coulomb branch operator with the largest scaling dimension. These expressions along with the Shapere--Tachikawa relation determine all the central charges of the UV SCFT.

\subsubsection{\label{subsubsec:IR_anomaly_matching}IR central charges from anomaly matching}

We detail here the computation of flavour central charges for theories with partially reduced punctures. The most involved aspect of the calculation is identifying the contribution of the various free hypermultiplets that arise under the corresponding nilpotent Higgsing. These are to be subtracted from the UV central charges in order to recover those of the irreducible IR SCFT.

The difference between $a$ and $c$ central charges is determined by the cubic $\mathfrak{u}(1)_r$ anomaly, which is unbroken by Higgs branch RG flows so the anomaly can be matched by simply subtracting the contributions of free hypermultiplets \cite{Giacomelli:2020ryy},
\begin{equation}
    48(a^{\rm UV}-c^{\rm UV}) = 48(a^{\rm IR}-c^{\rm IR}) + 2(\#\text{ of Nambu--Goldstone modes})~.
\end{equation}
Here the number of free hypermultiplets is, alternatively, the quaternionic dimension of the Higgs branch stratum associated to the Higgsing in question. For a general nilpotent Higgsing associated to the partition $[Y]$, the number of Nambu--Goldstone modes associated with flat directions for the moment map operators is given by (\emph{cf.} \cite{Tachikawa:2015bga}),
\begin{equation}
    \#\text{ of Nambu--Goldstone modes} = \frac12\left\{N^2-\sum_i s_i^2\right\} \,,\,\,\, \text{where}\,\,\, {}^tY=[s_1,s_2,\dots]~.
\end{equation}
Meanwhile, the flavour symmetry of the IR theory is given by $\mathfrak g_Y$ as above, and the flavour central charge of any of the $\mathfrak{su}(l_i)$ subalgebras can be expressed as
\begin{equation}\label{eq:ksulredpunc}
    k^{\rm IR}_{\mathfrak{su}(l_i)} = I_{\mathfrak{su}(l_i)\hookrightarrow\mathfrak{su}(N)}\,k^{\rm UV}_{\mathfrak{su}(N)} + k_{\mathfrak{su}(l_i)}^{\rm NG}~,
\end{equation}
where $k_{\mathfrak{su}(l_i)}^{\rm NG}$ is the contribution from the massless Nambu--Goldstone modes to the above 't Hooft anomaly in the IR and $I_{\mathfrak{su}(l_i)\hookrightarrow\mathfrak{su}(N)}$ is the embedding index. To understand the former contribution, we take a closer look at the $\mathfrak{su}(N)$ representation theory.

Let us consider the Higgsing to a specific regular puncture with partition
\begin{equation}
[Y]=[N^{l_N},\dots,\alpha^{l_\alpha},\dots,1^{l_1}]~.
\end{equation}
The fundamental representation of $\mathfrak{su}(N)$ then decomposes under the remaining global symmetry according to
\begin{equation}
    \text{fund}_{\mathfrak{su}(N)} \longrightarrow \left(V_{\frac{\alpha-1}{2}},l_\alpha\right)\bigoplus_{\alpha\neq i=1}^N\,\left(V_{\frac{i-1}{2}},l_i\right)~,
\end{equation}
where the first entry, $V_j$, is the spin-$j$ representation of the embedded $\mathfrak{su}(2)$, while the second entry, $l_i$, denotes the fundamental representation of $\mathfrak{su}(l_i)$ and the trivial representation of the remaining $\mathfrak{su}(l_{j\neq i})\subset\mathfrak g_Y$ subalgebras. We will follow this notation for the rest of this work.

With respect to the moment map operator, one should consider instead the decomposition of the adjoint representation. This is obtained by exploiting the identity $\text{fund}\otimes\overline{\text{fund}} = \text{adj}\oplus \text{sing}$, giving
\begin{equation}
    \begin{split}
        \text{adj}_{\mathfrak{su}(N)}\oplus(V_0,1) \longrightarrow \left(V_{\frac{\alpha-1}{2}}\otimes V_{\frac{\alpha-1}{2}},l_\alpha\otimes\overline{l_\alpha}\right) &\bigoplus_{\substack{i=1\\i\neq\alpha}}^N \left(V_{\frac{\alpha-1}{2}}\otimes V_{\frac{i-1}{2}},l_\alpha\otimes\overline{l_i}\oplus l_i\otimes\overline{l_\alpha}\right) \\
        &\bigoplus_{\substack{i,j=1\\i,j\neq\alpha}}^N \left(V_{\frac{i-1}{2}}\otimes V_{\frac{j-1}{2}},l_i\otimes\overline{l_j}\oplus l_j\otimes\overline{l_i}\right)~.
    \end{split}
\end{equation}
The tensor product of $\mathfrak{su}(2)$ representations decomposes as follows:
\begin{equation}
    V_{\frac{i-1}{2}} \otimes V_{\frac{j-1}{2}} = \bigoplus_{\frac{|i-j|}{2}}^{\frac{i+j-2}{2}} V_k~.
\end{equation}
The highest weight component in each of the resulting $\mathfrak{su}(2)$ representations, say $V_{\frac{M-1}{2}}$, becomes massive under the nilpotent vev and is integrated out leaving behind $M-1$ components which rearrange into $\frac{M-1}{2}$ Nambu--Goldstone modes in the IR whose contribution we want to count. Hence, we find that the total number of Nambu--Goldstone modes in the IR is
\begin{equation}
    \frac12\sum_{k=\frac{|i-j|}{2}}^{\frac{i+j-2}{2}} 2k = \frac{(i+j-|i-j|)(i+j-2+|i-j|)}{8} \eqqcolon n_{i,j}~.
\end{equation}
This result along with the explicit decomposition of the adjoint representation into representations of $\mathfrak g_Y$ tells us the exact number of Nambu--Goldstone modes that transform in various $\mathfrak{su}(l_\alpha)$ subalgebra representations.

A Nambu--Goldstone mode transforming in a representation $\mathcal R$ of $\mathfrak{su}(l_i)$ contributes $-4T_{\mathfrak{su}(l_\alpha)}(\mathcal R)$ to $k^{\rm IR}_{\mathfrak{su}(l_i)}$, where $T_{\mathfrak{su}(l_\alpha)}(\mathcal R)$ is the Dynkin index of the representation. For $\mathfrak{su}(l)$, we have
\begin{equation}
    T_{\mathfrak{su}(l)}(\text{sing}) = 0\;,\quad T_{\mathfrak{su}(l)}(\text{fund}) = \frac12\;,\quad T_{\mathfrak{su}(l)}(\text{adj})=T_{\mathfrak{su}(l)}(\text{fund}\otimes \overline{\text{fund}})=l~.
\end{equation}
Moreover, the Dynkin index distributes over direct sum of representations
\begin{equation}
T(\mathcal{R}_i\oplus\mathcal{R}_j)=T(\mathcal{R}_i)+T(\mathcal{R}_j)~.
\end{equation}
Therefore we sum up all these contribution to explicitly write down $k_i^{\rm NG}$ as follows:
\begin{equation}
    \begin{split}
        k_{\mathfrak{su}(l_i)}^{\rm NG} = -4n_{\alpha,\alpha}T_{\mathfrak{su}(l_\alpha)}\left(l_\alpha\otimes\overline{l_\alpha}\right) &- 4\sum_{\alpha\neq i=1}^N n_{\alpha,i}T_{\mathfrak{su}(l_\alpha)}(l_{\alpha}\otimes\overline{l_i}\oplus l_i\otimes\overline{l_\alpha}) \\
        &- 4\sum_{\alpha\neq i,j=1}^N n_{i,j} T_{\mathfrak{su}(l_\alpha)}\left(l_i\otimes\overline{l_j}\oplus l_j\otimes\overline{l_i}\right)~.
    \end{split}
\end{equation}

This can be simplified by using the above expressions for the Dynkin index
\begin{equation}
    \begin{split}
        k_{\mathfrak{su}(l_i)}^{\rm NG} = -4\left(n_{\alpha,\alpha}l_\alpha + \sum_{\alpha\neq i=1}^N n_{\alpha,i}\left(\frac12 \times l_i+l_i\times\frac12\right) + 0 \right)~,
    \end{split}
\end{equation}
where we have used the fact that $l_\alpha\otimes l_i$ implies that there are a total of $l_i$ Nambu--Goldstone modes in the fundamental representation of $\mathfrak{su}(l_\alpha)$ to calculate the Dynkin index, therefore $T_{\mathfrak{su}(l_\alpha)}\left(l_\alpha\otimes l_i\right)=\frac12\times l_i$. This can be simplified to obtain the following simple expression
\begin{equation}\label{eq:kGBcont}
    k_{\mathfrak{su}(l_i)}^{\rm NG} = -4\sum_{i=1}^N n_{\alpha,i}\,l_i = -\sum_{i=1}^N \frac{(\alpha+i-|\alpha-i|)(\alpha+i-2+|\alpha-i|)}{2} l_i~.
\end{equation}
Combining with \eqref{eq:ksulredpunc} this determines the flavour central charges of the IR theory.

\subsection{\label{subsec:AD3d}3d quivers: mirrors and direct reductions}

A key tool in the analysis of Argyres--Douglas SCFTs characterised by irregular Hitchin systems is the description of related three-dimensional theories as Lagrangian quiver gauge theories. In particular, the three-dimensional reductions of these theories admit Lagrangian quiver mirror descriptions, which can be taken as a physical counterpart to the statement that the corresponding moduli spaces of irregular Higgs bundles can be related to certain Nakajima quiver varieties \cite{Nakajima:1994nid,boalch2008irregular,boalch2020diagrams}. Notably, for many Argyres--Douglas theories the three-dimensional mirror admits a further Lagrangian mirror dual. These are therefore (non-mirror) dual descriptions of the three-dimensional reductions of the theories in question. In this section we review the state of the art of assigning Lagrangian quiver descriptions to the 3d mirrors of Argyres--Douglas SCFTs of type $A$ and, for type I, also to the 3d reductions .

\subsubsection{\label{subsubsec:AD3dmirrgen}3d mirrors of \texorpdfstring{$\ADAN{p}{b}{N}{[1^N]}$}{Dpb(slN,Y)} theories}

We first recall the Lagrangian unitary quivers that are mirror dual to the IR limit of the circle compactification of Argyres--Douglas theories $\ADAN{p}{b}{N}{[1^N]}$. We present the results of \cite{Xie:2021ewm} for the general $b$ case as well as the specialisation to the type I case $b=N$ (previously studied in \cite{Giacomelli:2020ryy}) as a simple algorithm to construct the 3d mirrors.\footnote{See also \cite{Beratto:2020wmn,Kang:2022zsl} for alternative 3d mirrors of some of the Argyres--Douglas theories discussed here that can also be realised in class $\mathcal{S}$ without irregular singularities \cite{Beem:2020pry}. The 3d mirrors of Argyres--Douglas theories of different types and with twisted punctures have been discussed instead in \cite{Carta:2021whq,Xie:2021ewm,Carta:2021dyx,Carta:2022spy,Carta:2022fxc}.} We split our discussion into two cases: $p\geqslant N$ and $p<N$. We restrict throughout to the case of maximal regular puncture $[Y]=[1^N]$. Reduced punctures can be obtained from the case $[Y]=[1^N]$ via the methods quiver subtraction and the ``decay and fission algorithm" \cite{Bourget:2023dkj,Bourget:2024mgn,Bourget:2021siw}, to which we will return later.

\begin{figure}
    \centering
    \begin{tikzpicture}[scale=1.8,every node/.style={scale=1.3},font=\scriptsize]
    \node[gauge] (gl0) at (0,0) {$1$};
    \node[gauge] (gl1) at (0.5,0.5) {$2$};
    \node[] (gl2) at (1,1) {$\cdots$};
    \node[gauge] (gl3) at (1.5,1.5) {$N-2$};
    \node[gauge] (gl4) at (2.25,2.25) {$N-1$};
    \node[gauge] (gr4) at (4.25,2.25) {\scriptsize$N-nm$};
    \node[gauge] (gr3) at (5,1.5) {\tiny$N-nm-1$};
    \node[] (gr2) at (5.5,1) {$\cdots$};
    \node[gauge] (gr1) at (6,0.5) {$2$};
    \node[gauge] (gr0) at (6.5,0) {$1$};
    \node[gauge] (gu0) at (2.25,3.25) {$1$};
    \node[gauge] (gu1) at (4.25,3.25) {$1$};
    \node[gauge] (gu2) at (3.25,4.25) {$1$};
    \draw (gl0)--(gl1)--(gl2)--(gl3)--(gl4)--(gr4)--(gr3)--(gr2)--(gr1)--(gr0);
    \draw (gu0)--(gu1);
    \draw (gu0)--(gu2);
    \draw (gu2)--(gu1);
    \draw (gl4)--(gu0);
    \draw (gl4)--(gu1);
    \draw (gl4)--(gu2);
    \draw (gr4)--(gu0);
    \draw (gr4)--(gu1);
    \draw (gr4)--(gu2);
    \node[] at (2.1,2.65) {$n$};
    \node[] at (2.6,2.6) {$n$};
    \node[] at (2.85,2.4) {$n$};
    \node[] at (4.4,2.65) {$k$};
    \node[] at (3.9,2.6) {$k$};
    \node[] at (3.65,2.4) {$k$};
    \node[] at (2.6,3.85) {$nk$};
    \node[] at (3.9,3.85) {$nk$};
    \node[] at (3.25,3.4) {$nk$};
    \node[] at (3.25,0.5) {$\widetilde H_{\text{free}} = \frac12m(n-1)(k-1)$};
    \node[] at (3.25,0.2) {hypermultiplets};
    \end{tikzpicture}
    \caption{\label{fig:3dmirrnkp}The 3d $\mathcal{N}=4$ quiver gauge theory that is mirror dual to the circle reduction of the $\ADAN{p}{b}{N}{[1^N]}$ Argyres--Douglas theory for $p>b$. The number of $\mathfrak{u}(1)$ gauge nodes in the complete graph is in general $m$, but for simplicity in the drawing we depict the case of $m=3$. The numbers labelling certain edges denote the multiplicity of the corresponding hypermultiplet, where the absence of an explicit label implies unit multiplicity.}
\end{figure}

The 3d mirrors of the above Argyres--Douglas theory with $p\geqslant N$ can be constructed as follows:
\begin{enumerate}
    \item Start with a complete graph with $m$ vertices and place a $\mathfrak{u}(1)$ gauge node at each vertex. Each edge is assigned multiplicity $nk$, where we recall that $n=\tfrac{b}{m}$ and $k=q-n=\tfrac{p-b}{m}$.
    \item Construct the Tail A (each node is a unitary gauge algebra of the specified rank)
    \begin{equation}
    \begin{tikzpicture}[scale=1.2,every node/.style={scale=1.2},font=\scriptsize]
    \node[gauge] (g0) at (0,0) {$N-2$};
    \node (g1) at (1.5,0) {$\cdots$};
    \node[gauge] (g2) at (3,0) {$\,2\,$};
    \node[gauge] (g3) at (4.5,0) {$\,1\,$};
    \node[gauge] (f0) at (-1.5,0) {$N-1$};
    \draw (g0)--(g1)--(g2)--(g3);
    \draw (g0)--(f0);
    \end{tikzpicture}
    \end{equation}
    and connect the $\mathfrak{u}(N-1)$ node to every node in the complete graph by an edge with multiplicity $n$.
    \item Construct another Tail B
    \begin{equation}
    \begin{tikzpicture}[scale=1.2,every node/.style={scale=1.2},font=\scriptsize]
    \node[gauge] (g0) at (0,0) {$N-nm-1$};
    \node (g1) at (1.5,0) {$\cdots$};
    \node[gauge] (g2) at (3,0) {$\,2\,$};
    \node[gauge] (g3) at (4.5,0) {$\,1\,$};
    \node[gauge] (f0) at (-2.5,0) {$N-nm$};
    \draw (g0)--(g1)--(g2)--(g3);
    \draw (g0)--(f0);
    \end{tikzpicture}
    \end{equation}
    and connect the $\mathfrak{u}(N-nm)$ node to every node in the complete graph by an edge with multiplicity $k$.
    \item Finally, connect the the two tails by joining the $\mathfrak{u}(N-1)$ gauge node of Tail A to the $\mathfrak{u}(N-nm)$ gauge node of Tail B by an edge with multiplicity $1$.
    \item Include an additional set of $\widetilde H_{\text{free}}=\frac12m(n-1)(k-1)$ free hypermultiplets.
\end{enumerate}

The general quiver of this form is as in Figure \ref{fig:3dmirrnkp}. It is immediate to see that all the nodes in the two tails apart from the ones connected to the complete graph are balanced. Moreover, the $\mathfrak{u}(N-1)$ gauge node in Tail A that connects to the complete graph is balanced as well. The same is not true for the node on Tail B that connects to the complete graph. We therefore expect that the $\mathfrak u(1)$ topological symmetries from these balanced nodes get enhanced to $\mathfrak{su}(N)\oplus\mathfrak{su}(N-nm)$ while the remaining unbalanced nodes provide a $\mathfrak{u}(1)^m$ symmetry (after factoring out the overall trivially acting $\mathfrak{u}(1)$), as expected from four-dimensional physics. Since the $\mathfrak{su}(N)$ flavour symmetry associated with the regular puncture $[Y]=[1^N]$ is coming from a $T(\mathfrak{su}(N))$ tail \cite{Gaiotto:2008ak} attached to the complete graph, the 3d mirror for cases with a reduced regular puncture can be easily obtained by substituting this $T(\mathfrak{su}(N))$ with the appropriate $T_{[1^N]}^{\sigma=[Y]}(\mathfrak{su}(N))$ tail.

\begin{figure}
    \centering
        \begin{tikzpicture}[scale=3.2,every node/.style={scale=1.3},font=\scriptsize]
    \node[gauge] (gl0) at (0,0) {$1$};
    \node[gauge] (gl1) at (0.5,0) {$2$};
    \node[] (gl2) at (1,0) {$\cdots$};
    \node[gauge] (gl3) at (1.5,0) {$N-2$};
    \node[gauge] (gl4) at (2.25,0) {$N-1$};
    \node[gauge] (gu0) at (2.75,0.5) {$1$};
    \node[gauge] (gu1) at (2.75,-0.5) {$1$};
    \node[gauge] (gu2) at (3.25,0) {$1$};
    \draw (gl0)--(gl1)--(gl2)--(gl3)--(gl4);
    \draw (gu0)--(gu1);
    \draw (gu0)--(gu2);
    \draw (gu2)--(gu1);
    \draw (gl4)--(gu0);
    \draw (gl4)--(gu1);
    \draw (gl4)--(gu2);
    \node[] at (2.4,0.25) {$n$};
    \node[] at (2.55,0.05) {$n$};
    \node[] at (2.4,-0.25) {$n$};
    \node[] at (3.12,0.25) {$nk$};
    \node[] at (2.85,-0.15) {$nk$};
    \node[] at (3.12,-0.25) {$nk$};
    \node[] at (1,-0.5) {$\widetilde H_{\text{free}} = \frac12m(n-1)(k-1)$};
    \node[] at (1,-0.7) {hypermultiplets};
    \end{tikzpicture}
    \caption{The 3d $\mathcal{N}=4$ quiver theory that is mirror dual to the circle reduction of the type I Argyres--Douglas $\ADAI{p}{N}{[1^N]}$ theory for $p>N$. The number of $\mathfrak{u}(1)$ gauge nodes in the complete graph is in general $m$, but for simplicity in the drawing we depict the case of $m=3$.}
    \label{fig:3dmirrIkp}
\end{figure}

Specialising to the case $b=N$, the 3d mirror for the case of maximal regular puncture is given in Figure \ref{fig:3dmirrIkp}. This case was discussed in \cite{Giacomelli:2020ryy} and can be obtained as a specialisation of the one in Figure \ref{fig:3dmirrnkp}. In this case Tail B is trivial and the symmetry is reduced to $\mathfrak{su}(N)\oplus\mathfrak{u}(1)^{m-1}$, again matching expectations from four dimensions.

Next we turn to the 3d mirrors of the $\ADAN{p}{b}{N}{[1^N]}$ theories with $p<N$. We first define $q=n+k$ and two positive integers $\mu,\nu$ such that
\begin{equation}
    n = \mu q + \nu~,\qquad 0\leqslant \nu < q~.
\end{equation}
The 3d mirror can be constructed as follows:
\begin{enumerate}
    \item Start with a complete graph with $m$ vertices with a $\mathfrak{u}(1)$ gauge node at each vertex such that each edge has multiplicity $\nu(q-\nu)$.
    \item Construct the Tail A
    \begin{equation}
    \begin{tikzpicture}[scale=1.2,every node/.style={scale=1.2},font=\scriptsize]
    \node[gauge] (g0) at (0,0) {$N-\mu-2$};
    \node (g1) at (1.5,0) {$\cdots$};
    \node[gauge] (g2) at (3,0) {$\,2\,$};
    \node[gauge] (g3) at (4.5,0) {$\,1\,$};
    \node[gauge] (f0) at (-2.5,0) {$N-\mu-1$};
    \draw (g0)--(g1)--(g2)--(g3);
    \draw (g0)--(f0);
    \end{tikzpicture}
    \end{equation}
    and connect the $\mathfrak{u}(N-\mu-1)$ gauge node to every node in the complete graph by an edge with multiplicity $\nu$.
    \item Construct another Tail B
    \begin{equation}
    \begin{tikzpicture}[scale=1.4,every node/.style={scale=1},font=\scriptsize]
    \node[gauge] (g0) at (0,0) {$\mu(p-1)+N-nm$};
    \node[gauge] (g1) at (3,0) {$(\mu-1)(p-1)+N-nm$};
    \node (g2) at (5,0) {$\cdots$};
    \node[gauge] (g3) at (7,0) {$(p-1)+N-nm$};
    \node[gauge] (m2) at (7,-0.8) {$N-nm$};
    \node[gauge] (m3) at (5,-0.8) {$N-nm-1$};
    \node (m4) at (3.5,-0.8) {$\cdots$};
    \node[gauge] (m5) at (2.5,-0.8) {$\,2\,$};
    \node[gauge] (m6) at (1.5,-0.8) {$\,1\,$};
    \draw (g0)--(g1)--(g2)--(g3)--(m2)--(m3)--(m4)--(m5)--(m6);
    \end{tikzpicture}
    \end{equation}
    and connect the $\mathfrak{u}\left(\mu(p-1)+N-nm\right)$ gauge node to every node in the complete graph by an edge with multiplicity $q-\nu$.
    \item Finally, connect the the two tails by joining the $\mathfrak{u}(N-\nu-1)$ gauge node of Tail A to the $\mathfrak{u}\left(\mu(p-1)+N-nm\right)$ gauge node of Tail B by an edge with multiplicity one.
    \item Include an additional set of $\widetilde H_{\text{free}}=  \frac12m(\nu-1)(q-\nu-1)$ free hypermultiplets.
\end{enumerate}
\begin{figure}
    \centering
        \begin{tikzpicture}[scale=1.6,every node/.style={scale=1.2},font=\scriptsize]
    \node[gauge] (gl0) at (0,0) {$1$};
    \node[gauge] (gl1) at (0.5,0.5) {$2$};
    \node[] (gl2) at (1,1) {$\cdots$};
    \node[gauge] (gl3) at (1.5,1.5) {\tiny$N-\mu-2$};
    \node[gauge] (gl4) at (2.25,2.25) {\tiny$N-\mu-1$};
    \node[gauge] (gr4) at (4.25+0.5,2.25) {\tiny$\mu(p-1)+N-nm$};
    \node[gauge] (gr3) at (5+0.5,1.5) {\tiny$(\mu-1)(p-1)+N-nm$};
    \node[] (gr2) at (5.5+0.5,1) {$\cdots$};
    \node[gauge] (gr1) at (6+0.5,0.5) {\tiny$p-1+N-nm$};
    \node[gauge] (gr0) at (6.5+0.5,0) {\tiny$N-nm$};
    \node[gauge] (grm1) at (7+0.5,-0.5) {\tiny$N-nm-1$};
    \node[] (grm2) at (7.5+0.5,-1) {$\cdots$};
    \node[gauge] (grm3) at (8+0.5,-1.5) {$2$};
    \node[gauge] (grm4) at (8.5+0.5,-2) {$1$};
    \node[gauge] (gu0) at (2.25,3.25+0.5) {$1$};
    \node[gauge] (gu1) at (4.25+0.5,3.25+0.5) {$1$};
    \node[gauge] (gu2) at (3.25+0.25,4.25+0.5) {$1$};
    \draw (gl0)--(gl1)--(gl2)--(gl3)--(gl4)--(gr4)--(gr3)--(gr2)--(gr1)--(gr0)--(grm1)--(grm2)--(grm3)--(grm4);
    \draw (gu0)--(gu1);
    \draw (gu0)--(gu2);
    \draw (gu2)--(gu1);
    \draw (gl4)--(gu0);
    \draw (gl4)--(gu1);
    \draw (gl4)--(gu2);
    \draw (gr4)--(gu0);
    \draw (gr4)--(gu1);
    \draw (gr4)--(gu2);
    \node[] at (2.1,2.65) {\tiny$\nu$};
    \node[] at (2.6,2.6) {\tiny$\nu$};
    \node[] at (2.85,2.4) {\tiny$\nu$};
    \node[] at (4.7+0.4,2.65+0.1) {\tiny$q-\nu$};
    \node[] at (4.15,2.5+0.4) {\tiny$q-\nu$};
    \node[] at (3.4+0.4,2.475+0.05) {\tiny$q-\nu$};
    \node[] at (2.3,3.85+0.5) {\tiny$\nu(q-\nu)$};
    \node[] at (4.3+0.4,3.85+0.5) {\tiny$\nu(q-\nu)$};
    \node[] at (3.25+0.25,3.1+0.5) {\tiny$\nu(q-\nu)$};
    \node[] at (3.25+0.25,-0.5) {$\widetilde H_{\text{free}} = \frac12m(\nu-1)(q-\nu-1)$};
    \node[] at (3.25+0.25,-0.8) {hypermultiplets};
\end{tikzpicture}
    \caption{The 3d $\mathcal{N}=4$ quiver theory that is mirror dual to the circle reduction of the $\ADAN{p}{b}{N}{[1^N]}$ Argyres--Douglas theory for $p<b$. The number of $\mathfrak{u}(1)$ gauge nodes in the complete graph is in general $m$, but for simplicity in the drawing we depict the case of $m=3$.}
    \label{fig:3dmirrnkn}
\end{figure}

The general 3d mirror quiver of this form is as in the Figure \ref{fig:3dmirrnkn}. Once again, all the nodes in Tail A are balanced as the node connected to the complete graph has $N-\mu-2+m\times\nu+\mu(mq-1)+N-nm=2(N-\mu-1)$ fundamental hypermultiplets. On the other hand for Tail B, all the nodes in the tail starting from the $\mathfrak{u}(N-nm-1)$ gauge node are balanced. This immediately gives rise to an enhanced symmetry $\mathfrak{su}(N-nm)$, as expected from four dimensions. Although the next $\mathfrak{u}(N-nm)$ gauge node is unbalanced, the adjacent $\mathfrak{u}\left(p-1+N-nm\right)$ node is balanced which can be easily checked by observing that $N-nm+2(p-1)+N-nm = 2(p-1+N-nm)$. All the subsequent nodes of Tail B are similarly balanced, including the gauge node that connects to the complete graph, since
\begin{equation}
    \begin{split}
        (\mu-1)(p-1)+N-nm+N-\mu-1+m(q-\nu) &= \mu(p-1)+2N-nm-m\nu-1 \\
        &= 2(\mu(p-1)+N-nm)~.
    \end{split}
\end{equation}
As this sequence of $\mu$ balanced gauge nodes is connected to the sequence of $N-\mu-1$ balanced gauge nodes of Tail A, they together give rise to an enhanced $\mathfrak{su}(N)$ symmetry. This matches the four-dimensional flavour symmetry coming from the regular puncture $[Y]=[1^N]$, although unlike the case with $p\geqslant N$ this symmetry is distributed over the two tails. Because of this, studying the 3d mirror of the theories labelled with a different partition $[Y]$ can be slightly more complicated than in the case $p<N$. (We discuss the required procedure later.) Finally, the remaining unbalanced nodes give rise to the $\mathfrak{u}(1)^m$ symmetry.

\begin{figure}
    \centering
        \begin{tikzpicture}[scale=1.9,every node/.style={scale=1.35},font=\scriptsize]
    \node[gauge] (gl0) at (0.55,0) {$1$};
    \node[gauge] (gl1) at (0.95,0.5) {$2$};
    \node[] (gl2) at (1.35,1) {$\cdots$};
    \node[gauge] (gl3) at (1.75,1.5) {\tiny$N-\mu-2$};
    \node[gauge] (gl4) at (2.25,2.25) {\tiny$N-\mu-1$};
    \node[gauge] (gr4) at (4.25+0.5,2.25) {\tiny$\mu(p-1)$};
    \node[gauge] (gr3) at (5+0.5,1.5) {\tiny$(\mu-1)(p-1)$};
    \node[] (gr2) at (5.5+0.5,1) {$\cdots$};
    \node[gauge] (gr1) at (6+0.5,0.5) {\tiny$2(p-1)$};
    \node[gauge] (gr0) at (6.5+0.5,0) {\tiny$p-1$};
    \node[gauge] (gu0) at (2.25,3.25+0.5) {$1$};
    \node[gauge] (gu1) at (4.25+0.5,3.25+0.5) {$1$};
    \node[gauge] (gu2) at (3.25+0.25,4.25+0.5) {$1$};
    \draw (gl0)--(gl1)--(gl2)--(gl3)--(gl4)--(gr4)--(gr3)--(gr2)--(gr1)--(gr0);
    \draw (gu0)--(gu1);
    \draw (gu0)--(gu2);
    \draw (gu2)--(gu1);
    \draw (gl4)--(gu0);
    \draw (gl4)--(gu1);
    \draw (gl4)--(gu2);
    \draw (gr4)--(gu0);
    \draw (gr4)--(gu1);
    \draw (gr4)--(gu2);
    \node[] at (2.1,2.65) {\tiny$\nu$};
    \node[] at (2.6,2.6) {\tiny$\nu$};
    \node[] at (2.85,2.4) {\tiny$\nu$};
    \node[] at (4.7+0.4,2.65+0.1) {\tiny$q-\nu$};
    \node[] at (4.15,2.5+0.4) {\tiny$q-\nu$};
    \node[] at (3.4+0.5,2.475+0.05) {\tiny$q-\nu$};
    \node[] at (2.3,3.85+0.5) {\tiny$\nu(q-\nu)$};
    \node[] at (4.2+0.5,3.85+0.5) {\tiny$\nu(q-\nu)$};
    \node[] at (3.25+0.25,3.1+0.5) {\tiny$\nu(q-\nu)$};
    \node[] at (3.25+0.25,-0.6) {$\widetilde H_{\text{free}} = \frac12m(\nu-1)(q-\nu-1)$};
    \node[] at (3.25+0.25,-0.9) {hypermultiplets};
    \end{tikzpicture}
    \caption{The 3d $\mathcal{N}=4$ quiver theory that is mirror dual to the circle reduction of the $\ADAI{p}{N}{[1^N]}$ Argyres--Douglas theory for $p<N$. The number of $\mathfrak{u}(1)$ gauge nodes in the complete graph is in general $m$, but for simplicity in the drawing we depict the case of $m=3$.}
    \label{fig:3dmirrIkn}
\end{figure}

For the type I case $b=N$ with $[Y]=[1^N]$ and $p<N$, we represent the three-dimensional mirror quiver in Figure \ref{fig:3dmirrIkn} \cite{Giacomelli:2020ryy}. The first part of Tail B became trivial and the only surviving part is what combines with Tail A to give the $\mathfrak{su}(N)$ symmetry. The full symmetry of the quiver is then $\mathfrak{su}(N)\oplus\mathfrak{u}(1)^{m-1}$, again as expected from four dimensions.

\subsubsection{\label{subsubsec:ADI3d}3d reduction of Type I theories}

In general, there is no expectation that the direct 3d reduction of a non-Lagrangian theory in four dimensions will admit a (non-mirror) dual Lagrangian description. (Equivalently, the three-dimensional mirror may not admit a Lagrangian mirror dual.) Indeed, for most regular class $\mathcal{S}$ theories no such Lagrangian realisation is expected, even though a Lagrangian 3d mirror is known \cite{Benini:2010uu}. Nevertheless, for type I and type II theories the 3d reduction does indeed admit a (non-mirror) dual Lagrangian description \cite{Closset:2020afy,Giacomelli:2020ryy}.\footnote{See also \cite{Dey:2021rxw} for an alternative Lagrangian description.} Here we recall these Lagrangians in type I.

For convenience, following \cite{Giacomelli:2020ryy}, we define the following parameters (recalling also \eqref{eq:bmnpq}):
\begin{equation}\label{eq:par3dred}
    x=\bigg\lfloor\frac{N}{p}\bigg\rfloor\,,\quad m_A=q(x+1)-n\,,\quad m_B=n-qx~.
\end{equation}
We also treat the cases $p>N$ and $p<N$ separately. For $p=N$ the theory is Lagrangian already in four dimensions and coincides with the specialisation of both cases.

\medskip

\paragraph{$b=N$ and $p\geqslant N$.} The quiver in this case is as follows:
\begin{equation}\label{3dbNkg0}
\begin{tikzpicture}[scale=3.5,every node/.style={scale=1},font=\scriptsize]
    \node[flavor] (p0) at (0.2,0) {$N$};
    \node[gauge] (p1) at (0.5,0) {$N-1$};
    \node[gauge] (p2) at (1,0) {$N-2$};
    \node[] (p3) at (1.4,0) {$\cdots$};
    \node[gauge] (p4) at (1.8,0) {$N-n+1$};
    \node[gauge] (p5) at (2.3,0) {$N-n$};
    \node[] (p6) at (2.7,0) {$\cdots$};
    \node[] at (2.7,-0.2) {$\underbrace{\hspace{13em}}_{\displaystyle q-n}$};
    \node[gauge] (p7) at (3.1,0) {$N-n$};
    \node[gauge] (p8) at (3.6,0) {$\mathfrak{su}(N-n)$};
    \draw (p0)--(p1)--(p2)--(p3)--(p4)--(p5)--(p6)--(p7)--(p8);

    \node[gauge] (m1) at (3.6,-0.5) {$N-n-1$};
    \node[gauge] (m2) at (3.1,-0.5) {$N-n-2$};
    \node[] (m3) at (2.7,-0.5) {$\cdots$};
    \node[gauge] (m4) at (2.3,-0.5) {$N-2n+1$};
    \node[gauge] (m5) at (1.8,-0.5) {$N-2n$};
    \node[] (m6) at (1.4,-0.5) {$\cdots$};
    \node[] at (1.4,-0.7) {$\underbrace{\hspace{13em}}_{\displaystyle q-n}$};
    \node[gauge] (m7) at (1,-0.5) {$N-2n$};
    \node[gauge] (m8) at (0.5,-0.5) {$\mathfrak{su}(N-2n)$};
    \draw (m1)--(m2)--(m3)--(m4)--(m5)--(m6)--(m7)--(m8);
    \draw (m1)--(p8);

    \node[] (c) at (0.5,-0.8) {$\vdots$};
    \draw (c)--(m8);
    
    \node[gauge] (l0) at (0.5,-1.1) {$\mathfrak{su}(2n)$};
    \node[gauge] (l1) at (1,-1.1) {$2n-1$};
    \node[gauge] (l2) at (1.5,-1.1) {$2n-2$};
    \node[] (l3) at (1.9,-1.1) {$\cdots$};
    \node[gauge] (l4) at (2.3,-1.1) {$n+1$};
    \node[gauge] (l5) at (2.8,-1.1) {$n$};
    \node[] (l6) at (3.2,-1.1) {$\cdots$};
    \node[] at (3.2,-0.9) {$\overbrace{\hspace{11em}}^{\displaystyle q-n}$};
    \node[gauge] (l7) at (3.6,-1.1) {$n$};
    \draw (l0)--(l1)--(l2)--(l3)--(l4)--(l5)--(l6)--(l7);
    \draw (c)--(l0);

    \node[gauge] (t0) at (3.6,-1.6) {$\mathfrak{su}(n)$};
    \node[gauge] (t1) at (3.1,-1.6) {$n-1$};
    \node[gauge] (t2) at (2.6,-1.6) {$n-2$};
    \node[] (t3) at (2.2,-1.6) {$\cdots$};
    \node[gauge] (t4) at (1.8,-1.6) {$2$};
    \node[gauge] (t5) at (1.3,-1.6) {$1$};
    \draw (t0)--(t1)--(t2)--(t3)--(t4)--(t5);
    \draw (t0)--(l7);
    
\end{tikzpicture}
\end{equation}
and there are an additional $\tilde{H}_{\text{free}} = \frac{1}{2}m(n-1)(q-n-1)$ free twisted hypermultiplets. We are now specifying some nodes corresponding to special unitary gauge algebras, as opposed to the usual unitary ones.

In the coprime case $m=1$, the 4d theory has no exactly marginal deformations. This manifests as the absence of special unitary gauge nodes in the 3d quiver, and we end up with just the usual quiver for the $T(\mathfrak{su}(N))$ theory,

\begin{equation}\label{3dbNkg0GCD1}
\begin{tikzpicture}[scale=1.2,every node/.style={scale=1.2},font=\scriptsize]
    \node[gauge] (g0) at (0,0) {$N-1$};
    \node (g1) at (1.5,0) {$\cdots$};
    \node[gauge] (g2) at (3,0) {$\,2\,$};
    \node[gauge] (g3) at (4.5,0) {$\,1\,$};
    \node[flavor] (f0) at (-1.5,0) {$\,N\,$};
    \draw (g0)--(g1)--(g2)--(g3);
    \draw (g0)--(f0);
\end{tikzpicture}
\end{equation}

with an additional $\tilde{H}_{\text{free}}$ free twisted hypermultiplets. The $p$ dependence at this level is completely encoded in the number of free twisted hypers.

\medskip

\paragraph{$b=N$ and $p\leqslant N$.} The quiver in this case is as follows:\footnote{If $p=N$, the final node is $\mathfrak{su}(1)$ rather than $\mathfrak{u}(1)$. In this way we recover the same quiver as is given in \eqref{3dbNkg0} for $p=N$.}
\begin{equation}\label{3dbNkl0}
\resizebox{\textwidth}{!}{
\begin{tikzpicture}[scale=5.4,every node/.style={scale=0.95},font=\scriptsize]
    \node[flavor] (p0) at (0.2,0) {$N$};
    \node[gauge] (p1) at (0.5,0) {$N-(x+1)$};
    \node[gauge] (p2) at (1,0) {$N-2(x+1)$};
    \node[] (p3) at (1.4,0) {$\cdots$};
    \node[gauge] (p4) at (1.8,0) {$N-m_B(x+1)$};
    \node (p5) at (2.3,0) {$\Xi[N-n]$};
    \node[gauge] (p6) at (2.8,0) {$\mathfrak{su}(N-n)$};
    \draw (p0)--(p1)--(p2)--(p3)--(p4)--(p5)--(p6);

    \node[gauge] (m1) at (2.8,-0.3) {$N-n-(x+1)$};
    \node[gauge] (m2) at (2.3,-0.3) {$N-n-2(x+1)$};
    \node[] (m3) at (1.9,-0.3) {$\cdots$};
    \node[gauge] (m4) at (1.5,-0.3) {$N-n-m_B(x+1)$};
    \node[] (m5) at (1,-0.3) {$\Xi[N-2n]$};
    \node[gauge] (m6) at (0.5,-0.3) {$\mathfrak{su}(N-2n)$};
    \draw (m1)--(m2)--(m3)--(m4)--(m5)--(m6);
    \draw (m1)--(p6);

    \node[] (c) at (0.5,-0.45) {$\vdots$};
    \draw (c)--(m6);
    
    \node[gauge] (l0) at (0.5,-0.6) {$\mathfrak{su}(2n)$};
    \node[gauge] (l1) at (1,-0.6) {$2n-(x+1)$};
    \node[gauge] (l2) at (1.5,-0.6) {$2n-2(x+1)$};
    \node[] (l3) at (1.9,-0.6) {$\cdots$};
    \node[gauge] (l4) at (2.3,-0.6) {$2n-m_B(x+1)$};
    \node[] (l5) at (2.8,-0.6) {$\Xi[n]$};
    \draw (l0)--(l1)--(l2)--(l3)--(l4)--(l5);
    \draw (c)--(l0);

    \node[gauge] (t0) at (2.8,-0.9) {$\mathfrak{su}(n)$};
    \node[gauge] (t1) at (2.3,-0.9) {$n-(x+1)$};
    \node[gauge] (t2) at (1.8,-0.9) {$n-2(x+1)$};
    \node[] (t3) at (1.4,-0.9) {$\cdots$};
    \node[gauge] (t4) at (1,-0.9) {$n-m_B(x+1)$};
    \node (t5) at (0.5,-0.9) {$\Xi[0]$};
    \draw (t0)--(t1)--(t2)--(t3)--(t4)--(t5);
    \draw (t0)--(l5);
\end{tikzpicture}
}
\end{equation}
with an additional $\tilde{H}_{\text{free}} = \frac{1}{2}m (n-qx-1)(q(1+x)-n -1)$ free twisted hypermultiplets. Here $\Xi[Y]$ indicates the following sequence of unitary gauge nodes:
    \begin{equation}
    \begin{tikzpicture}[scale=1.2,every node/.style={scale=1.2},font=\scriptsize]
    \node[gauge] (g0) at (0,0) {$Y+(m_A-2) x $};
    \node (g1) at (1.5,0) {$\cdots$};
    \node[gauge] (g2) at (3,0) {$Y+2x$};
    \node[gauge] (g3) at (4.5,0) {$Y+x$};
    \node[gauge] (f0) at (-2.5,0) {$Y+(m_A-1) x $};
    \draw (g0)--(g1)--(g2)--(g3);
    \draw (g0)--(f0);
    \end{tikzpicture}
    \end{equation}

It is again illustrative to consider the case $m=1$, where again the four-dimensional theory has no exactly marginal deformations and the 3d quiver has no special unitary gauge nodes,
\begin{equation}\label{3dbNkl0GCD1}
\begin{tikzpicture}[scale=1.2,every node/.style={scale=1},font=\scriptsize]
    \node[gauge] (g0) at (0,0) {$N-(x+1)$};
    \node (g1) at (1.4,0) {$\cdots$};
    \node[gauge] (g2) at (3,0) {$N-m_B(x+1)$};
    \node[gauge] (g3) at (5,0) {$(m_A-1)x$};
    \node[gauge] (g4) at (6.7,0) {$(m_A-2)x$};
    \node (g5) at (8,0) {$\cdots$};
    \node[gauge] (g6) at (8.9,0) {$\,x\,$};
    \node[flavor] (f0) at (-1.5,0) {$\,N\,$};
    \draw (g0)--(g1)--(g2)--(g3)--(g4)--(g5)--(g6);
    \draw (g0)--(f0);
\end{tikzpicture}
\end{equation}
This coincides with the $T^\sigma_\rho(\mathfrak{su}(N))$ theory \cite{Gaiotto:2008ak} (see Appendix \ref{app:Trhosigma} for a review) with partitions specified as follows,
\begin{equation}\label{eq:rhoIkl0}
    \rho=[(x+1)^{m_B},x^{m_A}],\qquad \sigma=[1^N]~,
\end{equation}
with an additional $\tilde{H}_{\text{free}}$ free twisted hypermultiplets. 

Observe that from the 3d reduction of a generic type I theory with $m=1$ we do not find general $T^\sigma_\rho(\mathfrak{su}(N))$ theories, but only those with $\rho$ of the form $[1^N]$ or that of \eqref{eq:rhoIkl0}.

\subsection{\label{subsec:ADVOA}VOAs and Higgs branches in type I}

For any 4d $\mathcal N=2$ SCFT, $\mathcal T$, we denote the associated vertex operator algebra (VOA) by $\mathbb{V}(\mathcal T)$ \cite{Beem:2013sza}. We recall that for any continuous global symmetry (algebra) $\mathfrak g$ of $\mathcal{T}$ with flavour central charge $k_{4d}$, the associated VOA includes an affine current subalgebra $V^{k_{2d}}(\mathfrak g)\to\mathbb{V}(\mathcal{T})$ where $k_{2d} = -\frac{1}{2}k_{4d}$. Furthermore, the \emph{associated variety} $X_{\mathbb{V}(\mathcal T)}$ is conjecturally identified with the Higgs branch of $\mathcal T$ \cite{Beem:2017ooy}. Here we review some basic facts about this 4d/2d dictionary in the case of coprime type I Argyres--Douglas theories.

There is a simple and universal characterisation of the VOAs associated to type I theories with coprime values of $N$ and $p$ (so in particular $p=q=N+k$ and $\gcd(N,k)=1$). For these theories, the associated VOAs are described more or less entirely in terms of their flavour symmetries. Indeed, for such theories with maximal regular puncture the associated VOAs have been identified as simple affine Kac--Moody VOAs at boundary admissible level \cite{Xie:2016evu} (see also \cite{Buican:2015ina,Buican:2015hsa,Cordova:2015nma,Buican:2015tda,Buican:2017uka} for results using the Schur limit of the superconformal index),\footnote{An admissible level for a simply laced Lie algebra is a rational number of the form $k=-h^\vee + \frac{p}{q}$ with $\gcd(p,q)=1$ and $p\geqslant h^\vee$. Boundary admissible levels are then the levels with $p=h^\vee$, as is the case for these examples.}
\begin{equation}
    \mathbb{V}\left[\ADAIF{q}{N}\right] = V_{k_{2d}}(\mathfrak{sl}_N) \;,\quad k_{2d} = -N+\frac{N}{q}~.
\end{equation}
The associated varieties of (simple) affine current algebras at admissible levels are the closures of certain nilpotent (coadjoint) orbits in $\mathfrak{sl}_N^\ast$, which are determined by the parameter $q$ appearing in the level \cite{Arakawa2015associated}. In particular, for $q\geqslant N$ the associated variety is given by the closure of the principal nilpotent orbit (the full nilpotent cone),
\begin{equation}
    X_{V_{k_{2d}}(\mathfrak{sl}_N)} = \overline{\mathbb O_{\text{prin}}} = \mathcal{N}_{\mathfrak{sl}_N}~.
\end{equation}
This coincides with the Higgs branch of the $T(\mathfrak{su}(N))$ theory \eqref{3dbNkg0GCD1} that arises upon circle reduction to three dimensions (see Appendix \ref{app:Trhosigma}). 

Alternatively, for $q<N$ the associated variety is given a different nilpotent orbit closure,
\begin{equation}\label{eq:MqlN}
    X_{V_{k_{2d}}(\mathfrak{sl}_N)} = \overline{\mathbb O_q}~,
\end{equation}
where we denote by $\mathbb O_q$ the orbit corresponding to the partition $[q^x,s]$ of $N$ with $s<q$.\footnote{We connect with the notation of \eqref{eq:par3dred} for $\gcd(N,p)=1$ by writing $N=xq+m_B$, from which it is clear that $s=m_B$.} This is the transpose of the partition $\rho$ in \eqref{eq:rhoIkl0}, so of course \eqref{eq:MqlN} coincides with the Higgs branch of the 3d reduction \eqref{3dbNkl0GCD1} (see Appendix \ref{app:Trhosigma}).

For theories with a partially reduced regular puncture, the corresponding VOAs are obtained by the relevant quantised Drinfel'd--Sokolov reduction of the corresponding simple affine vertex algebra \cite{Beem:2014rza}, which leads to the identification
\begin{equation}
    \mathbb{V}\left[\ADAI{q}{N}{[Y]}\right] = \mathcal W_{k_{2d}}(\mathfrak{sl}_N,\mathfrak f_Y) \;,\quad k_{2d} = -N+\frac{N}{q}~,
\end{equation}
where $\mathcal W_{k_{2d}}(\mathfrak{sl}_N,\mathfrak f_Y)$ is a (generalised) simple affine $\mathcal W$-algebra. By virtue of this description via Drinfel'd--Sokolov reduction, the associated variety is the intersection of the Slodowy slice $\mathcal S_{\mathfrak f_Y}$ transverse to $\mathfrak f_Y$ with the associated variety of $V_{k_{2d}}(\mathfrak{sl}_N)$ \cite[Cor 6.2]{Arakawa:2018egx}, \emph{i.e.},
\begin{equation}
    X_{\mathcal W_{k_{2d}}(\mathfrak{sl}_N,\mathfrak f_Y)} = X_{V_{k_{2d}}(\mathfrak{sl}_N)} \;\cap\; \mathcal S_{\mathfrak f_Y}~.
\end{equation}
Consequently, for $q>N$, the Higgs branch of $\ADAI{q}{N}{[N]}$ is a point (so the theory is \emph{un-Higgsable}). This implies that the associated vertex operator algebra---which is the principal $\mathcal W$-algebra in this case---is $C_2$ co-finite. When $q=N+1$ these turn out to be trivial theories (with $c=0$), but generally they are non-trivial interacting theories.

For $q<N$, on the other hand, the Higgs branch of $\ADAI{q}{N}{[q^x,s]}$ will be zero-dimensional and the associated non-principal affine $\mathcal W$-algebra will be $C_2$ co-finite. These un-Higgsable theories will be empty theories if $s=1$ or $s=q-1$, otherwise they will be nontrivial and interacting. Furthermore, since additional Higgsing is impossible, a nilpotent Higgsing of $\ADAIF{q}{N}$ by $[Y]$ that is larger than (in the sense of corresponding to a nilpotent orbit that is not contained in the closure of that of) $[q^x,s]$ is not possible.

We comment here in passing that also in the non-coprime case $\gcd(N,p)\neq 1$, for $p<N$ the Argyres--Douglas theory $\ADAI{p}{N}{[p^x,s]}$ with $s<N$ is again un-Higgsable. We return to derive this point in Subsection \ref{subsec:typeIEMC}, where it will be a consequence of the same property in the coprime case along with the general result \eqref{eq:typeIwEMD}. Unlike the coprime case, we do not have a simple VOA understanding of this fact. (Though for a subset of cases, the proposal of \cite{Creutzig:2018lbc} may shed some light on the issue.)

\section{\label{sec:typeI}Equivalences of type I AD theories}

In this section we investigate equivalences and relationships between type I Argyres--Douglas SCFTs. We initially restrict our attention to theories with $\gcd(N,p)=1$ (the \emph{coprime case}), so in particular these are theories with no $\mathcal N=2$ EMDs. Within this setting, we first recall several previously proposed identifications between superficially distinct AD theories. We interpret and provide further evidence for these identifications using three-dimensional techniques to relate the 3d $\mathcal N=4$ quiver gauge theories describing the circle reductions of the theories in question. We further conjecture and provide evidence for a more general set of identifications amongst coprime type I theories from which the aforementioned identifications are recovered as special cases. We find a simple and elegant presentation of two principal equivalences as well as provide a vertex algebraic perspective on these results.

In the second part, we consider the non-coprime case, including examples with $\mathcal N=2$ exactly marginal deformations. We propose a general realisation of these theories as a linear sequence of special unitary conformal gaugings of coprime type I Argyres--Douglas theories, sometimes with additional hypermultiplets charged under the gauge symmetry. We first analyse some explicit examples where we provide evidence for the claimed identification based on matching of Coulomb branch operators, central charges, and 3d reductions. Building on these examples, we then propose a general statement of equivalences. We provide a simple diagrammatic way to understand the somewhat elaborate general statement, for which we performed the aforementioned checks in a large number of cases. We then use a restriction of the general statement to provide additional equivalences involving non-coprime theories without EMDs. Finally, we propose the extensions of the two aforementioned primary equivalences to type I theories \emph{with} $\mathcal N=2$ exactly marginal deformations.

\subsection{\label{subsec:typeIisowoEMD}Equivalences in the coprime case}

Throughout this subsection we restrict to the case $m=\gcd(N,p)=1$ and so $q=p$---these are all theories without $\mathcal{N}=2$ exactly marginal deformations. Perhaps the best known instance of an equivalence between these type I theories is the identification of the $(A_{N-1},A_{k-1})$ and $(A_{k-1},A_{N-1})$ theories \cite{Cecotti:2010fi}. This is expressed in our notation as follows:
\begin{equation}\label{eq:iso1}
    \ADAI{q}{N}{[N]} \cong \ADAI{q}{k}{[k]}~.
\end{equation}
This equivalence is especially transparent in the geometric engineering construction of these theories where it amounts to a simple rewriting of the isolated hypersurface singularity that defines these Argyres--Douglas theories; it is a particular instance of the more general set of equivalences between $(G,G')$ and $(G',G)$ Argyres--Douglas theories constructed via geometric engineering \cite{Cecotti:2010fi}.

Another general family of equivalences takes the following form:
\begin{equation}\label{eq:isoql}
    \ADAI{q}{N}{\left[Y\right]} \cong \ADAI{q}{ql+N}{\left[q^l,Y\right]}~,\qquad l=0,1,2,\ldots~.
\end{equation}
To our knowledge, this equivalence first appeared in \cite{Xie:2019yds} for the case of full regular puncture $[Y]=[1^N]$. The evidence provided there was the matching of the flavour symmetry central charges and of the Coulomb branch spectrum. We provide additional evidence for this conjecture in the form of matching the 3d reductions of both the theories in \eqref{eq:isoql} and this perspective will allow us to generalise it. Although we are only considering type I theories without EMDs, a simple extension of this equivalence also holds true for type I theories with EMDs \eqref{eq:isopl}. We will explore this in Subsection \ref{subsec:typeIisowEMD}.

An immediate observation in this equivalence is that the na\"ive flavour symmetry algebras on the two sides do not agree. In particular, there is an extra factor of $\mathfrak{su}(l)$ on the right hand side since the irregular singularity on both sides contributes the same flavour symmetry. The crucial observation in \cite{Xie:2019yds} was that for $[Y]=[1^N]$, the flavour central charge associated to this subalgebra vanishes and hence it decouples from the total symmetry algebra.

For generic $[Y]=[N^{l_N},\dots,1^{l_1}]$, the flavour central charge of this subalgebra can be computed by considering the SCFT on the right in \eqref{eq:isoql} as the IR limit of the nilpotent Higgsing by $[q^l,Y]$, as discussed in Subsection \ref{subsec:ADflavourcc}. The flavour central charge of the UV symmetry is
\begin{equation}
    k_{\mathfrak{su}(ql+N)}^{\rm UV} = 2\frac{(q-1)(ql+N)}{q}~,
\end{equation}
while the contribution from the Nambu--Goldstone modes to the 4d flavour central charge of the infrared $\mathfrak{su}(l)$ symmetry from \eqref{eq:kGBcont} can be expressed as
\begin{equation}
    k_{\mathfrak{su}(l)}^{\rm NG} = -\sum_{i=1}^{ql+N} \frac{l_i}{2} \big(q+i-|q-i|\big)\big(q+i-2+|q-i|\big)~.
\end{equation}
Now given that the largest entry in the partition $[Y]$ is $q$ (see the discussion at the end of the previous section), we have that $l_i=0$ for all $i>q$. Therefore the above expression can be simplified to
\begin{equation}
    k_{\mathfrak{su}(l)}^{\rm NG} = -2(q-1)\sum_{i=1}^q i l_i = -2(q-1)(ql+N)~.
\end{equation}
It is then immediately clear that the flavour central charge of the $\mathfrak{su}(l)$ subalgebra vanishes,
\begin{equation}\label{eq:zeroflavcc}
    k_{\mathfrak{su}(l)}^{\rm IR} = q\,k_{\mathfrak{su}(ql+N)}^{\rm UV} + k_{\mathfrak{su}(l)}^{\rm NG} = 0~,
\end{equation}
where we have used that $I_{\mathfrak{su}(l)\hookrightarrow\mathfrak{su}(ql+N)}=q$.

This check of the equivalence \eqref{eq:isoql} can be further augmented by explicitly matching the $a$ and $c$ central charges, as well as the Coulomb branch spectrum. The strongest check however is that the 3d reductions of the two Argyres--Douglas SCFTs in \eqref{eq:isoql} match. 

We have already reviewed in Subsection \ref{subsubsec:ADI3d} that these 3d reductions are just linear quiver gauge theories with unitary and special unitary gauge nodes, where the number of special unitary gauge nodes is equal to $m-1$. Hence, for type I theories with $m=1$, the 3d reduction has only unitary gauge nodes and corresponds to some $T^\sigma_\rho(\mathfrak{su}(N))$ theory, which can be very simply realised by a brane setup with a number of D3--branes ending on NS5--branes with transverse D5--branes \cite{Hanany:1996ie,Gaiotto:2008ak} (see Appendix \ref{app:Trhosigma}). We observe that the brane setup of the 3d reductions of the type I theory on the r.h.s.\ of \eqref{eq:isoql} is identical to the brane setup of the 3d reduction of the theory on the l.h.s.\ of \eqref{eq:isoql} along with some decoupled five-branes. Specifically, for the case of the maximal puncture $[Y]=[1^N]$ these are $(q-N)$ NS5--branes and $l$ D5--branes for $q>N$ and only $l$ D5--branes for $q<N$. By decoupled we mean that they do not intersect with any D3--brane in the system and thus they do not provide any new degrees of freedom to the theory living on the D3--branes. Hence, the 3d theories described by the two brane systems are equivalent.

Let us see more explicitly how this works. We will assume that $[Y]=[1^N]$. This is because the associated $\mathfrak{su}(N)$ symmetry is manifest on both sides of the equivalence and so we can obtain the result for generic $[Y]$ by the same nilpotent Higgsing on both sides. Moreover, for simplicity we only present the case $q>N$, since the derivation for $q<N$ is conceptually analogous. In order to determine the 3d reduction and the associated brane setup for the theory $\ADAI{q}{ql+N}{[q^l,1^N]}$ on the r.h.s.\ of \eqref{eq:isoql}, we start from the one for the theory with the full regular puncture $\ADAIF{q}{ql+N}$ found in \cite{Closset:2020afy,Giacomelli:2020ryy} that we reviewed in Subsection \ref{subsubsec:ADI3d}. Using that for $q>N$,
\begin{equation}
    x=\left\lfloor\frac{ql+N}{q}\right\rfloor=\left\lfloor l+\frac{N}{q}\right\rfloor=l~,
\end{equation}
we have that the 3d reduction is given by the $T^\sigma_\rho[\mathfrak{su}(ql+N)]$ theory with $\sigma=[1^{ql+N}]$ and $\rho=[(l+1)^N,l^k]$. This implies that the 3d reduction of the theory $\ADAI{q}{ql+N}{[q^l,1^N]}$ we are interested in corresponds instead to
\begin{equation}
    T^\sigma_\rho[\mathfrak{su}(ql+N)]\,,\quad \sigma=\left[q^l,1^N\right]\,,\quad \rho=\left[(l+1)^N,l^k\right]~.
\end{equation}
The associated brane setup is given by a stack of $ql+N$ D3--branes stretched between $N+l$ D5--branes and $q=N+k$ NS5--branes, where the entries of $\sigma$ correspond to the net number of D3--branes ending on each D5--brane while the entries of $\rho$ are the net number of D3--branes ending on each NS5--brane
\begin{equation}\hspace{-0.35cm}
\resizebox{\textwidth}{!}{
        \begin{tikzpicture}[scale=0.85,every node/.style={scale=1},font=\scriptsize]
        \draw[thick,red] (0,2)--(0,0);
        \draw[thick,red] (1,2)--(1,0);
        \draw[thick,red] (2,2)--(2,0);
        \draw[thick,red] (3,2)--(3,0);
        \draw[thick,red] (4,2)--(4,0);
        \draw[thick,red] (5,2)--(5,0);
        \draw[thick,red] (6,2)--(6,0);
        \draw[thick,red] (7,2)--(7,0);
        \draw[thick,red] (8,2)--(8,0);
        \draw[thick,red] (9,2)--(9,0);
        \node at (-1,1) {\color{blue}\small $\bm\otimes$};
        \node at (-2,1) {\color{blue}\small $\bm\otimes$};
        \node at (-3,1) {\color{blue}\small $\bm\otimes$};
        \node at (-4,1) {\color{blue}\small $\bm\otimes$};
        \node at (-5,1) {\color{blue}\small $\bm\otimes$};
        \node at (-6,1) {\color{blue}\small $\bm\otimes$};
        \node at (-7,1) {\color{blue}\small $\bm\otimes$};
        \node at (-8,1) {\color{blue}\small $\bm\otimes$};
        \node at (-9,1) {\color{blue}\small $\bm\otimes$};
        \draw (-9,1)--(-7,1);
        \node at (-6.5,1) {$\cdots$};
        \draw (-6,1)--(-3,1);
        \node at (-2.5,1) {$\cdots$};
        \draw (-2,1)--(1,1);
        \node at (1.5,1) {$\cdots$};
        \draw (2,1)--(6,1);
        \node at (6.5,1) {$\cdots$};
        \draw (7,1)--(9,1);
        \node at (0.5,0.7) {\tiny$(N\text{-}1)$};
        \node at (0.5,0.4) {\tiny$(l\text{+}1)$};
        \node at (0.5,0.1) {\tiny$\text{+}kl$};
        \node at (2.5,0.7) {\tiny$2(l\text{+}1)$};
        \node at (2.5,0.4) {\tiny$\text{+}kl$};
        \node at (3.5,0.7) {\tiny$l\text{+}1$};
        \node at (3.5,0.4) {\tiny$\text{+}kl$};
        \node at (4.5,0.7) {\tiny$kl$};
        \node at (5.5,0.7) {\tiny$(k\text{-}1)l$};
        \node at (7.5,0.7) {\tiny$2l$};
        \node at (8.5,0.7) {\tiny$l$};
        \node at (-0.5,0.7) {\tiny$ql\text{+}N$};
        \node at (-1.5,0.7) {\tiny$q(l\text{-}1)$};
        \node at (-1.5,0.4) {\tiny$\text{+}N$};
        \node at (-3.5,0.7) {\tiny$q\text{+}N$};
        \node at (-4.5,0.7) {\tiny$N$};
        \node at (-5.5,0.7) {\tiny$N\text{-}1$};
        \node at (-7.5,0.7) {\tiny$2$};
        \node at (-8.5,0.7) {\tiny$1$};
    \end{tikzpicture}
    }
\end{equation}
In the drawing, each red vertical line represents an NS5--brane, each blue circled cross represents a D5--brane and each black horizontal line stretched between pairs of five-branes represents a set of D3--branes, whose multiplicity is specified by the number below them.

We now move the $l$ D5--branes, on which $q$ D3--branes end, to the very right. This can be done using a sequence of Hanany--Witten moves \cite{Hanany:1996ie}, according to which every time a D5--brane crosses an NS5--brane, one D3--brane stretched between them gets annihilated. Hence, we see that we can move these $l$ D5--branes across all the NS5--branes and reach a configuration where there is no D3--brane ending on them anymore. The application of this sequence of Hanany--Witten moves also leaves no D3--brane attached to the $k=q-N$ NS5--branes on the very right
\begin{equation}\hspace{-0.3cm}
        \begin{tikzpicture}[scale=1.1,every node/.style={scale=1.3},font=\scriptsize]
        \node at (5.5,-0.2) {$\underbrace{\color{white}AAAAA}_{k}$};
        \node at (7.5,0.7) {$\underbrace{\color{white}AAAAA}_{l}$};
        \draw[thick,red] (0,2)--(0,0);
        \draw[thick,red] (1,2)--(1,0);
        \draw[thick,red] (2,2)--(2,0);
        \draw[thick,red] (3,2)--(3,0);
        \draw[thick,red] (4,2)--(4,0);
        \draw[thick,red] (5,2)--(5,0);
        \draw[thick,red] (6,2)--(6,0);
        \node at (-1,1) {\color{blue}\small $\bm\otimes$};
        \node at (-2,1) {\color{blue}\small $\bm\otimes$};
        \node at (-3,1) {\color{blue}\small $\bm\otimes$};
        \node at (-4,1) {\color{blue}\small $\bm\otimes$};
        \node at (-5,1) {\color{blue}\small $\bm\otimes$};
        \node at (7,1) {\color{blue}\small $\bm\otimes$};
        \node at (8,1) {\color{blue}\small $\bm\otimes$};
        \draw (-5,1)--(-3,1);
        \node at (-2.5,1) {$\cdots$};
        \draw (-2,1)--(1,1);
        \node at (1.5,1) {$\cdots$};
        \draw (2,1)--(4,1);
        \node at (5.5,1) {$\cdots$};
        \node at (7.5,1) {$\cdots$};
        \node at (0.5,0.7) {\tiny$N\text{-}1$};
        \node at (2.5,0.7) {\tiny$2$};
        \node at (3.5,0.7) {\tiny$1$};
        \node at (-0.5,0.7) {\tiny$N$};
        \node at (-1.5,0.7) {\tiny$N\text{-}1$};
        \node at (-3.5,0.7) {\tiny$2$};
        \node at (-4.5,0.7) {\tiny$1$};
    \end{tikzpicture}
\end{equation}

\noindent We thus see that in this final configuration $q-N$ NS5--branes and $l$ D5--branes are completely decoupled. The remaining part of the brane system instead describes the $T(\mathfrak{su}(N))$ theory, which as we have seen in Subsection \ref{subsubsec:ADI3d}, corresponds to the 3d reduction of the $\ADAIF{q}{N}$ theory for $q>N$. One can also easily check that for both of the theories involved in the equivalence the number of twisted hypermultiplets is the same and it is given by
\begin{align}\label{eq:freehyptypeIequiv}
    \widetilde H_{\text{free}}=\frac{(N-1)(q-N-1)}{2}~,
\end{align}
where we used the fact that this number does not change after partial closure of the regular puncture. This shows that the 3d reductions of the $\ADAIF{q}{N}$ and the $\ADAI{q}{ql+N}{[q^l,1^N]}$ theories are the same 3d $\mathcal N=4$ theories.

A natural question to ask at this point is whether some more equivalences can be obtained by adding different combinations of decoupled NS5--branes and D5--branes. This will always give equivalent $T^\sigma_\rho(\mathfrak{su}(N))$ quiver theories in 3d, but it is not always guaranteed that the resulting partitions $\sigma$ and $\rho$ are associated with the reduction of some type I AD theory. In particular, we have seen in Subsection \ref{subsubsec:ADI3d} that only some specific $\rho$ can show up from the reduction of a type I AD theory, specifically, $\rho=[1^N]$ for $q>N$ and $\rho=[(x+1)^{m_B},x^{m_A}]$ for $q<N$.

We have found another equivalence of type I AD theories that can be understood as adding only decoupled NS5--branes to the brane setup of the corresponding 3d theories. For the case with maximal puncture $[Y]=[1^N]$, it amounts to adding $q-N$ decoupled NS5--branes for $q>N$, and no decoupled five-branes for $q<N$ but rather rearranging those already present by moving all D5--branes on the left or the right of the NS5--branes. This leads to the following new equivalence between type I theories:
\begin{equation}\label{eq:iso2}
    \ADAI{q}{N}{[Y]} \cong \ADAI{q}{qL-N}{[Y^c]}~,
\end{equation}
where
\begin{equation}
    [Y^c] = \left[(q-1)^{l_1},(q-2)^{l_2},\dots,(q-N)^{l_N}\right]~.
\end{equation}
The flavour symmetry algebras in this equivalence match trivially while one can check case by case that the central charges and the Coulomb branch spectrum agree on both sides of \eqref{eq:iso2}. 

Let us see more in detail how this duality can be understood in 3d as adding $q-N$ decoupled NS5--branes to the brane system when $q>N$ and $[Y]=[1^N]$, for which the equivalence simplifies to
\begin{equation}\label{eq:iso2full}
    \ADAIF{q}{N} \cong \ADAI{q}{(q-1)N}{[(q-1)^N]}~.
\end{equation}
Once again, a completely analogous derivation works for $q<N$ and for a generic $[Y]$ but with a different number of decoupled five-branes.

As before, to determine the quiver describing the 3d reduction of the theory on the r.h.s.\ of the equivalence \eqref{eq:iso2full}, we first consider the one for the theory with the full regular puncture $\ADAIF{q}{(q-1)N}$. Using that for $q>N$ we have that
\begin{align}
    x=\left\lfloor\frac{(q-1)N}{q}\right\rfloor=\left\lfloor N-\frac{N}{q}\right\rfloor=N-1~,
\end{align}
we can see that the 3d reduction is given by the $T^\sigma_\rho[\mathfrak{su}((q-1)N)]$ theory with $\sigma=[1^{(q-1)N}]$ and $\rho=[N^{q-N},(N-1)^N]$. This implies that the 3d reduction of the theory on the r.h.s.\ of \eqref{eq:iso2full} we are interested in, instead corresponds to,
\begin{align}
    T^\sigma_\rho[\mathfrak{su}((q-1)N)]\,,\quad \sigma=[(q-1)^N]\,,\quad \rho=[N^{q-N},(N-1)^N]~.
\end{align}
The associated brane setup is given by a stack of $(q-1)N$ D3--branes stretched between $N$ D5--branes and $q$ NS5--branes.
\begin{equation}\hspace{-0.3cm}
\resizebox{\textwidth}{!}{
        \begin{tikzpicture}[scale=0.95,every node/.style={scale=1},font=\scriptsize]
        \draw[thick,red] (0,2)--(0,0);
        \draw[thick,red] (1,2)--(1,0);
        \draw[thick,red] (2,2)--(2,0);
        \draw[thick,red] (3,2)--(3,0);
        \draw[thick,red] (4,2)--(4,0);
        \draw[thick,red] (5,2)--(5,0);
        \draw[thick,red] (6,2)--(6,0);
        \draw[thick,red] (7,2)--(7,0);
        \draw[thick,red] (8,2)--(8,0);
        \draw[thick,red] (9,2)--(9,0);
        \draw[thick,red] (10,2)--(10,0);
        \draw[thick,red] (11,2)--(11,0);
        \node at (-1,1) {\color{blue}\small $\bm\otimes$};
        \node at (-2,1) {\color{blue}\small $\bm\otimes$};
        \node at (-3,1) {\color{blue}\small $\bm\otimes$};
        \node at (-4,1) {\color{blue}\small $\bm\otimes$};
        \node at (-5,1) {\color{blue}\small $\bm\otimes$};
        \draw (-5,1)--(-3,1);
        \node at (-2.5,1) {$\cdots$};
        \draw (-2,1)--(2,1);
        \node at (2.5,1) {$\cdots$};
        \draw (3,1)--(8,1);
        \node at (8.5,1) {$\cdots$};
        \draw (9,1)--(11,1);
        \node at (0.5,0.7) {\tiny$(q\text{-}2)N$};
        \node at (1.5,0.7) {\tiny$(q\text{-}3)N$};
        \node at (3.5,0.7) {\tiny$N(N\text{+}1)$};
        \node at (4.5,0.7) {\tiny$N^2$};
        \node at (5.5,0.7) {\tiny$N(N\text{-}1)$};
        \node at (6.5,0.7) {\tiny$(N\text{-}1)^2$};
        \node at (7.5,0.7) {\tiny$(N\text{-}1)$};
        \node at (7.5,0.4) {\tiny$(N\text{-}2)$};
        \node at (9.5,0.7) {\tiny$2(N\text{-}1)$};
        \node at (10.5,0.7) {\tiny$N\text{-}1$};
        \node at (-0.5,0.7) {\tiny$(q\text{-}1)N$};
        \node at (-1.5,0.7) {\tiny$(q\text{-}1)$};
        \node at (-1.5,0.4) {\tiny$(N\text{-}1)$};
        \node at (-3.5,0.7) {\tiny$2(q\text{-}1)$};
        \node at (-4.5,0.7) {\tiny$q\text{-}1$};
    \end{tikzpicture}
    }
\end{equation}

If we try to move all the D5--branes to the right side using Hanany--Witten moves, we obtain the following brane setup
\begin{equation}\hspace{-0.3cm}
        \begin{tikzpicture}[scale=1.1,every node/.style={scale=1.3},font=\scriptsize]
        \node at (0.5,-0.2) {$\underbrace{\color{white}AAAAA}_{q-N}$};
        \draw[thick,red] (0,2)--(0,0);
        \draw[thick,red] (1,2)--(1,0);
        \draw[thick,red] (2,2)--(2,0);
        \draw[thick,red] (3,2)--(3,0);
        \draw[thick,red] (4,2)--(4,0);
        \draw[thick,red] (5,2)--(5,0);
        \draw[thick,red] (6,2)--(6,0);
        \node at (7,1) {\color{blue}\small $\bm\otimes$};
        \node at (8,1) {\color{blue}\small $\bm\otimes$};
        \node at (9,1) {\color{blue}\small $\bm\otimes$};
        \node at (10,1) {\color{blue}\small $\bm\otimes$};
        \node at (11,1) {\color{blue}\small $\bm\otimes$};
        \node at (0.5,1) {$\cdots$};
        \draw (2,1)--(4,1);
        \node at (4.5,1) {$\cdots$};
        \draw (5,1)--(8,1);
        \node at (8.5,1) {$\cdots$};
        \draw (9,1)--(11,1);
        \node at (2.5,0.7) {\tiny$1$};
        \node at (3.5,0.7) {\tiny$2$};
        \node at (5.5,0.7) {\tiny$N\text{-}1$};
        \node at (6.5,0.7) {\tiny$N$};
        \node at (7.5,0.7) {\tiny$N\text{-}1$};
        \node at (9.5,0.7) {\tiny$2$};
        \node at (10.5,0.7) {\tiny$1$};
    \end{tikzpicture}
\end{equation}

\noindent As mentioned previously, we notice that there are $q-N$ decoupled NS5--branes. The interacting branes instead describe the $T(\mathfrak{su}(N))$ theory, which corresponds to the 3d reduction of the $\ADAIF{q}{N}$ theory for $q>N$. In addition to this quiver gauge theory, one can easily check that the 3d reduction of both of the theories involved in the equivalence also contain the same number of free twisted hypers, which is given by \eqref{eq:freehyptypeIequiv} in this case as well. This shows that the 3d reductions of the $\ADAIF{q}{N}$ and the $\ADAI{q}{(q-1)N}{[(q-1)^N]}$ theories coincide.

It is useful here to note that the special case $[Y]=[N]$ of this equivalence reproduces the well-known equivalence mentioned at the beginning of this subsection \eqref{eq:iso1}. Moreover, in \cite{Xie:2019yds} another set of equivalences between type I theories was proposed, that was termed ``level-rank dualities''. In our notation these are of the form
\begin{equation}\label{eq:isospecialcaseXie}
    \ADAI{q}{q-N}{[q-N-l,1^l]} \cong \ADAI{q}{ql+N}{[(q-1)^l,N+l]}~.
\end{equation}
for $N$ and $q-N$ co-prime. It is immediately clear that this is a special case of \eqref{eq:iso2} for hook-type partitions of the form $[Y]=[q-N-l,1^l]$. Therefore, our general equivalence \eqref{eq:iso2} can be viewed as a significant generalisation of these type of ``level-rank dualities''.

We also mention a set of equivalences specific to $A_1$ type Argyres--Douglas theories that can be expressed as
\begin{equation}\label{eq:TypeISU2Equiv}
    \begin{split}
        \ADAIF{2\kappa+2}{2} &\cong \ADAI{\kappa+1}{\kappa+2}{[\kappa,1^2]} \\
        \ADAI{2\kappa+2}{2}{[2]} &\cong \ADAI{\kappa+1}{\kappa}{[\kappa-1,1]}~.
    \end{split}
\end{equation}
These can once again be verified by computing the central charges and the Coulomb branch spectrum. The 3d reductions for these equivalences turn out to be identical without adding any decoupled branes and are therefore not detected in the discussion above. Furthermore, this equivalence directly explains the known identification of the sub-sub-regular and sub-regular affine $\mathcal W$-algebras as the associated vertex operator algebras of the top and bottom SCFTs on the left, respectively \cite{Beem:2017ooy,Creutzig:2017qyf}. However these are still not independent equivalences as they can be seen as special cases of \eqref{eq:typeIwEMD}. We will expand on the details in Subsection \ref{subsec:gauging_generalities}.

The two general equivalences we have presented here have an elegant and simple pictorial depiction at the level of the Young tableaux that describes the partition $[Y]$. Before describing this, we note that the theory $\ADAI{q}{N}{[Y]}$ is uniquely specified by the partition $[Y]$ and the integer $q$. Since the value of $N$ can be recovered from $[Y]$ by counting the total number of boxes in the associated Young tableaux.

The first equivalence \eqref{eq:isoql} can be depicted as adding $l$, an arbitrary positive integer, columns of height $q$ (shown here for $l=3$, $q=7$, and $[Y]=[5,3,2^2,1]$)
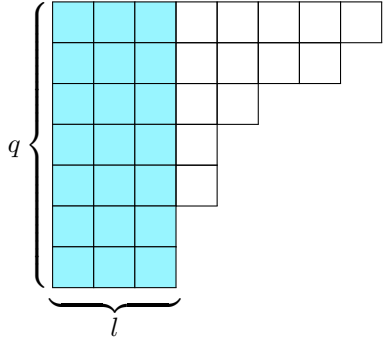
\begin{center}
    \ytableausetup{baseline,boxframe=normal}
    \begin{tabular}{r@{}l}
    \raisebox{-9.6ex}{$q\left\{\vphantom{\begin{array}{c}~\\[20ex] ~
    \end{array}}\right.$} &
    \begin{ytableau}
        *(CadetBlue1)&*(CadetBlue1)&*(CadetBlue1)& & & & & \\
        *(CadetBlue1)&*(CadetBlue1)&*(CadetBlue1)& & & & \\
        *(CadetBlue1)&*(CadetBlue1)&*(CadetBlue1)& & \\
        *(CadetBlue1)&*(CadetBlue1)&*(CadetBlue1)& \\
        *(CadetBlue1)&*(CadetBlue1)&*(CadetBlue1)& \\
        *(CadetBlue1)&*(CadetBlue1)&*(CadetBlue1)\\
        *(CadetBlue1)&*(CadetBlue1)&*(CadetBlue1)\\
    \end{ytableau}\\[-2ex]
    &\hspace{-0.1em}$\underbrace{\hspace{4.9em}}_{\displaystyle l}$
    \end{tabular}
\end{center}

\noindent This gives us the Young tableaux of the partition specifying the theory on the r.h.s.\ of the equivalence \eqref{eq:isoql}. Since the value of $q$ is the same for both of the theories involved in the equivalence, this specifies completely the theory on the r.h.s. Alternatively, by reading the equivalence from right to left, one can understand it as the statement that any type I theory whose Young tableaux specifying the regular puncture has a sub-diagram which is a rectangle formed by $l$ columns of height $q$ can be ``simplified" by removing such columns and appropriately re-adjusting the value of $N$.

For the second equivalence \eqref{eq:iso2}, the Young tableaux $[Y^c]$ of the theory on the r.h.s.\ of \eqref{eq:iso2} is instead obtained from $[Y]$ by first drawing a rectangle of height $q$ and width $L$ and then taking the complement of $[Y]$ inside this rectangle (shown here for $q=7$ and $[Y]=[5,3,2^2,1]$ which implies that $L=5$)

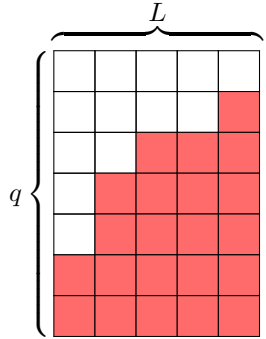
\begin{center}
    \ytableausetup{baseline,boxframe=normal}
    \begin{tabular}{r@{}l}
    \vspace{-0.12cm}&\hspace{-0.05em}$\overbrace{\hspace{7.9em}}^{\displaystyle L}$\\
    \raisebox{-9.6ex}{$q\left\{\vphantom{\begin{array}{c}~\\[20ex] ~
    \end{array}}\right.$} &
    \begin{ytableau}
        \textcolor{white}{1} & & & & \\
        & & & &*(IndianRed1) \\
        & &*(IndianRed1)&*(IndianRed1)&*(IndianRed1) \\
        &*(IndianRed1)&*(IndianRed1)&*(IndianRed1)&*(IndianRed1) \\
        &*(IndianRed1)&*(IndianRed1)&*(IndianRed1)&*(IndianRed1) \\
        *(IndianRed1)&*(IndianRed1)&*(IndianRed1)&*(IndianRed1)&*(IndianRed1) \\
        *(IndianRed1)&*(IndianRed1)&*(IndianRed1)&*(IndianRed1)&*(IndianRed1) \\
    \end{ytableau}
    \end{tabular}
\end{center}

\noindent Once again, the new Young tableaux and the fact that the value of $q$ is the same on both sides of the equivalence allow us to uniquely determine the theory on the r.h.s.\ of \eqref{eq:iso2}.

From this way of depicting the two equivalences, it becomes evident that the first equivalence \eqref{eq:isoql} can be obtained from the second one \eqref{eq:iso2}, which is thus more general and fundamental. This can be seen by applying the transformation of the Young tableaux described above corresponding to the second equivalence \eqref{eq:iso2} to either $[Y]$ or $[q^l,Y]$ reproduces the same partition $[Y^c]$ and thus the same theory. Another way to see this is to apply the second equivalence \eqref{eq:iso2} to the same theory but viewing the partition labelling its regular puncture as $[Y]$ and as $[Y,0^l]$, to obtain two seemingly different but actually equivalent theories, which reproduces the first equivalence \eqref{eq:iso1}. From this point of view, it is clear that the $\mathfrak{su}(l)$ symmetry that is na\"ively present in the theory on the r.h.s.~of \eqref{eq:iso1} from the part $q^l$ of the partition actually acts trivially, since upon application of \eqref{eq:iso2} it is mapped to the trivial entries $0^l$.

\subsection{\label{subsec:VOAtypeI}VOA perspective}

An immediate consequence of these equivalences amongst type I Argyres--Douglas theories is that the associated vertex operator algebras should be isomorphic. For the cases with $N$ and $p$ coprime, \emph{i.e.}\ $q=p=N+k$, the associated vertex operator algebras are affine $\mathcal W$-algebras (see Subsection \ref{subsec:ADVOA}) and these equivalences can be nicely understood in this context. 

Corresponding to \eqref{eq:iso1} we expect the following isomorphisms:
\begin{equation}
    \mathcal W^{-N+\frac{N}{q}}\left(\mathfrak{sl}_N,\mathfrak f_{[Y]}\right) \cong \mathcal W^{-(N+ql)+\frac{N+ql}{q}}\left(\mathfrak{sl}_{N+ql},\mathfrak f_{[q^l,Y]}\right)~.
\end{equation}
Indeed, these equivalences amount to the statement that for $\mathfrak{sl}_N$ affine Kac--Moody VOAs, the boundary admissible levels $k=-N+\frac{N}{q}$ are \emph{collapsing} for the nilpotent $f_{[Y]}$ of type $[Y]=[q^l,1^{N-ql}]$; this collapsing level statement was first conjectured in relation to these Argyres--Douglas theories in \cite{Xie:2019yds} and a proof has been given in \cite{Arakawa:2021ogm}.

Perhaps more interesting is the equivalence \eqref{eq:iso2}. At the level of vertex algebras, this gives the isomorphism
\begin{eqnarray}\label{eq:conjugation_duality}
    \mathcal W^{-N+\frac{N}{q}}\left(\mathfrak{sl}_N,\mathfrak f_{[Y]}\right) \cong \mathcal W^{-(qL-N)+\frac{qL-N}{q}}\left(\mathfrak{sl}_{qL-N},\mathfrak f_{[Y^c]}\right)~,
\end{eqnarray}
where $[Y^c]$ is the complement of $[Y]$ in the sense discussed above.

Combined with the collapsing level result for boundary admissible levels, this family of isomorphisms should follow from a basic isomorphism where $[Y]=[1^N]$ and $Y^c=[(q-1)^N]$. This is an isomorphism between a boundary admissible level affine Kac--Moody VOA and a \emph{rectangular $\mathcal{W}$-algebra} of size $N\times (q-1)$ at the specified level:
\begin{equation}
    V_{-N+\frac{N}{q}}\left(\mathfrak{sl}_N\right) = \mathcal{W}^{-(ql-N)+\frac{ql-N}{q}}\left(\mathfrak{sl}_{ql-N},f_{[(q-1)^N]}\right)~.
\end{equation}

This equivalence (and its generalisation to arbitrary rectangular partitions) can be understood by embedding of these vertex algebras in the matrix-extended $\mathcal{W}_{1+\infty}$ algebras of \cite{Eberhardt:2019xmf}. We recall that for each fixed positive integer $N$, these form a two-parameter (called $n$ and $\kappa$ in \emph{loc. cit.}) family of infinitely strongly generated vertex operator algebras $\mathcal{W}_{1+\infty}(\kappa,n;N)$ extending a $\mathfrak{gl}_N$ affine Kac--Moody vertex subalgebra at level $k=n\kappa$.\footnote{For computational reasons, the matrix extended $\mathcal{W}_{1+\infty}$ algebra is defined with $\mathfrak{gl}_N$ global symmetry; the reduction to $\mathfrak{sl}_N$ is straightforward upon factoring out a decoupled overall $\mathcal{W}_{1+\infty}$.} Along various codimension-one subspaces in $(n,\kappa)$ space, this VOA develops null states and upon taking a quotient reduces to simpler algebras; these ``truncation curves'' were argued in \emph{loc. cit.} to take the general form
\begin{equation}
    n\kappa-\alpha\kappa+\beta(\kappa+N)-\gamma+\delta=0~,
\end{equation}
where $(\alpha,\beta,\gamma,\delta)\in\mathbb{Z}_{\geqslant0}^4$ are non-negative integers and $\alpha\beta=\gamma\delta=0$.

The vertex algebras arising upon taking the simple quotient along these truncation curves were characterised in \emph{loc. cit.} in terms of a gluing procedure for corner vertex algebras, see \cite{Gaiotto:2017euk,Prochazka:2017qum}. However, for simple cases these can be identified with easier-to-describe vertex algebras. For the cases $\beta=\gamma=\delta=0$, $\alpha=M_1$ and $\alpha=\gamma=\delta=0$, $\beta=M_2$, we propose to identify the following truncations,\footnote{As far as we are aware, this remains a conjecture. The identification is motivated first by the structure of null states (at these levels, the $\mathfrak{gl}_N$-valued strong generators at level $N+1$ are becoming null, and we expect so are all the higher generators). Also, the intersection of one of these truncation curves (say $\alpha=N$, others vanishing) and the truncation curves giving rise to Grassmannian coset vertex algebras ($\beta=1$, $\delta=M$, others vanishing) reproduce the results of \cite{Creutzig:2018pts} on the isomorphism of rectangular $\mathcal{W}$-algebras with Grassmannian cosets.}
\begin{equation}
\begin{split}
    \mathcal{W}_{1+\infty}(\kappa,M_1;N)&\Surjrightarrow \mathcal{W}^{-M_1 N+N+\kappa}\left(\mathfrak{gl}_{M_1 N},f_{[M_1^N]}\right)~,\\
    \mathcal{W}_{1+\infty}\left(-\tfrac{M_2 N}{M_2+n},n;N\right)&\Surjrightarrow\mathcal{W}^{-M_2 N+\frac{M_2 N}{M_2 + n}}\left(\mathfrak{gl}_{M_2 N},f_{[M_2^N]}\right)~.
\end{split}
\end{equation}
These truncation curves thus give rise to the rectangular $\mathcal{W}$ algebras; indeed the two sets of truncation curves are equivalent, being exchanged under the duality symmetry $(n,\kappa)\leftrightarrow(-N-\kappa,-\frac{n\kappa}{N+\kappa})$ of $\mathcal{W}(\kappa,n;N)$.

We can now consider the double truncations that arise at the intersection of these curves. That is, we can consider vertex algebras that can be simultaneously realised as rectangular $\mathcal{W}$-algebras of size $N\times M_1$ and $N\times M_2$. At these double truncation curves, we have $n=M_1$, $\kappa = - \tfrac{M_2 N}{M_2+N}$, which leads to an isomorphism
\begin{equation}
    \mathcal{W}^{-M_1 N+\frac{M_1 N}{M_1+M_2}}\left(\mathfrak{gl}_{M_1 N},f_{[M_1^N]}\right) \cong
    \mathcal{W}^{-M_2 N+\frac{M_2 N}{M_1+M_2}}\left(\mathfrak{gl}_{M_2 N},f_{[M_2^N]}\right)~.
\end{equation}
This is precisely the vertex algebra version of the duality \eqref{eq:conjugation_duality} where $[Y]$ is rectangular (and so $[Y^c]$ is also rectangular).

Some comments are in order. First, this argument for duality of vertex algebras applies regardless of the coprimality constraint (which in this case amounts to $(M_1 N,M_1+M_2) \equiv (M_2 N,M_1+M_2) = 1$). For the actual AD SCFTs, in the non-coprime case we expect the relevant vertex algebras to be extensions of these rectangular $\mathcal{W}$-algebras; isomorphism of those extensions would be interesting to describe more explicitly. 

Additionally, this isomorphism of rectangular $\mathcal{W}$-algebras is the generalisation to $N>1$ of a nice observation that those ${\mathcal W}_N$ principal $\mathcal{W}$-algebras that arise from the $(A_{N-1},A_{k-1})$-type Argyres--Douglas SCFTs (here with $N$ and $k$ coprime) are precisely the $\mathcal{W}_N$ algebras that arise at the intersection of $\mathcal{W}_N$ truncation curves in the (non-matrix-extended) $\mathcal{W}_{1+\infty}$ algebra.\footnote{CB would like to thank Tom\'{a}\v{s} Proch\'{a}zka for first explaining this point to him.} It would be very interesting to understand whether the double-truncation property of these VOAs sheds any light on other good (conjectural) properties of those VOAs which arise in four dimensional SCFTs.

\subsection{\label{subsec:typeIEMC}Generalised gauging description of the non-coprime case}

In the setting of theories of class $\mathcal S$ characterised by only regular punctures, SCFTs with exactly marginal deformations are realised as generalised gauge theories with ``matter'' consisting of trinion theories, which correspond to three-punctured spheres and have no exactly marginal deformations. These gauge theory descriptions are understood as corresponding to degeneration limits of the underlying UV curve, with different decompositions of the Riemann surface into three-punctured spheres interpreted as different S-duality frames.

Considering type I Argyres--Douglas theories, the irregular singularity on the Riemann sphere obstructs a similar result. One may still expect to be able to express type I Argyres--Douglas theories with exactly marginal deformations as various gaugings of 4d $\mathcal N=2$ SCFTs without exactly marginal deformations. However the building blocks \emph{a priori} need not themselves be type I Argyres--Douglas theories. In \cite{Closset:2020afy,Giacomelli:2020ryy} it was shown that one can express type I theories with exactly marginal deformations constructible via geometric engineering as gaugings of a type I Argyres--Douglas theory and another unidentified 4d $\mathcal N=2$ theory.

Another approach to this issue was explored in \cite{Xie:2017vaf,Xie:2017aqx}, which associates an auxiliary Riemann sphere (with the number of marked points equal to the number of exactly marginal deformations plus three) to each Argyres--Douglas theory. One is then able to consider various decompositions of this auxiliary Riemann sphere into three-punctured spheres. The construction is evidently designed so that the theories associated to the auxiliary three-punctured spheres do not have any exactly marginal deformations, and are given the moniker ``Arygres--Douglas matter theories''.\footnote{More precisely, in \cite{Xie:2017vaf,Xie:2017aqx} the Arygres--Douglas matter theories are defined to have no exactly marginal deformation and a non-abelian flavour symmetry, which can be gauged to generate theories with exactly marginal deformations.}

This gives a conjectural description of type I Argyres--Douglas theories as gaugings of various Argyres--Douglas matter theories. However, not all of the Argyres--Douglas matter theories are type I Argyres--Douglas theories. We propose an explicit expression for any type I Argyres--Douglas theory with EMDs as conformal gaugings of other type I Argyres--Douglas theories without any EMDs.\footnote{Some special cases of our general result have also appeared in \cite{Buican:2014hfa}, specifically for the theories $\ADAI{8}{4}{[4]}\cong\mathcal{T}_{2,\frac{3}{2},\frac{3}{2}}$ and $\ADAI{4}{6}{[2^3]}\cong\mathcal{T}_{2,3,\frac{3}{2},\frac{3}{2}}$ in their notation.} For the cases with reduced regular puncture, our expression is especially useful as it explicitly characterises the effect of partially closing the regular puncture on the involved type I theories and their gaugings. Further, our result expresses type I theories in terms of other type I theories and is self-contained, and thus can be viewed as a special S-type duality frame that relates type I theories to other type I theories. Our result can also be considered as an extension of the one of \cite{Closset:2020afy,Giacomelli:2020ryy} to a generic regular puncture. Indeed, in the case of maximal regular puncture the two coincide, since we are able to identify the previously unknown 4d theories involved in the gauging in \cite{Closset:2020afy,Giacomelli:2020ryy} as type I Argyres--Douglas theories with a partially closed regular puncture.

\subsubsection{\label{subsec:typeIEMCex}Examples}

Here we perform a detailed analysis of several examples that demonstrate important aspects of the general case. We consider four examples in total, three of which have only one exactly marginal deformation and one that has two.

\subsubsection*{Example 1: $\ADAIF{6}{4}$}

This theory has $m={\rm gcd}(6,4)=2$, and it has only one exactly marginal deformation. The Coulomb branch spectrum of this theory, as reviewed in Subsection \ref{subsec:CBspecCM} following \cite{Xie:2012hs}, is
\begin{equation}\label{eq:gaugingex1CB}
\left\{\frac43,\frac43,\frac53,2,\frac73,\frac83,\frac{10}{3}\right\}~.
\end{equation}
The regular puncture is maximal, so the $c$ and $a$ central charges can be obtained from \eqref{eq:cfullpunc} and by using \eqref{eq:ShapereTachikawa},
\begin{equation}
    c = \frac{37}{6}~,\qquad a = \frac{47}{8}~.
\end{equation}
The flavour symmetry algebra is $\mathfrak{su}(4)\oplus\mathfrak{u}(1)$, with $\mathfrak{su}(4)$ coming from the regular puncture. This symmetry is manifest in the quiver for the 3d reduction, which is given as follows (\emph{cf.} \ref{subsec:AD3d}).

\begin{equation}\label{eq:gaugingex1quiv}
    \begin{tikzpicture}[scale=1.2,every node/.style={scale=1.2},font=\scriptsize]
    \node[gauge] (g0) at (0,0) {$\,3\,$};
    \node[gauge] (g1) at (1,0) {$\,2\,$};
    \node[gauge] (g2) at (2,0) {$\mathfrak{su}(2)$};
    \node[gauge] (g3) at (3,0) {$\,1\,$};
    \node[flavor] (f0) at (-1,0) {$\,4\,$};
    \draw (g0)--(g1)--(g2)--(g3);
    \draw (g0)--(f0);
\end{tikzpicture}
\end{equation}

We observe that \eqref{eq:gaugingex1quiv} can be obtained by an $\mathfrak{su}(2)$ gauging of two sub-quivers. The first one is
\begin{equation}
    \begin{tikzpicture}[scale=1.2,every node/.style={scale=1.2},font=\scriptsize]
    \node[gauge] (g0) at (0,0) {$\,3\,$};
    \node[gauge] (g1) at (1,0) {$\,2\,$};
    \node[flavor] (f0) at (-1,0) {$\,4\,$};
    \node[flavor] (f1) at (2,0) {$\,2\,$};
    \draw (g0)--(g1);
    \draw (g0)--(f0);
    \draw (g1)--(f1);
\end{tikzpicture}
\end{equation}
and can be identified with the $T^\sigma_\rho[\mathfrak{su}(10)]$ theory with $\sigma=[2^4,1^2]$ and $\rho=[4,3^2]$. Such a theory corresponds to the quiver for the 3d reduction of the $\ADAI{3}{10}{[2^4,1^2]}$, since it can be obtained as the $[1^{10}]\to[2^4,1^2]$ Higgs branch flow from the $T^\sigma_\rho[\mathfrak{su}(10)]$ theory with $\sigma=[1^{10}]$ and $\rho=[4,3^2]$, which instead corresponds to the 3d reduction of the theory with the maximal puncture $\ADAIF{3}{10}$ (\emph{cf.} \ref{subsec:AD3d}). The second sub-quiver is simply that of $T[\mathfrak{su}(2)]$,
\begin{equation}
    \begin{tikzpicture}[scale=1.2,every node/.style={scale=1.2},font=\scriptsize]
    \node[gauge] (g0) at (0,0) {$\,1\,$};
    \node[flavor] (f0) at (-1,0) {$\,2\,$};
    \draw (g0)--(f0);
\end{tikzpicture}
\end{equation}
which is the 3d reduction of $\ADAIF{3}{2}$.

This observation suggests that the $\ADAIF{6}{4}$ theory can be expressed as the gauging of the diagonal $\mathfrak{su}(2)$ symmetries of the $\ADAI{3}{10}{[2^4,1^2]}$ and the $\ADAIF{3}{2}$ theories
\begin{equation}\label{eq:gaugingex1}
    \ADAIF{6}{4} \cong \ADAI{3}{10}{[2^4,1^2]} \longleftarrow \mathfrak{su}(2) \longrightarrow \ADAIF{3}{2}~.
\end{equation}
A partial version of this first appeared in \cite[Eq.\ (3.36)]{Giacomelli:2020ryy} and \cite[Eq.\ (2.25)]{Closset:2020afy}, but here we explicitly identify the unknown theory dubbed $\mathcal{D}_3(4,2)$ in those references as the type I AD theory with a partially closed regular puncture $\ADAI{3}{10}{[2^4,1^2]}$. Indeed, the flavour symmetry of $\ADAI{3}{10}{[2^4,1^2]}$ is $\mathfrak g_F^{(1)}=\mathfrak{su}(4)\oplus\mathfrak{su}(2)\oplus \mathfrak{u}(1)$ while that of $\ADAIF{3}{2}$ is $\mathfrak g_F^{(2)}=\mathfrak{su}(2)$. The diagonal $\mathfrak{su}(2)$ subalgebra of $\mathfrak{su}(2)\oplus\mathfrak{su}(2) \subset \mathfrak g_F^{(1)}\oplus\mathfrak g_F^{(2)}$ is gauged such that after gauging the remaining flavour symmetry is exactly $\mathfrak{su}(4)\oplus\mathfrak{u}(1)$, as expected.

We can perform additional checks for the statement \eqref{eq:gaugingex1} on top of the matching of the 3d reduction explained above. The Coulomb branch spectra of the theories on the right hand side of \eqref{eq:gaugingex1} are $\left\{\frac43,\frac53,\frac73,\frac83,\frac{10}{3}\right\}$ and $\left\{\frac43\right\}$, respectively. These, along with an operator with dimension 2 coming from the $\mathcal N=2$ vector multiplet involved in the gauging, reproduce the Coulomb branch spectrum \eqref{eq:gaugingex1CB} of the original SCFT. It is also possible to compute the $c$ and $a$ central charges from \eqref{eq:gaugingex1}, and we have the match
\begin{equation}\label{eq:ccmatchex1}
    \left(\frac{37}{6},\frac{47}{8}\right) = \left(\frac{31}{6},\frac{115}{24}\right) + 3\left(\frac16,\frac{5}{24}\right) + \left(\frac12,\frac{11}{24}\right)~,
\end{equation}
where we are combining the central charges of $\ADAI{3}{10}{[2^4,1^2]}$, a 4d $\mathcal N=2$ vector multiplet in the adjoint of $\mathfrak{su}(2)$, and $\ADAIF{3}{2}$.

For this to be a consistent interpretation, the $\mathfrak{su}(2)$ gauging must be conformal. The vanishing of the beta function is equivalent to the vanishing of the triangle anomaly $\mathrm{Tr} R\mathfrak{g}\mathfrak{g}$, where $\mathfrak{g}=\mathfrak{su}(2)$ is the gauge algebra. The contributions of the two type I theories to this anomaly are $k_{\mathfrak{su}(2)}^{(1)}$ and $k_{\mathfrak{su}(2)}^{(2)}$, respectively, and conformality requires
\begin{equation}
    \mathrm{Tr} R\,\mathfrak{su}(2)^2\propto k_{\mathfrak{su}(2)}^{(1)} + k_{\mathfrak{su}(2)}^{(2)} - 4 h^\vee_{\mathfrak{su}(2)} = k_{\mathfrak{su}(2)}^{(1)} + k_{\mathfrak{su}(2)}^{(2)}-4\times 2=0~.
\end{equation}
The flavour central charges can be computed by using equations \eqref{eq:ksuNfullpunc}, \eqref{eq:ksulredpunc}, and \eqref{eq:kGBcont}, and we have
\begin{equation}
    k_{\mathfrak{su}(2)}^{(1)} = \frac{16}{3} \;,\quad k_{\mathfrak{su}(2)}^{(2)} = \frac83~,
\end{equation}
from which we see that the gauging is indeed conformal.

\subsubsection*{Example 2: $\ADAI{6}{4}{[3,1]}$}

This theory has one exactly marginal deformation as well. The Coulomb branch spectrum of this theory is
\begin{equation}\label{eq:gaugingex2CB}
\left\{\frac43,\frac43,\frac53,2\right\} 
\end{equation}
Since the regular puncture is not maximal, the $c$ central charge is obtained by realising this theory as the IR limit of the Higgs branch flow of the form $[3,1]$ starting from the UV theory $\ADAIF{6}{4}$. The central charges thus computed are
\begin{equation}
    c = \frac73~,\qquad a = \frac94~.
\end{equation}
The flavour symmetry is ${\mathfrak u}(1)\oplus{\mathfrak u}(1)$, where the regular puncture contributes only a single ${\mathfrak u}(1)$. 

The 3d reduction in this case is trickier to work out and we will not need it explicitly. We first consider the quiver for the 3d reduction of the theory with the full regular puncture $\ADAIF{6}{4}$ given in \eqref{eq:gaugingex1quiv}. We want to execute a nilpotent Higgsing $[1^4]\to[3,1]$. This is non-trivial due to the presence of the special unitary gauge node.\footnote{Some aspects of 3d quivers with mixed unitary and special unitary gauge nodes have been studied for example in \cite{Giacomelli:2020ryy,Bourget:2021jwo,Dey:2020hfe}.} However, we can exploit the fact that in three dimensions one obtains an $\mathfrak{su}(p)$ gauge algebra by starting from a theory with a $\mathfrak{u}(p)$ gauge algebra and gauging its $\mathfrak{u}(1)$ topological symmetry with a twisted vector multiplet. This produces a $\mathfrak{u}(1)$ baryonic symmetry that can be gauged to go back to the $\mathfrak{u}(p)$ gauge theory. The quiver in \eqref{eq:gaugingex1quiv} can then be obtained from a similar linear quiver gauge theory but with only unitary nodes by gauging the $\mathfrak{u}(1)$ topological symmetry of the $\mathfrak{u}(2)$ node on the right:
\begin{equation}
    \begin{tikzpicture}[scale=1.2,every node/.style={scale=1.2},font=\scriptsize]
    \node[gauge] (g0) at (0,0) {$\,3\,$};
    \node[gauge] (g1) at (1,0) {$\,2\,$};
    \node[gauge] (g2) at (2,0) {$\,2\,$};
    \node[gauge] (g3) at (3,0) {$\,1\,$};
    \node[flavor] (f0) at (-1,0) {$\,4\,$};
    \draw (g0)--(g1)--(g2)--(g3);
    \draw (g0)--(f0);
\end{tikzpicture}
\end{equation}
This is an \emph{ugly} theory as some of the gauge nodes have a number of fundamental hypers which is twice the number of colours minus one. The fundamental monopole operators for these nodes then have free scaling dimension $\Delta=\frac{1}{2}$ and give rise to a decoupled free sector in the IR. This decoupling can be characterised by the following infrared duality \cite{Gaiotto:2008ak},
\begin{align}\label{uglyduality}
&\text{$\mathfrak{u}(N)$ $+$ $2N-1$ flavours} \longleftrightarrow \,\,\text{$\mathfrak{u}(N-1)$ $+$ $2N-1$ flavours + 1 twisted hypermultiplet}~. 
\end{align}
By applying this duality until there are no ugly nodes, we obtain the quiver for $T[\mathfrak{su}(4)]$ with $1+1$ additional twisted hypers, where we are distinguishing the one that is charged under the $\mathfrak{u}(1)$ topological symmetry that we have to gauge to go back to \eqref{eq:gaugingex1quiv} from the one that is not. At this point we can implement the Higgsing $[1^4]\to[3,1]$, which results in the $T^\sigma_\rho[\mathfrak{su}(4)]$ theory with $\sigma=[3,1]$ and $\rho=[1^4]$
\begin{equation}\label{eq:gaugingex2quiv}
\begin{tikzpicture}[scale=1.2,every node/.style={scale=1.2},font=\scriptsize]
    \node[gauge] (g0) at (0,0) {$\,1\,$};
    \node[gauge] (g1) at (1,0) {$\,1\,$};
    \node[gauge] (g2) at (2,0) {$\,1\,$};
    \node[flavor] (f0) at (-1,0) {$\,1\,$};
    \node[flavor] (f1) at (3,0) {$\,1\,$};
    \draw (g0)--(g1)--(g2);
    \draw (g0)--(f0);
    \draw (g2)--(f1);
\end{tikzpicture}
\end{equation}
to which we add the $1+1$ twisted hypermultiplets. Crucially, this Higgsing only affects the Higgs branch flavour symmetry, so it does not interfere with the topological symmetry that we need to gauge. Hence we expect the two operations to commute and by gauging this topological symmetry in the theory after the Higgsing we should get the 3d reduction of the $\ADAI{6}{4}{[3,1]}$ theory. However, we actually do not need to perform this last step, since it will be sufficient to work with the quiver \eqref{eq:gaugingex2quiv} for our purposes.

We propose that the $\ADAI{6}{4}{[3,1]}$ theory can be expressed as the following gauging of two type I theories:
\begin{eqnarray}\label{eq:gaugingex2}
        \ADAI{6}{4}{[3,1]} \cong \ADAI{3}{4}{[2,1^2]} \longleftarrow &\mathfrak{su}(2)& \longrightarrow \ADAIF{3}{2}~,\nn\\
        &|&\nn\\
        &\fbox{1}&
\end{eqnarray}
where the lower square box indicates that there is one hypermultiplet in the fundamental representation of the $\mathfrak{su}(2)$ gauge algebra. Note that both the theories on the r.h.s.\ of \eqref{eq:gaugingex2} have no exactly marginal deformations, so the only exactly marginal deformation comes from the $\mathfrak{su}(2)$ gauging. Furthermore, the flavour symmetry of $\ADAI{3}{4}{[2,1^2]}$ is $\mathfrak g_{F}^{(1)}=\mathfrak{su}(2)\oplus \mathfrak{u}(1)$ while that $\ADAIF{3}{2}$ is $\mathfrak g_{F}^{(2)}=\mathfrak{su}(2)$. The diagonal $\mathfrak{su}(2)$ subalgebra of $\mathfrak{su}(2)\oplus\mathfrak{su}(2) \subset \mathfrak g_{F}^{(1)}\oplus\mathfrak g_{F}^{(2)}$ is gauged such that after gauging the remaining flavour symmetry is $\mathfrak{u}(1)$. However, the hypermultiplet in the fundamental representation of the gauge algebra $\mathfrak{su}(2)$ contributes another $\mathfrak{u}(1)$ and therefore the total flavour symmetry algebra is ${\mathfrak u}(1)\oplus{\mathfrak u}(1)$, as expected.

The Coulomb branch spectra of the theories on the right are $\left\{\frac43,\frac53\right\}$ and $\left\{\frac43\right\}$. These, along with an operator with dimension 2 coming from the $\mathcal N=2$ vector multiplet involved in the gauging, reproduce the Coulomb branch spectrum of our original SCFT \eqref{eq:gaugingex2CB}.
It is also possible to compute the $c$ and $a$ central charges from the r.h.s.\ of the expression
\begin{equation}\label{eq:ccmatchex2}
    \left(\frac73,\frac94\right) = \left(\frac76,\frac{13}{12}\right) + 3\left(\frac16,\frac{5}{24}\right) + \left(\frac12,\frac{11}{24}\right) + 2\left(\frac{1}{12},\frac{1}{24}\right)~.
\end{equation}
The numbers of the r.h.s.\ of \eqref{eq:ccmatchex2} correspond to the contribution of $\ADAI{3}{4}{[2,1^2]}$, a 4d $\mathcal N=2$ vector multiplet, $\ADAIF{3}{2}$, and the fundamental 4d $\mathcal N=2$ hypermultiplet, respectively. The factors multiplying the contributions of the supermultiplets correspond to the dimensions of their representations under the gauge algebra.

The conformality of this gauging can be confirmed once again by computing the anomaly mentioned in the previous example. Since we also have a fundamental 4d $\mathcal N=2$ hypermultiplet in \eqref{eq:gaugingex2}, we have the expression
\begin{equation}
    \mathrm{Tr} R\,\mathfrak{su}(2)^2 \propto k_{\mathfrak{su}(2)}^{(1)} - 4h^\vee_{\mathfrak{su}(2)} + k_{\mathfrak{su}(2)}^{(2)} + 4T_{\mathfrak{su}(2)}(\text{fund})=k_{\mathfrak{su}(2)}^{(1)} - 4\times2 + k_{\mathfrak{su}(2)}^{(2)} + 4\times\frac12~,
\end{equation}
where the last term captures the contribution of the 4d $\mathcal N=2$ hypermultiplet. The 4d flavour central charges of the $\mathfrak{su}(2)$ subalgebra of $\mathfrak g_F^{(1)}$ and $\mathfrak g_F^{(2)}$, respectively, are given by
\begin{equation}
    k_{\mathfrak{su}(2)}^{(1)} = \frac{10}{3} \;,\quad k_{\mathfrak{su}(2)}^{(2)} = \frac83~,
\end{equation}
computed by using \eqref{eq:ksuNfullpunc}, \eqref{eq:ksulredpunc}, and \eqref{eq:kGBcont}. This implies that the above anomaly and hence also the beta function of the gauge coupling vanish, so that this gauging is conformal. It must be noted here that the extra fundamental $\mathcal N=2$ hypermultiplet is crucial for the gauging to be conformal and for ensuring that the 4d central charges of the original theory are reproduced from its description as the gauging of type I theories in \eqref{eq:ccmatchex2}.

In order to show the matching of the 3d reductions we proceed as follows. First, we want to determine the 3d reduction of the theory $\ADAI{3}{4}{[2,1^2]}$, the first theory on the left side of \eqref{eq:gaugingex2}. For this, we start as usual from the one for the theory with the full regular puncture $\ADAIF{3}{4}$, which using the results reviewed in Subsection \ref{subsec:AD3d} is given by the $T^\sigma_\rho[\mathfrak{su}(4)]$ theory with $\sigma=[1^4]$ and $\rho=[2,1^2]$. Then we consider the HB Higgsing $[1^4]\to[2,1^2]$, which gives the $T^\sigma_\rho[\mathfrak{su}(4)]$ theory with $\sigma=\rho=[2,1^2]$ whose corresponding quiver is
\begin{equation}
    \begin{tikzpicture}[scale=1.2,every node/.style={scale=1.2},font=\scriptsize]
    \node[gauge] (g0) at (0,0) {$\,1\,$};
    \node[gauge] (g1) at (1,0) {$\,1\,$};
    \node[flavor] (f0) at (-1,0) {$\,2\,$};
    \node[flavor] (f1) at (2,0) {$\,1\,$};
    \draw (g0)--(g1);
    \draw (g0)--(f0);
    \draw (g1)--(f1);
\end{tikzpicture}
\end{equation}
Then, we consider the 3d reduction of $\ADAIF{3}{2}$, the second theory on the r.h.s.\ of \eqref{eq:gaugingex2}, which is simply given by
\begin{equation}
    \begin{tikzpicture}[scale=1.2,every node/.style={scale=1.2},font=\scriptsize]
    \node[gauge] (g0) at (0,0) {$\,1\,$};
    \node[flavor] (f0) at (-1,0) {$\,2\,$};
    \draw (g0)--(f0);
\end{tikzpicture}
\end{equation}
Gauging the diagonal combination of the two $\mathfrak{su}(2)$ symmetries of these two quivers with one fundamental hypermultiplet as prescribed in \eqref{eq:gaugingex2}, we get
\begin{equation}
    \begin{tikzpicture}[scale=1.2,every node/.style={scale=1.2},font=\scriptsize]
    \node[gauge] (g0) at (0,0) {$\,1\,$};
    \node[gauge] (g1) at (1,0) {$\,1\,$};
    \node[gauge] (g2) at (2,0) {$\mathfrak{su}(2)$};
    \node[gauge] (g3) at (3,0) {$\,1\,$};
    \node[flavor] (f0) at (-1,0) {$\,1\,$};
    \node[flavor] (f1) at (2,-1) {$\,1\,$};
    \draw (g0)--(g1)--(g2)--(g3);
    \draw (g0)--(f0);
    \draw (g2)--(f1);
\end{tikzpicture}
\end{equation}
In order to compare with \eqref{eq:gaugingex2quiv}, we consider a similar quiver but with unitary nodes only
\begin{equation}
    \begin{tikzpicture}[scale=1.2,every node/.style={scale=1.2},font=\scriptsize]
    \node[gauge] (g0) at (0,0) {$\,1\,$};
    \node[gauge] (g1) at (1,0) {$\,1\,$};
    \node[gauge] (g2) at (2,0) {$\,2\,$};
    \node[gauge] (g3) at (3,0) {$\,1\,$};
    \node[flavor] (f0) at (-1,0) {$\,1\,$};
    \node[flavor] (f1) at (2,-1) {$\,1\,$};
    \draw (g0)--(g1)--(g2)--(g3);
    \draw (g0)--(f0);
    \draw (g2)--(f1);
\end{tikzpicture}
\end{equation}
from which we can obtain the previous quiver by gauging the topological symmetry of the $\mathfrak{u}(2)$ node. This theory has several ugly nodes, which we can dualise using the duality \eqref{uglyduality}. The result is precisely the quiver \eqref{eq:gaugingex2quiv} with $1+1$ twisted hypermultiplets, where again we distinguish between those that are charged under the topological symmetry that we need need to gauge to go back to the actual theory that describes the 3d reduction of the AD theory from those that are not. This precisely matches the result we obtained for the $\ADAI{6}{4}{[3,1]}$ theory, as expected.

In principle, the gauging result \eqref{eq:gaugingex2} for non-maximal regular puncture can be derived from the one for the corresponding theory with the maximal puncture \cite{Closset:2020afy,Giacomelli:2020ryy} by studying the effect of the Higgsing. This explains how the hypermultiplet in the fundamental representation of the $\mathfrak{su}(2)$ gauge algebra arises. One can study the Higgsing in steps consisting of several minimal transitions, and we will focus on the last minimal transition. Namely, we start from the theory $\ADAI{6}{4}{[2^2]}$ and assume the result, which follows from the general one we will present momentarily, that this can be expressed as
\begin{eqnarray}
        \ADAI{6}{4}{[2^2]} \cong \ADAI{3}{4}{[1^4]} \longleftarrow &\mathfrak{su}(2)& \longrightarrow \ADAI{3}{2}{[1^2]}~,
\end{eqnarray}
where for the theory $\ADAI{3}{4}{[1^4]}$ an $\mathfrak{su}(2)$ subalgebra of its $\mathfrak{su}(4)$ symmetry is gauged, while its $\mathfrak{su}(2)$ commutant corresponds to the global symmetry carried by the maximal puncture in the full $\ADAI{6}{4}{[2^2]}$ theory. We then want to study the result of the transition $[2^2]\to[3,1]$. This in particular breaks the $\mathfrak{su}(2)$ global symmetry down to $\mathfrak{u}(1)$. In terms of the $\ADAI{3}{4}{[1^4]}$ component, this corresponds to closing its maximal puncture as $[1^4]\to [2,1^2]$, breaking its $\mathfrak{su}(4)$ symmetry down to $\mathfrak{su}(2)\oplus \mathfrak{u}(1)$, where we recall that the $\mathfrak{su}(2)$ part is gauged while the $\mathfrak{u}(1)$ is the residual global symmetry. In the Higgsing, Nambu--Goldstone bosons charged under both the $\mathfrak{su}(2)$ and the $\mathfrak{u}(1)$ part are produced. Specifically, under the breaking $\mathfrak{su}(4)\to \mathfrak{su}(2)\oplus \mathfrak{su}(2)\oplus\mathfrak{u}(1)$ we have the decomposition
\begin{align}
    { 15}\to \left(V_0,{3}\right)^0\oplus \left(V_{\frac{1}{2}},{ 2}\right)^{+1}\oplus \left(V_{\frac{1}{2}},{ 2}\right)^{-1}\oplus \left(V_1,{ 1}\right)^0
\end{align}
and the two $(V_{\frac{1}{2}},{ 2})^{\pm1}$ components correspond to a hypermultiplet in the fundamental of the $\mathfrak{su}(2)$. Since this $\mathfrak{su}(2)$ symmetry is gauged, this field is not decoupled and remains as the fundamental hypermultiplet in \eqref{eq:gaugingex2}.

\subsubsection*{Example 3: $\ADAI{8}{6}{[5,1]}$}

This theory has one exactly marginal deformation as well. The Coulomb branch spectrum is as follows,
\begin{equation}\label{eq:gaugingex3CB}
    \left\{\frac54,\frac54,\frac32,\frac32,\frac74,2,\frac94\right\}~.
\end{equation}
Since the regular puncture is not maximal, the $c$ central charge has to be obtained by considering this theory as the IR limit of the Higgs branch flow of type $[5,1]$ from the UV theory $\ADAIF{8}{6}$. The central charges computed this way are given by
\begin{equation}
    c = \frac{25}{6}~,\qquad a = \frac{49}{12}~.
\end{equation}
The flavour symmetry algebra of this theory is $\mathfrak{u}(1)\oplus\mathfrak{u}(1)$, where the regular puncture contributes only a single $\mathfrak{u}(1)$.

Once again, studying the 3d reduction is tricky since we have to consider a HB vev in a quiver with mixed unitary and special unitary gauge nodes. We proceed as before and start from the 3d reduction of the theory $\ADAIF{8}{6}$ with the full regular puncture
\begin{equation}
    \begin{tikzpicture}[scale=1.2,every node/.style={scale=1.2},font=\scriptsize]
    \node[gauge] (g0) at (0,0) {$\,5\,$};
    \node[gauge] (g1) at (1,0) {$\,4\,$};
    \node[gauge] (g2) at (2,0) {$\,3\,$};
    \node[gauge] (g3) at (3,0) {$\mathfrak{su}(3)$};
    \node[gauge] (g4) at (4,0) {$\,2\,$};
    \node[gauge] (g5) at (5,0) {$\,1\,$};
    \node[flavor] (f0) at (-1,0) {$\,6\,$};
    \draw (g0)--(g1)--(g2)--(g3)--(g4)--(g5);
    \draw (g0)--(f0);
\end{tikzpicture}
\end{equation}
We then turn the $\mathfrak{su}(3)$ gauge node into a $\mathfrak{u}(3)$ node by gauging the $\mathfrak{u}(1)$ symmetry of this model
\begin{equation}
    \begin{tikzpicture}[scale=1.2,every node/.style={scale=1.2},font=\scriptsize]
    \node[gauge] (g0) at (0,0) {$\,5\,$};
    \node[gauge] (g1) at (1,0) {$\,4\,$};
    \node[gauge] (g2) at (2,0) {$\,3\,$};
    \node[gauge] (g3) at (3,0) {$\,3\,$};
    \node[gauge] (g4) at (4,0) {$\,2\,$};
    \node[gauge] (g5) at (5,0) {$\,1\,$};
    \node[flavor] (f0) at (-1,0) {$\,6\,$};
    \draw (g0)--(g1)--(g2)--(g3)--(g4)--(g5);
    \draw (g0)--(f0);
\end{tikzpicture}
\end{equation}
Now we apply the duality \eqref{uglyduality} several times until we get a quiver with no ugly nodes. This reproduces the $T[\mathfrak{su}(6)]$ theory with $1+2$ twisted hypers, where we are again distinguishing those that are charged under the topological symmetry that we need to gauge to go back to the original theory with the special unitary node from those that are not. Closing the regular puncture $[1^6]\to[5,1]$ gives us the following quiver theory:
\begin{equation}\label{eq:gaugingex3quiv}
    \begin{tikzpicture}[scale=1.2,every node/.style={scale=1.2},font=\scriptsize]
    \node[gauge] (g0) at (0,0) {$\,1\,$};
    \node[gauge] (g1) at (1,0) {$\,1\,$};
    \node[gauge] (g2) at (2,0) {$\,1\,$};
    \node[gauge] (g3) at (3,0) {$\,1\,$};
    \node[gauge] (g4) at (4,0) {$\,1\,$};
    \node[flavor] (f0) at (-1,0) {$\,1\,$};
    \node[flavor] (f1) at (5,0) {$\,1\,$};
    \draw (f0)--(g0)--(g1)--(g2)--(g3)--(g4)--(g5)--(f1);
\end{tikzpicture}
\end{equation}
with the $1+2$ twisted hypers. In order to obtain the 3d reduction of the $\ADAI{8}{6}{[5,1]}$ we should now gauge back the topological symmetry, since we can exploit the fact that the procedure of gauging and the flow triggered by the HB vev commute. Nevertheless, as before it is enough for us to deal with the quiver in \eqref{eq:gaugingex3quiv}.

We propose that the $\ADAI{8}{6}{[5,1]}$ theory can be expressed as the following gauging of two type I theories,
\begin{equation}\label{eq:gaugingex3}
    \begin{split}
        \ADAI{8}{6}{[5,1]} \cong \ADAI{4}{5}{[3,1^2]} \longleftarrow &\mathfrak{su}(2) \longrightarrow \ADAIF{4}{3}~.
    \end{split}
\end{equation}
Note that both the theories on the r.h.s.\ of \eqref{eq:gaugingex3} have no exactly marginal deformations. Furthermore, the flavour symmetry of $\ADAI{4}{5}{[3,1^2]}$ is $\mathfrak g_F^{(1)}=\mathfrak{su}(2)\oplus\mathfrak{u}(1)$ while of $\ADAIF{4}{3}$ is $\mathfrak g_F^{(2)}=\mathfrak{su}(3)\supset\mathfrak{su}(2)\oplus\mathfrak{u}(1)$. The diagonal $\mathfrak{su}(2)$ subalgebra of $\mathfrak{su}(2)\oplus\mathfrak{su}(2) \subset \mathfrak g_F^{(1)}\oplus\mathfrak g_F^{(2)}$ is gauged such that after gauging the remaining flavour symmetry is $\mathfrak{u}(1)\oplus\mathfrak{u}(1)$, as expected.

The Coulomb branch spectra of the theories on the right are $\left\{\frac54,\frac32,\frac74\right\}$ and $\left\{\frac54,\frac32,\frac94\right\}$. These, along with an operator with dimension 2 coming from the $\mathcal N=2$ vector multiplet involved in the gauging, reproduce the Coulomb branch spectrum of our original SCFT \eqref{eq:gaugingex3CB}.
It is also possible to compute the $c$ and $a$ central charges from the r.h.s.\ of the expression
\begin{equation}\label{eq:ccmatchex3}
    \left(\frac{25}{6},\frac{49}{12}\right) = \left(\frac53,\frac{19}{12}\right) + 3\left(\frac16,\frac{5}{24}\right) + \left(2,\frac{15}{8}\right)~.
\end{equation}
The numbers of the r.h.s.\ of \eqref{eq:ccmatchex3} correspond to the contribution of $\ADAI{4}{5}{[3,1^2]}$, a 4d $\mathcal N=2$ vector multiplet, and $\ADAIF{4}{3}$, respectively. The factor multiplying the contribution of the vector multiplet corresponds to the dimension of the adjoint representation under the gauge algebra.

The conformality of this gauging can be confirmed once again by computing the anomaly
\begin{equation}
    \mathrm{Tr} R\,\mathfrak{su}(2)^2 \propto k_{\mathfrak{su}(2)}^{(1)} - 4\times2 + k_{\mathfrak{su}(2)}^{(2)}~,
\end{equation}
where $k_{\mathfrak{su}(2)}^{(1)}$ and $k_{\mathfrak{su}(2)}^{(2)}$ are the 4d flavour central charges of the $\mathfrak{su}(2)$ subalgebras of $\mathfrak g_F^{(1)}$ and $\mathfrak g_F^{(2)}$, respectively. These flavour charges can be computed using \eqref{eq:ksuNfullpunc}, \eqref{eq:ksulredpunc}, and \eqref{eq:kGBcont} to be
\begin{equation}
    k_{\mathfrak{su}(2)}^{(1)} = \frac72 \;,\quad k_{\mathfrak{su}(2)}^{(2)} = \frac92~.
\end{equation}
This implies that the above anomaly and hence the beta function of the gauge coupling vanish, which ensures that this gauging is conformal. 

In order to match the 3d reduction we proceed as before. We start from the 3d theory describing the circle reduction of $\ADAIF{4}{5}$, which is given by the $T^\sigma_\rho[\mathfrak{su}(5)]$ theory with $\sigma=[1^5]$ and $\rho=[2,1^3]$. Closing the puncture $[1^5]\to[3,1^2]$ thus gives us the $T^\sigma_\rho[\mathfrak{su}(5)]$ theory with $\sigma=[3,1^2]$ and $\rho=[2,1^3]$, which is the 3d reduction of the first theory on the r.h.s.\ of \eqref{eq:gaugingex3} and whose quiver is
\begin{equation}
    \begin{tikzpicture}[scale=1.2,every node/.style={scale=1.2},font=\scriptsize]
    \node[gauge] (g0) at (0,0) {$\,1\,$};
    \node[gauge] (g1) at (1,0) {$\,1\,$};
    \node[gauge] (g2) at (2,0) {$\,1\,$};
    \node[flavor] (f0) at (-1,0) {$\,2\,$};
    \node[flavor] (f1) at (3,0) {$\,1\,$};
    \draw (f0)--(g0)--(g1)--(g2)--(f1);
\end{tikzpicture}
\end{equation}
On the other hand, the 3d reduction of the second theory on the r.h.s.\ of \eqref{eq:gaugingex3} involved in the gauging, $\ADAIF{4}{3}$, is
\begin{equation}
    \begin{tikzpicture}[scale=1.2,every node/.style={scale=1.2},font=\scriptsize]
    \node[gauge] (g0) at (0,0) {$\,2\,$};
    \node[gauge] (g1) at (1,0) {$\,1\,$};
    \node[flavor] (f0) at (-1,0) {$\,3\,$};;
    \draw (f0)--(g0)--(g1);
\end{tikzpicture}
\end{equation}
We now gauge a diagonal combination of the $\mathfrak{su}(2)$ symmetries of these two quivers, where for the second one we decompose $\mathfrak{su}(3)\to \mathfrak{su}(2)\oplus \mathfrak{u}(1)$ which amounts to splitting the flavours as $3=2+1$
\begin{equation}
    \begin{tikzpicture}[scale=1.2,every node/.style={scale=1.2},font=\scriptsize]
    \node[gauge] (g0) at (0,0) {$\,1\,$};
    \node[gauge] (g1) at (1,0) {$\,1\,$};
    \node[gauge] (g2) at (2,0) {$\,1\,$};
    \node[gauge] (g3) at (3,0) {$\mathfrak{su}(2)$};
    \node[gauge] (g4) at (4,0) {$\,2\,$};
    \node[gauge] (g5) at (5,0) {$\,1\,$};
    \node[flavor] (f0) at (0,1) {$\,1\,$};
    \node[flavor] (f1) at (4,1) {$\,1\,$};
    \draw (f0)--(g0)--(g1)--(g2)--(g3)--(g4)--(g5);
    \draw (g4)--(f1);
\end{tikzpicture}
\end{equation}
Then, we turn the $\mathfrak{su}(2)$ gauge node into $\mathfrak{u}(2)$
\begin{equation}
    \begin{tikzpicture}[scale=1.2,every node/.style={scale=1.2},font=\scriptsize]
    \node[gauge] (g0) at (0,0) {$\,1\,$};
    \node[gauge] (g1) at (1,0) {$\,1\,$};
    \node[gauge] (g2) at (2,0) {$\,1\,$};
    \node[gauge] (g3) at (3,0) {$\,2\,$};
    \node[gauge] (g4) at (4,0) {$\,2\,$};
    \node[gauge] (g5) at (5,0) {$\,1\,$};
    \node[flavor] (f0) at (0,1) {$\,1\,$};
    \node[flavor] (f1) at (4,1) {$\,1\,$};
    \draw (f0)--(g0)--(g1)--(g2)--(g3)--(g4)--(g5);
    \draw (g4)--(f1);
\end{tikzpicture}
\end{equation}
and we keep applying the duality \eqref{uglyduality} until we get a quiver with no ugly nodes left. The result is exactly \eqref{eq:gaugingex3quiv} with $1+2$ free twisted hypermultiplets, where the 1 is charged under the topological symmetry that we would need to gauge to go back to the actual 3d reduction of the AD theory.

It is useful once again to try and understand how the gauging result \eqref{eq:gaugingex3} can be obtained from the one for the theory with the maximal puncture by Higgsing the $\mathfrak{su}(6)$ flavour symmetry. This in particular allows us to understand how the global symmetry carried by the regular puncture of the theory $\ADAI{8}{6}{[5,1]}$ gets distributed between the two theories that are gauged in \eqref{eq:gaugingex3}. Again, we start considering the $\ADAI{8}{6}{[4,2]}$ theory, from which the theory $\ADAI{8}{6}{[5,1]}$ we are interested in can be obtained with a single minimal transition. The HB of the former theory is isomorphic to an $A_3$ singularity described by the algebraic relation
\begin{align}
    xy=z^4~,
\end{align}
where $x$, $y$, $z$ are certain chiral operators in the theory that satisfy such relation. The Higgsing we are interested in corresponds to giving a vev to, for example, $x$ but not to $y$ and $z$
\begin{align}\label{eq:vevs}
    \langle x \rangle \neq 0\,,\qquad \langle y \rangle = \langle z \rangle = 0~.
\end{align}

As we will see from the general result that we will present momentarily, the $\ADAI{8}{6}{[4,2]}$ theory can be expressed as
\begin{eqnarray}\label{eq:gaugingex3bis}
        \ADAI{8}{6}{[4,2]} \cong \ADAI{4}{5}{[2,1^3]} \longleftarrow &\mathfrak{su}(3)& \longrightarrow \ADAI{4}{3}{[1^3]}~,\nn\\
        &|&\nn\\
        &\fbox{1}&
\end{eqnarray}
We would then like to understand how the operator $x$ is realised in this description of the theory and what is the effect of giving it a vev. For this we focus on the theory $\ADAI{4}{5}{[2,1^3]}$. This can be obtained from the theory $\ADAI{4}{5}{[1^5]}$ with a nilpotent vev. We should then investigate the branching rule for the moment map operator in the adjoint representation of the original $\mathfrak{su}(5)$ symmetry under the symmetry breaking pattern $\mathfrak{su}(5)\to \mathfrak{su}(3)\oplus \mathfrak{su}(2)\oplus \mathfrak{u}(1)$
\begin{align}
    { 24}\to \left(V_0,{ 8}\right)^{0}\oplus\left(V_0,{ 1}\right)^{0}\oplus\left(V_{\frac{1}{2}},{ 3}\right)^{+1}\oplus\left(V_{\frac{1}{2}},\overline{ 3}\right)^{-1}\oplus\left(V_1,{ 1}\right)^{0}~.
\end{align}
The $\left(V_{\frac{1}{2}},{ 3}\right)^{+1}$ and $\left(V_{\frac{1}{2}},\overline{ 3}\right)^{-1}$ components give rise to two chiral operators that are respectively in the fundamental and anti-fundamental representation of the gauged $\mathfrak{su}(3)$ symmetry, which we denote by $W^a$ and $\tilde{W}_a$ where $a=1,2,3$ is the colour index . The aforementioned operators $x$ and $y$ are constructed taking the following gauge invariant combinations of $W^a$ and $\tilde{W}_a$ with the chirals $q^a$ and $\tilde{q}_a$ inside the fundamental hypermultiplet in \eqref{eq:gaugingex3bis}:
\begin{align}
    x=W^a \tilde{q}_a\,,\qquad y=\tilde{W}_a q^a~.
\end{align}
The vevs \eqref{eq:vevs} can then be implemented by
\begin{align}
    \langle W^1\rangle \neq 0\,,\qquad \langle \tilde{q}_1\rangle \neq 0~,
\end{align}
while all other components do not have a vev. This has the effect of Higgsing the $\mathfrak{su}(3)$ gauge algebra down to $\mathfrak{su}(2)$. Moreover, the original hypermultiplet in the fundamental representation of $\mathfrak{su}(3)$ disappears, since one component acquired a vev while the other components recombine with the broken generators of the gauge symmetry into long multiplets (see \emph{e.g.}~\cite{Agarwal:2014rua}). Finally, the vevs have the effect of Higgsing the theory $\ADAI{4}{5}{[2,1^3]}$ to $\ADAI{4}{5}{[3,1^2]}$. The other theory $\ADAI{4}{3}{[1^3]}$ is instead not affected by the vev, but now only an $\mathfrak{su}(2)$ subalgebra of its $\mathfrak{su}(3)$ symmetry is gauged due to the mechanism of gauge-flavor locking which locked the broken $\mathfrak{u}(1)$ part of the original $\mathfrak{su}(3)$ gauge algebra with a combination of the $\mathfrak{u}(1)$ flavour symmetries under which the operator $x$ that acquired a vev was charged. Overall, we can see that we recovered \eqref{eq:gaugingex3}.

\subsubsection*{Example 4: $\ADAI{8}{4}{[2,1^2]}$}

As a last example, we consider a theory with two exactly marginal deformations. The Coulomb branch spectrum of this theory is
\begin{equation}\label{eq:gaugingex4CB}
    \left\{\frac32,\frac32,\frac32,2,2,\frac52,\frac52,3\right\}~.
\end{equation}
Furthermore, since the regular puncture is not maximal in this case, the $c$ central charge has to obtained by considering this theory as the IR limit of the RG flow triggered by a nilpotent Higgsing of the form $[2,1^2]$ from the UV theory $\ADAIF{8}{4}$. The central charges computed this way turn out to be
\begin{equation}
    c = \frac{27}{4} \quad,\; a = \frac{13}{2}~.
\end{equation}
The flavour symmetry algebra of this theory is $\mathfrak{su}(2)\oplus\mathfrak{u}(1)^4$, where the regular puncture contributes $\mathfrak{su}(2)\oplus\mathfrak{u}(1)$.

We propose that this theory can be expressed as the following gauging of two type I theories,
\begin{eqnarray}\label{eq:gaugingex4part}
        \ADAI{8}{4}{[2,1^2]} \cong \ADAIF{2}{5} \longleftarrow &\mathfrak{su}(3)& \longrightarrow \ADAIF{6}{3}~.\nn\\
        &|&\nn\\
        &\fbox{1}&
\end{eqnarray}
Note that only one of the theories on the r.h.s.\ of \eqref{eq:gaugingex4part} has no exactly marginal deformations. In particular, $\ADAIF{6}{3}$ has one exactly marginal deformation. The additional exactly marginal deformation is then carried by the $\mathfrak{su}(3)$ gauging. The flavour symmetry of $\ADAIF{2}{5}$ is $\mathfrak g_F^{(1)}=\mathfrak{su}(5)\supset\mathfrak{su}(3)\oplus\mathfrak{su}(2)\oplus\mathfrak{u}(1)$ while the flavour symmetry of $\ADAIF{6}{3}$ is $\mathfrak g_F^{(2)}=\mathfrak{su}(3)\oplus\mathfrak{u}(1)^2$. The residual flavour symmetry algebra after gauging the diagonal $\mathfrak{su}(3)$ is $\mathfrak{su}(2)\oplus\mathfrak{u}(1)^4$, where once again the additional $\mathfrak{u}(1)$ is from the hypermultiplet in the fundamental of $\mathfrak{su}(3)$, which matches the flavour symmetry of the original theory we are considering .

The Coulomb branch spectra of the theories on the r.h.s.\ of \eqref{eq:gaugingex4part} are $\left\{\frac32,\frac52\right\}$ and $\left\{\frac32,\frac32,2,\frac52\right\}$. Since the gauge algebra now is $\mathfrak{su}(3)$, the Coulomb branch operators built from the 4d $\mathcal N=2$ vector multiplet are of dimensions 2 and 3. Summing these, we reproduce the Coulomb branch spectrum \eqref{eq:gaugingex4CB}.
It is also possible to compute the $c$ and $a$ central charges from the r.h.s.\ of the expression
\begin{equation}\label{eq:ccmatchex4part}
    \left(\frac{27}{4},\frac{13}{2}\right) = \left(2,\frac74\right) + 8\left(\frac16,\frac{5}{24}\right) + \left(\frac{19}{6},\frac{71}{24}\right) + 3\left(\frac{1}{12},\frac{1}{24}\right)~.
\end{equation}
The numbers of the r.h.s.\ of \eqref{eq:ccmatchex4part} correspond to the contribution of $\ADAIF{2}{5}$, a 4d $\mathcal N=2$ vector multiplet, $\ADAIF{6}{3}$, and the 4d $\mathcal N=2$ hypermultiplet in the fundamental of $\mathfrak{su}(3)$, respectively. The factors multiplying the contributions of the supermultiplets correspond to the dimensions of their respective representations under the gauge algebra.

The conformality of this gauging can be confirmed once again by computing the anomaly
\begin{equation}
    \mathrm{Tr} R\,\mathfrak{su}(3)^2 \propto k_{\mathfrak{su}(3)}^{(1)} - 4\times3 + k_{\mathfrak{su}(3)}^{(2)} +4\times\frac12~,
\end{equation}
where $k_{\mathfrak{su}(3)}^{(1)}$ and $k_{\mathfrak{su}(3)}^{(2)}$ are the 4d flavour central charges of the $\mathfrak{su}(3)$ subalgebras of $\mathfrak g_F^{(1)}$ and $\mathfrak g_F^{(2)}$, respectively. These can be computed to be
\begin{equation}
    k_{\mathfrak{su}(3)}^{(1)} = k_{\mathfrak{su}(3)}^{(2)} = 5~.
\end{equation}
This implies that the above anomaly and hence the beta function of the gauge coupling vanish, so that this gauging is conformal. It must be noted here that the extra fundamental $\mathcal N=2$ hypermultiplet is crucial for ensuring that the gauging is conformal and that the 4d central charges of the original theory are reproduced from its description as the gauging of type I theories as in \eqref{eq:ccmatchex4part}.

However, note that the second theory on the r.h.s.\ of \eqref{eq:gaugingex4part}, $\ADAIF{6}{3}$, is itself a type I theory with an exactly marginal deformation. Therefore we can further express it as a gauging of two type I Argyres--Douglas theories as follows,
\begin{equation}\label{eq:gaugingex4part2}
    \ADAIF{6}{3} \cong \ADAIF{2}{5} \longleftarrow \mathfrak{su}(2) \longrightarrow \ADAIF{4}{2}~.
\end{equation}
We have already seen that the theory on the l.h.s.\ of \eqref{eq:gaugingex4part2} has central charges $(c,a)=\left(\frac{19}{6},\frac{71}{24}\right)$, flavour symmetry $\mathfrak{su}(3)\oplus \mathfrak{u}(1)^2$, and Coulomb branch spectrum $\left\{\frac32,\frac32,2,\frac52\right\}$.

These can be reproduced from the $\mathfrak{su}(2)$ gauging of the type I theories without any exactly marginal deformations on the r.h.s.\ of \eqref{eq:gaugingex4part2}. In particular, the flavour symmetry algebra of $\ADAIF{2}{5}$ as before can be expressed as $\mathfrak g_F^{(1)}=\mathfrak{su}(5)\supset\mathfrak{su}(3)\oplus\mathfrak{su}(2)\oplus\mathfrak{u}(1)$ while that of $\ADAIF{4}{2}$ is $\mathfrak g_F^{(3)}=\mathfrak{su}(2)\oplus\mathfrak{u}(1)$ which leaves $\mathfrak{su}(3)\oplus\mathfrak{u}(1)^2$ after gauging the diagonal $\mathfrak{su}(2)$ subalgebra. Meanwhile, for the Coulomb branch spectrum we have that
\begin{equation}
    \left\{\frac32,\frac32,2,\frac52\right\} = \left\{\frac32,\frac52\right\} \cup \{2\} \cup \left\{\frac32\right\}~.
\end{equation}
Whereas for the central charges we verify that
\begin{equation}
    \left(\frac{19}{6},\frac{71}{24}\right) = \left(2,\frac74\right) + 3\left(\frac16,\frac{5}{24}\right) + \left(\frac23,\frac{7}{12}\right)~,
\end{equation}
where the numbers correspond to the contributions of $\ADAIF{2}{5}$, a 4d $\mathcal N=2$ vector multiplet, and $\ADAIF{4}{2}$, respectively.

We once again have to ensure that this $\mathfrak{su}(2)$ gauging here is conformal. Given that in this case
\begin{equation}
    k_{\mathfrak{su}(2)}^{(1)} = 5 \;,\quad k_{\mathfrak{su}(2)}^{(3)}= 3~,
\end{equation}
it immediately follows that the corresponding ABJ anomaly vanishes
\begin{equation}
    \mathrm{Tr} R\,\mathfrak{su}(2)^2 \propto k_{\mathfrak{su}(2)}^{(1)} - 4\times2 + k_{\mathfrak{su}(2)}^{(3)} = 0~.
\end{equation}
Therefore the $\mathfrak{su}(2)$ gauging here is conformal as well.

Summarising, we find that one can view the original theory, $\ADAI{8}{4}{[2,1^2]}$ as the conformal gaugings of three type I Argyres--Douglas theories without any exactly marginal couplings
\begin{eqnarray}\label{eq:gaugingex4}
        \ADAI{8}{4}{[2,1^2]} \cong \ADAIF{2}{5} \longleftarrow &\mathfrak{su}(3)& \longrightarrow \ADAIF{2}{5} \longleftarrow \mathfrak{su}(2) \longrightarrow \ADAIF{4}{2}~.\nn\\
        &|&\nn\\
        &\fbox{1}&
\end{eqnarray}
Interestingly, one can also check that
\begin{eqnarray}\label{eq:gaugingex4extra}
        \ADAI{4}{10}{[3^2,2,1^2]} \cong \ADAIF{2}{5} \longleftarrow &\mathfrak{su}(3)& \longrightarrow \ADAIF{2}{5}~.\nn\\
        &|&\nn\\
        &\fbox{1}&
\end{eqnarray}

All the above claims can be further checked by matching the 3d reduction of these theories. We avoid showing this explicitly here, since the derivation is conceptually equivalent to those of the previous examples but it is more convoluted. Moreover, one can also in principle repeat the Higgsing analysis, but again we refrain from presenting it explicitly here.

After this lengthy exposition through various representative examples, we will now state the general result for type I Argyres--Douglas theories with exactly marginal deformations. We have checked this general result in a large number of representative cases by performing the tests discussed for these four examples using \texttt{Mathematica}.

\subsubsection{\label{subsec:typeIEMCgen}General result}

Consider a general type I Argyres--Douglas theory, $\ADAI{p}{N}{[Y]}$, such that $m=\gcd(p,N)\neq1$. We also consider the most generic partition,
\begin{equation}
    [Y] = \left[N^{l_N},\dots,2^{l_2},1^{l_1}\right] \quad,\; L=\sum_{i=1}^Nl_i~.
\end{equation}
We define the related conjugate partition $[\widetilde Y]$ given the original partition $[Y]$ and $q=\frac{p}{m}$,
\begin{equation}\label{eq:Ytilde}
    [\widetilde Y] = \left[\left(q-1\right)^{l_1},\left(q-2\right)^{l_2},\dots,\left(q-N\right)^{l_N}\right]~,
\end{equation}
which for $m=1$ is equal to $[Y^c]$. The entries in the partition $[\widetilde Y]$ sum up to $qL-N$ but they need not be all positive. Thus we collect the positive entries as $[\widetilde Y_+]$ and the absolute values of the negative entries as $[\widetilde Y_-]$, which can themselves be considered as partitions of positive integers $\widetilde N_+$ and $\widetilde N_-$, respectively. Finally $[\widetilde Y]$ may also contain some zeroes and we denote this number by $\widetilde N_0$.

This partition $[\widetilde Y]$ can easily be expressed pictorially. We demonstrate this by using an example where $[Y]=[4^2,3,2,1^3]$ and $p=6$, therefore $N=16$, $m=2$, and $q=3$. Consider the Young tableaux corresponding to this partition
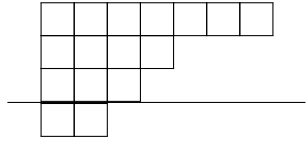
\begin{center}
    \ytableausetup{baseline,boxframe=normal,boxsize=1.2em}
    \begin{ytableau}
        \none & & & & & & & & \none \\
        \none & & & & & \none \\
        \none & & & & \none \\
        \hline
        \none & &
    \end{ytableau}
\end{center}
The horizontal line is placed after $q$ (in this case three) boxes and it splits $[Y]$ into two halves. First we note that there is exactly one column of height exactly equal to $q$, which determines $\widetilde N_0=1$. Next, we denote the partition below the line by $[\widetilde Y_-]$, which in this case is simply $[\widetilde Y_-]=[1^2]$. Finally, we consider the sub-diagram of this Young tableaux above the line and take its complement partition above the line and denote it by $[\widetilde Y_+]$, which in this case is $[\widetilde Y_+]=[2^3,1]$.
The following diagram makes these definitions evident:
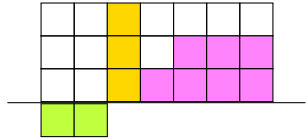
\begin{center}
    \ytableausetup{baseline,boxframe=normal,boxsize=1.2em}
    \begin{ytableau}
        \none & & & *(Gold1) & & & & & \none \\
        \none & & & *(Gold1) & & *(Orchid1) & *(Orchid1) & *(Orchid1) & \none \\
        \none & & & *(Gold1) & *(Orchid1) & *(Orchid1) & *(Orchid1) & *(Orchid1) & \none \\
        \hline
        \none & *(OliveDrab1) & *(OliveDrab1)
    \end{ytableau}
\end{center}
where the magenta partition denotes $[\widetilde Y_+]=[2^3,1]$ (after rotating it by $180^\circ$), the olive partition denotes $[\widetilde Y_-]=[1^2]$, and gold denotes the columns of height $q=3$, which are in number $\widetilde N_0=1$. 

Using these definitions, we now present the full conjecture that expresses any type I Argyres--Douglas theory as the conformal gauging of two type I Argyres--Douglas theories
\begin{equation}\label{eq:typeIwEMD}
    \begin{split}
        \ADAI{p}{N}{\left[Y\right]} \cong &\ADAI{q}{qL-\frac{N}{m}}{\left[\widetilde{Y}_+,1^{N-\frac{N}{m}-\widetilde{N}_-}\right]} \\
        &\hspace*{3cm}\Big\uparrow \\
        &\hspace*{1cm} \mathfrak{su}\left(N-\frac{N}{m}-\widetilde N_-\right) \text{\textemdash\textemdash}\; \fbox{$\widetilde N_0$} \\
        &\hspace*{3cm}\Big\downarrow \\
        &\ADAI{(m-1)q}{N-\frac{N}{m}}{\left[\widetilde Y_-,1^{N-\frac{N}{m}-\widetilde N_-}\right]}~.
    \end{split}
\end{equation}
This expression must be interpreted as stating that the original theory with exactly marginal deformations can be expressed as the conformal gauging of the two type I Argyres--Douglas theories. 

It is straightforward to see that the first theory on the r.h.s\ of \eqref{eq:typeIwEMD} does not contain any exactly marginal deformations as
\begin{equation}
    \gcd\left(q,\frac{N}{m}\right) = 1~.
\end{equation}
Meanwhile the other theory on the r.h.s.~of \eqref{eq:typeIwEMD} generically contains $(m-2)$ exactly marginal deformations as
\begin{equation}
    \gcd\left((m-1)q,N-\frac{N}{m}\right) = m-1~.
\end{equation}
Thus it has one less exactly marginal deformation than the original theory as each special unitary gauging is responsible for one exactly marginal deformation as we have noted in the examples in the previous subsection. However, since this theory itself is a type I Argyres--Douglas theory, we can apply this expression once again to the bottom theory on the r.h.s.\ of \eqref{eq:typeIwEMD}. Thus one eventually ends up with an expression for the type I theory on the left as a linear sequence of $m-1$ special unitary conformal gaugings of $m$ type I Argyres--Douglas theories without any exactly marginal deformations.\footnote{More general non-linear conformal gaugings of Argyres--Douglas theories have also been considered in the literature, see \emph{e.g.}~\cite{Closset:2020afy,Kang:2021lic,Kang:2022zsl,Carta:2023bqn}.}

We point out that a version of this result is present in \cite{Xie:2017vaf}. The presentation in \cite{Xie:2017vaf} is in terms of an auxiliary Riemann surface whose decomposition into pairs of pants, corresponding to AD theories without EMDs, contains the same information as \eqref{eq:typeIwEMD}. The advantage of \eqref{eq:typeIwEMD} over the auxiliary Riemann surface description is that the results here are expressed very explicitly. Further, the presentation here clearly expresses type I theories with exactly marginal deformations as gaugings of other type I theories without exactly marginal deformations, whereas in the auxiliary Riemann surface such a result is not immediately obvious as one has to use \eqref{eq:GentypeIiso} to re-express the theories associated to the pairs of pants as type I theories. Moreover, while the approach of \cite{Xie:2017vaf} correctly realises an AD theory with EMDs in terms of the gauging of AD theories without EMDs when the regular puncture is maximal, it still requires working out on a case by case basis the effect of the vev when we partially close the regular puncture. In particular, the Higgsing of the special unitary gauge algebras and the appearance of hypermultiplets charged under it that need to be worked out are instead already incorporated in our result \eqref{eq:typeIwEMD}. Finally, while the results in \cite{Xie:2017vaf} seem to be supported by matching of Coulomb branch spectrum and central charges, we provide an additional non-trivial check in terms of their 3d reductions.

We will now provide a pictorial representation of \eqref{eq:typeIwEMD}. For concreteness, we will consider the following sufficiently complex example to demonstrate this representation: $\ADAI{9}{30}{[8,6,5^2,3,1^3]}$. This Argyres--Douglas theory has $m=3$, $q=3$ and thus has two exactly marginal deformations. Therefore it is the gauging of three type I AD theories without exactly marginal deformations. The pictorial description can be constructed by splitting the Young tableaux representing $[Y]=[8,6,5^2,3,1^3]$ into $m=3$ sections in the following way:
\begin{align}\label{eq:ExYTSplit}
    \ytableausetup{baseline,boxframe=normal,boxsize=1.2em}
    \textbf{I }:\;&\ydiagram[]{4,4,4}\ydiagram[*(CadetBlue1)]{4,1,1} \nn \\
    \textbf{II }:\;&\ydiagram[]{1,1,1}\ydiagram[*(PaleGreen1)]{3,3,1} \nn \\
    \textbf{III }:\;&\ydiagram[*(IndianRed1)]{1,1}
\end{align}
The three sections \textbf{I}, \textbf{II}, and \textbf{III} will describe the three type I AD theories without exactly marginal deformations. We now explain how to read off the data for the AD theories corresponding to each section.

Starting with \textbf{I}, we look at the part of $[Y]$ that has columns of length at most $q=3$, coloured in blue in the figure above. The $\mathfrak{su}(r^{(1)})$ special unitary algebra that connects the first and the second theory is determined by the expression
\begin{equation}
    r^{(1)} = N-\frac{N}{m}\times1-\widetilde N_-^{(1)} = 30-\frac{30}{3}-12 = 8~,
\end{equation}
where $\widetilde N_-^{(1)}$ denotes the number of boxes below the first horizontal line. Additionally we note that there is one column of height exactly $q$ and hence there is one hypermultiplet in the fundamental of the gauged $\mathfrak{su}(8)$. Now add $r^{(1)}$ columns of height $q-1=2$ to the tableaux in blue and $r^{(1)}$ columns of height 1 to the tableaux in green as shown below:
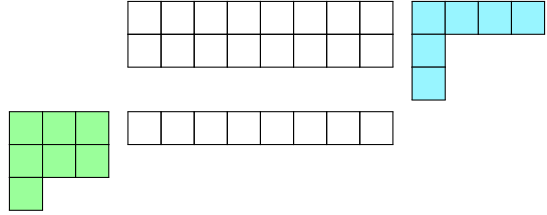
\begin{eqnarray}
    \ytableausetup{baseline,boxframe=normal,boxsize=1.2em}
    &\ydiagram[]{8,8}&\;\ydiagram[*(CadetBlue1)]{4,1,1} \nn \\
    \ydiagram[*(PaleGreen1)]{3,3,1}\;&\ydiagram[]{8}& \nn 
\end{eqnarray}
The partition that labels the AD theory \textbf{I} in this diagram is simply given by the boxes on the top line, $[3,2^8,1^3]$, and it is gauged to the theory in section \textbf{II} by an $\mathfrak{su}(8)$ gauging with a single hypermultiplet in its fundamental representation.

However since the original theory has two exactly marginal deformations we repeat this procedure sequentially. Now focusing on the gauging between section \textbf{II} and \textbf{III}, the $\mathfrak{su}(r^{(2)})$ special unitary gauge algebra that connects the second and the third theory is determined by
\begin{equation}
    r^{(2)} = N-\frac{N}{m}\times2-\widetilde N_-^{(2)} = 30-\frac{30}{3}\times2-2 = 8~.
\end{equation}
We repeat the above diagrammatic construction to obtain the following picture
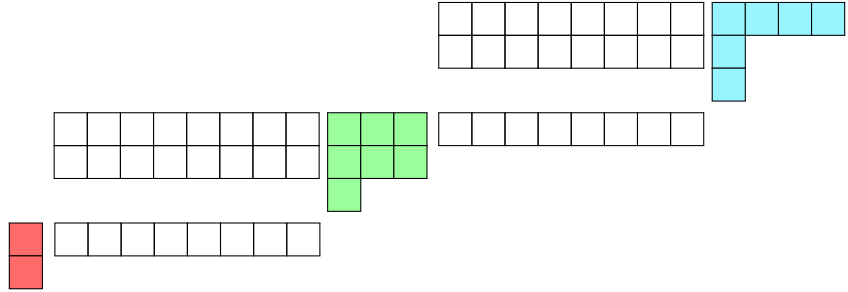
\begin{eqnarray}
    \ytableausetup{baseline,boxframe=normal,boxsize=1.2em}
    &\begin{ytableau}
        \none & \none & \none & \none & \none & \none & \none & \none & \none & \none
    \end{ytableau}&\ydiagram[]{8,8}\;\ydiagram[*(CadetBlue1)]{4,1,1} \nn \\
    &\ydiagram[]{8,8}\;\ydiagram[*(PaleGreen1)]{3,3,1}&\ydiagram[]{8} \nn \\
    \ydiagram[*(IndianRed1)]{1,1}&\hspace*{-1.4cm}\ydiagram[]{8} \nn
\end{eqnarray}
Once again the partition labelling the AD theory \textbf{II} is determined by the boxes on the second line, \emph{i.e.}, $[3,2^2,2^8,1^8]$. Further, there is another column of length $q=3$ in the second line and hence it contributes a single hypermultiplet in the fundamental of the gauged $\mathfrak{su}(8)$. Therefore this picture denotes the following gauging of three AD theories,
\begin{eqnarray}\label{eq:ExNonConjDiag}
    \ADAI{3}{22}{[3,2^8,1^3]} \longleftarrow &\mathfrak{su}(8)& \longrightarrow \ADAI{3}{31}{[3,2^{10},1^8]} \longleftarrow \mathfrak{su}(8) \longrightarrow \ADAI{3}{10}{[2,1^8]}~,\nn\\
    &|&\nn\hspace*{5.25cm}|\\
    &\fbox{1}&\hspace*{5.07cm}\fbox{1}
\end{eqnarray}
where we point out that all three theories have the same subscript $q=3$, which is the same as the $q$ for the original theory. Further we notice that all three type I AD theories in \eqref{eq:ExNonConjDiag} do not have any exactly marginal deformations and therefore we can apply \eqref{eq:isoql} and \eqref{eq:iso2} to obtain
\begin{eqnarray}
    \ADAI{3}{14}{[2^3,1^8]} \longleftarrow &\mathfrak{su}(8)& \longrightarrow \ADAI{3}{26}{[2^8,1^{10}]} \longleftarrow \mathfrak{su}(8) \longrightarrow \ADAI{3}{10}{[2,1^8]}~.\nn\\
    &|&\nn\hspace*{5cm}|\\
    &\fbox{1}&\hspace*{4.8cm}\fbox{1}
\end{eqnarray}
Note that here we leave the final theory untouched so that this expression exactly matches the one obtained by repeated applications of \eqref{eq:typeIwEMD}.

The above analysis can be done for a generic partition, and we summarise the steps involved here:
\begin{enumerate}
    \item Split the general Young tableaux into $m$ sections of height $q$.
    \item Isolate the sub-diagrams that contain columns that do not extend below a given section; these are the coloured boxes in \eqref{eq:ExYTSplit}.
    \item The $\mathfrak{su}(r^{(i)})$ gauge algebra that connects the $i$-th section to the $(i+1)$-th section is determined by
    \begin{equation}\label{eq:RkGauge}
        r^{(i)} = N-\frac{N}{m} i-\widetilde N_-^{(i)}~,
    \end{equation}
    where $\widetilde N_-^{(i)}$ is the total number of boxes below the $i$-th section.
    \item Add $r^{(i)}$ columns of height $q-1$ to the $i$-th section and $r^{(i)}$ columns of height $1$ to the $(i+1)$-th section.
    \item The number of hypermultiplets charged under the $i$-th gauged symmetry is equal to number of columns of length $q$ in the $i$-th section.
    \item The partitions labelling the individual theories can then be read off by looking at the corresponding sections.
\end{enumerate}
Thus this algorithm enables us to use diagrammatic manipulations to recover the full expression that can be obtained by iterations of \eqref{eq:typeIwEMD}.

The gauging formula \eqref{eq:typeIwEMD} can be exploited to obtain several interesting results about the type I Argyres--Douglas theories $\ADAI{p}{N}{\left[Y\right]}$. In the next subsection we will show how it encompasses some of the equivalences between coprime type I theories that we discussed in Subsection \ref{subsec:typeIisowoEMD}, while in Subsection \ref{subsec:typeIisowEMD} we will show how it can be used to extend some of such equivalences to the non-coprime case. Here instead we explain how the gauging formula \eqref{eq:typeIwEMD} allows us to motivate the statement we made at the end of Subsection \ref{subsec:ADVOA}, that for $p<N$ the Argyres--Douglas theory $\ADAI{p}{N}{[p^x,s]}$, where $s= N\;(\text{mod}\;p)$, is an un-Higgsable SCFT. Indeed, let us consider the hypothetical theory 
\begin{equation}
    \ADAI{p}{N}{\left[Y\right]}\,,\quad [Y]=[p+\alpha,\cdots]\,,\quad p<N\,,\quad \alpha>0\,,
\end{equation}
which, if it exists, would be obtained by Higgsing from the one with partition $\ADAI{p}{N}{[p^x,s]}$. Then by applying \eqref{eq:typeIwEMD} several times, since of course
\begin{equation}
    lq<p<p+\alpha\,,\qquad l=1,\cdots,m-1\,,
\end{equation}
such entry will end up in the partition $[Y']$ characterising the very last theory at the end of the sequence of $m-1$ gaugings, where it will become $(p+\alpha)-(m-1)q=q+\alpha$. However, this last theory would be a coprime type I theory of the form
\begin{equation}
    \ADAI{q}{\frac{N}{m}}{\left[Y'\right]}\,,\quad [Y']=[q+\alpha,\cdots]\,,\quad q<\frac{N}{m}\,,\quad \alpha>0
\end{equation}
since we are considering the case $p<N$, which is not allowed (see Subsection \ref{subsec:ADVOA}). Hence, $\ADAI{p}{N}{[p^x,s]}$ for $p<N$ is an un-Higgsable SCFT.

\subsubsection{\label{subsec:gauging_generalities}Recovering exceptional equivalences for isolated type I theories}

In the previous Subsection \ref{subsec:typeIEMC} we motivated and presented the result \eqref{eq:typeIwEMD}. Even though we only considered its utility for type I Argyres--Douglas theories with exactly marginal deformations, \eqref{eq:typeIwEMD} is still valid and gives non-trivial predictions also when applied to cases without any exactly marginal deformation. In this subsection we discuss the more general applications of the expression \eqref{eq:typeIwEMD} to Argyres--Douglas theories without any EMDs. This includes the coprime Argyres--Douglas theories as well as non-coprime theories that do not have any EMDs.

Considering the coprime cases with $\gcd(p,N)=1$ in \eqref{eq:typeIwEMD}, we note that the rank of the gauge algebra is trivial $\mathfrak{su}(0)$, since there cannot be any negative entries in \eqref{eq:Ytilde} for $m=1$. Further, the theory at the bottom of \eqref{eq:typeIwEMD} corresponds to the trivial $\mathfrak{sl}_0$ and is an artefact of applying \eqref{eq:typeIwEMD} to a theory with $m=1$. Therefore such an expression must be viewed as merely the equivalence between the following two theories ($p=q$ in this case),
\begin{equation}
     \ADAI{q}{N}{\left[Y\right]} \equiv \ADAI{q}{qL-N}{[Y^c]}
\end{equation}
for $[Y^c]$ determined by \eqref{eq:Ytilde}. This reproduces the equivalence \eqref{eq:iso2} between coprime type I Argyres--Douglas theories.

The condition that $m=\gcd(N,p)=1$ is actually only sufficient, but not necessary for the absence of exactly marginal deformations. A useful diagnostic for when a theory has no exactly marginal deformations, apart from explicitly checking that there is no dimension 2 operator in the Coulomb branch spectrum, is to notice when in \eqref{eq:typeIwEMD} we have a single gauge node which however is a trivial $\mathfrak{su}(1)$ or $\mathfrak{su}(0)$. One instance is for $N=2$ and $p=2\kappa+2$, in which case there are no exactly marginal deformations even though $m=2$. In fact, these are well studied Argyres--Douglas theories constructible via geometric engineering
\begin{equation}
    \ADAIF{2\kappa+2}{2} = (A_1,D_{2\kappa+2}) \;\,\quad \ADAI{2\kappa+2}{2}{[2]} = (A_1,A_{2\kappa-1})~.
\end{equation}

Applying \eqref{eq:typeIwEMD} to these theories allows us to recover the equivalences \eqref{eq:TypeISU2Equiv}. In the first case, we indeed obtain
\begin{equation}
    \begin{split}
        \ADAIF{2\kappa+2}{2} &\cong \ADAI{\kappa+1}{2\kappa+1}{[\kappa^2,1]} \cong \ADAI{\kappa+1}{\kappa+2}{[\kappa,1^2]}~,
    \end{split}
\end{equation}
where the second equality can be obtained by applying \eqref{eq:typeIwEMD} once again or equivalently by using \eqref{eq:iso2}. An immediate implication of this equivalence is that the Higgs branch of $\ADAIF{2\kappa+2}{2}$ is
\begin{equation}
    \overline{\mathbb O_{[\kappa+1,1]}} \;\cap\; \mathcal S_{\mathfrak f_{[\kappa,1^2]}}~.
\end{equation}
This matches the result obtained in \cite{Beem:2017ooy} for the Higgs branches of $(A_1,D_{2\kappa+2})$ Argyres--Douglas theories and confirms a non-trivial equivalence predicted by applying \eqref{eq:typeIwEMD} to a case with no exactly marginal deformation.

Further considering the case with the closed regular puncture, we have
\begin{equation}
    \begin{split}
        \ADAI{2\kappa+2}{2}{[2]} &\cong \ADAI{\kappa+1}{\kappa}{[\kappa-1,1]}\cong \ADAI{\kappa+1}{\kappa+2}{[\kappa,2]}~.
    \end{split}
\end{equation}
The corresponding Higgs branch, according to the second equality, can be expressed as the intersection of the Slodowy slice at $\mathfrak f_{[\kappa,2]}$ with the closure of the nilpotent orbit $\mathbb O_{[\kappa+1,1]}$ of $\mathfrak{sl}_{\kappa+2}$~,
\begin{equation}
    \overline{\mathbb O_{[\kappa+1,1]}} \;\cap\; \mathcal S_{\mathfrak f_{[\kappa,2]}}~.
\end{equation}
However, according to the first line, the Higgs branch of the $(A_1,A_{2\kappa-1})$ Argyres--Douglas theories, as was shown in \cite{Beem:2017ooy}, is the closure of the minimal nilpotent orbit of $\mathfrak{sl}_k$ which is an $A$-type Kleinian singularity. Therefore we have the following identification of symplectic singularities:
\begin{equation}
    \overline{\mathbb O_{[\kappa+1,1]}} \;\cap\; \mathcal S_{\mathfrak f_{[\kappa,2]}} \cong \mathbb C^2/\mathbb Z_\kappa~.
\end{equation}
This identification is already known from \cite{slodowy2006simple} and thus serves as an additional piece of evidence for the conjectured equivalences.\footnote{PS would like to thank Julius Grimminger for pointing this out.}

\subsection{\label{subsec:typeIisowEMD}Equivalences in the non-coprime case}

In Subsection \ref{subsec:typeIisowoEMD} we have seen some equivalences between type I AD theories for which $m=\gcd(N,p)=1$ and so with no exactly marginal deformations, namely \eqref{eq:isoql} and \eqref{eq:iso2}. Such equivalences actually hold even in cases with $m>1$, as we will argue in this subsection. 

Let us start from the version of the equivalence \eqref{eq:isoql} for $m=\gcd(N,p)\geq1$
\begin{equation}\label{eq:isopl}
    \ADAI{p}{N}{\left[Y\right]} \equiv \ADAI{p}{pl+N}{\left[p^l,Y\right]}~.
\end{equation}
As in the coprime case one can easily check that the extra $\mathfrak{su}(l)$ flavour symmetry on the l.h.s.\ has trivial central charge, since in \eqref{eq:zeroflavcc} the coprimality condition did not play any role. Moreover, this result can be obtained by combining the result for $m=1$ with the gauging formula \eqref{eq:typeIwEMD}. Indeed, if we apply \eqref{eq:typeIwEMD} to the theory on the r.h.s.~we get (recall that $p=mq$)
\begin{equation}
    \begin{split}
        \ADAI{mq}{mql+N}{\left[(mq)^l,Y\right]} \equiv &\ADAI{q}{qL-\frac{N}{m}}{\left[\widetilde{Y}_+,1^{N-\frac{N}{m}-\widetilde{N}_-}\right]}\\
        &\hspace*{3cm}\Big\uparrow\\
        &\hspace*{1cm} \mathfrak{su}\left(N-\frac{N}{m}-\widetilde N_-\right) \text{\textemdash\textemdash}\; \fbox{$\widetilde N_0$}\\
        &\hspace*{3cm}\Big\downarrow\\
        &\ADAI{(m-1)q}{(m-1)ql+N-\frac{N}{m}}{\left[\left((m-1)q\right)^l,\widetilde Y_-,1^{N-\frac{N}{m}-\widetilde N_-}\right]}~.
    \end{split}
\end{equation}
The top theory involved in the gauging is identical to the one that appears in the gauging formula of the $\ADAI{p}{N}{\left[Y\right]}$ theory, see \eqref{eq:typeIwEMD}, so we see that the extra entries $(mq)^l$ added to the partition only affect the bottom theory, but instead appear as $\left((m-1)q\right)^l$. We can iterate this reasoning by applying the gauging formula $m-1$ times and see that all the theories involved in the gaugings are exactly those that we have for the $\ADAI{p}{N}{\left[Y\right]}$ theory except for the very last one, to which we can apply the equivalence \eqref{eq:isoql} since it has no exactly marginal coupling
\begin{equation}
    \ADAI{q}{qL+\frac{N}{m}}{\left[q^l,\widetilde Y_-,1^{\frac{N}{m}-\widetilde N_-}\right]} \equiv \ADAI{q}{\frac{N}{m}}{\left[\widetilde Y_-,1^{\frac{N}{m}-\widetilde N_-}\right]}~.
\end{equation}
This is exactly the same theory that appears at the end of the sequences of gaugings that realise the $\ADAI{p}{N}{\left[Y\right]}$ theory, which shows that \eqref{eq:isoql} indeed holds for any $m=\gcd(N,p)$.

Let us now consider the version of the equivalence \eqref{eq:iso2} for $m=\gcd(N,p)\geq1$
\begin{equation}\label{eq:isoconj}
    \ADAI{p}{N}{[Y]} \cong \ADAI{p}{pL-N}{[Y^c]}~,
\end{equation}
where
\begin{equation}
    [Y^c] = \left[(p-1)^{l_1},(p-2)^{l_2},\dots,(p-N)^{l_N}\right]~.
\end{equation}
In this case we claim that this is a direct consequence of the gauging formula \eqref{eq:typeIwEMD}, in the sense that the description of the theories on the two sides of the equivalence in terms of conformal gaugings of theories with no exactly marginal deformations is identical. For simplicity, let us show this for a subclass of theories with the maximal puncture $[Y]=[1^N]$ and $m=2$, for which the conjectured equivalence reads (we also assume $q>1$ in the following)
\begin{equation}
    \ADAI{2q}{N}{[1^N]} \cong \ADAI{2q}{(2q-1)N}{[(2q-1)^N]}~.
\end{equation}
Applying the gauging formula on both sides, one obtains the same description in terms of a conformal gauging of two theories with no exactly marginal deformations (although the role of the two theories with no exactly marginal deformations is swapped between the two sides)
\begin{equation}
        \ADAI{q}{qN-\frac{N}{2}}{\left[(q-1)^N,1^{\frac{N}{2}}\right]} \longleftarrow \mathfrak{su}(N/2) \longrightarrow \ADAI{q}{\frac{N}{2}}{\left[1^{\frac{N}{2}}\right]}~.
\end{equation}
One can similarly show that the equivalence \eqref{eq:isoconj} is a direct consequence of the gauging formula \eqref{eq:typeIwEMD} for generic $m>2$ and $[Y]$.

Thus we see that the equivalences \eqref{eq:isoql} and \eqref{eq:iso2} for type I theories without any EMD can be simply extended to \eqref{eq:isopl} and \eqref{eq:isoconj} for type I theories with EMDs, respectively. We also note that the pictorial description of these two equivalences as removing columns of length $p$ and conjugating the partition in a box of height $p$ are still accurate for the cases with EMDs.

\section{\label{sec:iso}Restricting the landscape of AD theories to type I}

In this section, we consider more general Argyres--Douglas theories of the form $\ADAN{p}{b}{N}{[Y]}$ for generic $b$. The main result that we will present here is that all such theories can be described as type I Argyres--Douglas theories with a partially closed regular puncture and a collection of free 4d $\mathcal N=2$ hypermultiplets. 

We start by stating this conjecture in full generality and comment on some simple checks. We then consider some specialisations and potential symmetry enhancements. This discussion is supplemented by a couple of examples where we explicitly check that the Coulomb branch spectra and the 3d mirrors agree on both sides of the equivalence.

We then show that the 3d mirrors of these general Argyres--Douglas theories and of the conjectured corresponding type I theories are given by the same 3d $\mathcal N=4$ quiver gauge theory for generic $N,p$, and $b$. We do this for $Y=[1^N]$ since this $\mathfrak{su}(N)$ symmetry is manifest in both of the equivalent descriptions in the regular puncture and so the same result for generic $[Y]$ can be obtained by performing the same nilpotent Higgsing on both sides of the equivalence. Finally we explicitly derive the extra number of free hypermultiplets by carefully analysing the representation theory of $\mathfrak{su}(N)$ under a general nilpotent Higgsing.

\subsection{\label{subsec:isostate}Statement}

Consider the general Argyres--Douglas theories of the form
\begin{equation}
    \ADAN{p}{b}{N}{[Y]} \;,\quad [Y]=[N^{l_N},\dots,1^{l_1}]~.\nn
\end{equation}
The flavour symmetry, as reviewed in Subsection \ref{subsec:ADflavourcc}, consists of the $\mathfrak g_Y$ contribution \eqref{eq:regpuncsymm} from the regular puncture and an additional $\mathfrak{u}(N-b)\oplus\mathfrak{u}(1)^{m-1}$ contribution from the irregular singularity with $m=\gcd(b,p)$. We conjecture that these SCFTs can be equivalently described by a type I Argyres--Douglas theory as follows:
\begin{equation}\label{eq:GentypeIiso}
    \ADAN{p}{b}{N}{[Y]} \cong\; \ADAI{p}{(N-b)p+b}{[(p-1)^{N-b},Y]} \otimes H_{\text{free}}\text{ free hypermultiplets}~,
\end{equation}
where,
\begin{equation}\label{eq:GentypeIisofree}
    H_{\text{free}} = (N-b)\sum_{i=p}^N (i-p+1)\;l_i~.
\end{equation}
More precisely, there are $(N-b)(i-p+1)$ free hypermultiplets in the fundamental representation of every simple factor $\mathfrak{su}(l_i)\subset\mathfrak g_Y$. We particularly point out that the regular puncture in the type I description on the r.h.s.\ of \eqref{eq:GentypeIiso} is partially reduced even for $[Y]=[1^N]$.

The additional free hypermultiplets play a key role in ensuring the agreement of the central charges in \eqref{eq:GentypeIiso}. Note that the $c$ and $a$ central charges for the Argyres--Douglas theories on the l.h.s.\ of \eqref{eq:GentypeIiso} are not known for $b\neq N,N-1$. However, these can be determined via this conjecture as $H_{\text{free}}$ can be derived from a direct calculation. This is achieved by carefully accounting for the number of massless hypermultiplets that arise upon the nilpotent Higgsing of the manifest $\mathfrak{su}(N)$ symmetry on the two sides of \eqref{eq:GentypeIiso} with $[Y]=[1^N]$. We provide an explicit derivation of \eqref{eq:GentypeIisofree} via this reasoning in Subsection \ref{subsec:isofree}.

The flavour symmetry from the type I description in \eqref{eq:GentypeIiso} has the contribution $\mathfrak g_Y\,\oplus\,\mathfrak{u}(N-b)$ from the regular puncture and $\mathfrak{u}(1)^{m-1}$ from the irregular singularity since
\begin{equation}
    \gcd\Big((N-b)p+b,p\Big) = \gcd(b,p) = m~.
\end{equation}
This is in precise agreement with the symmetry of the theory on the l.h.s.\ of \eqref{eq:GentypeIiso}.
However, the $\mathfrak{u}(N-b)$ subalgebra from the irregular puncture on the l.h.s.\ of \eqref{eq:GentypeIiso} is manifest in the regular puncture on the type I side. Moreover, in special cases it is possible that $p-1$ is equal to one of the entries of $[Y]$. This leads to a symmetry enhancement of the $\mathfrak{su}(l_{p-1})\oplus\mathfrak{u}(N-b)$ subalgebra to $\mathfrak{su}(N-b+l_{p-1})$ that is manifest only in the type I description. 

To our knowledge, the $m=1$ version of this equivalence \eqref{eq:GentypeIiso} first appeared in \cite{Xie:2019yds} for cases with the maximal regular puncture $[Y]=[1^N]$. The evidence there was restricted to matching the flavour central charges and the Coulomb branch spectra. Here we conjecture this equivalence in full generality for $m\neq1$ and generic $[Y]$ along with an explicit characterisation of the number of free hypermultiplets associated to the latter. We further provide strong evidence for this conjecture in terms of an explicit agreement of the quivers describing the 3d mirrors.

A special case of the conjectured equivalence \eqref{eq:GentypeIiso} for $b=N-1$ states that any type II Argyres--Douglas theory is equivalent to a type I theory with a partially reduced regular puncture
\begin{equation}\label{eq:typeItypeIIeq}
    \ADAN{p}{N-1}{N}{[Y]} \cong \ADAI{p}{p+N-1}{[p-1,Y]} \otimes H_{\text{free}}\text{ free hypermultiplets}~,
\end{equation}
where,
\begin{equation}
    H_{\text{free}} =\sum_{i=p}^N (i-p+1)\;l_i~.
\end{equation}
Moreover, for the case of $b=N-1$ and no regular puncture, $[Y]=[N]$, we have
\begin{equation}
    \begin{split}
        \ADAN{p}{N-1}{N}{[N]} &\cong \ADAI{p}{p+N-1}{[p-1,N]}\\
        &\cong \ADAI{p}{p-N+1}{[p-N,1]}~,
    \end{split}
\end{equation}
where for the second equivalence we have used the type I result \eqref{eq:isoconj}. This reproduces the previously conjectured equivalence between type II theories without a regular puncture and the corresponding type I theories with a minimal regular puncture \cite{Xie:2013jc}.

We will now perform explicit checks for a couple of examples. These checks consist of matching the CB spectra and the 3d mirror quivers. We further verified the agreement of the CB spectra for a plethora of cases by implementing it in a \texttt{Mathematica} code. In Subsection \ref{subsec:iso3dmatch} we generalise the strategy used for matching the 3d mirrors used in the examples to obtain strong evidence for the equivalence \eqref{eq:GentypeIiso}.

\subsection{\label{subsec:isoex}Examples}

\subsubsection*{Example 5: $\ADAF{3}{4}{5}$}

This is a type II theory that has no $\mathcal N=2$ exactly marginal deformation. The Coulomb branch spectrum of this theory is
\begin{equation}\label{eq:CBSpecEx5}
    \left\{\frac43,\frac53,\frac73,\frac83,\frac{11}{3}\right\}~.
\end{equation}
Since the regular puncture is maximal, the central charges can be computed by using \eqref{eq:cfullpunc} and the Shapere--Tachikawa relation \eqref{eq:ShapereTachikawa} to be
\begin{equation}\label{eq:acex5}
    c = \frac{65}{12} \;,\quad a = 5~.
\end{equation}
The flavour symmetry algebra of this theory is $\mathfrak{su}(5)\oplus\mathfrak{u}(1)$. The quiver gauge theory that describes the 3d mirror of this Argyres--Douglas theory is given by
\begin{equation}\label{eq:ex5quiv1}
    \begin{tikzpicture}[scale=1.1,every node/.style={scale=1.2},font=\scriptsize]
    \node[gauge] (g0) at (0,0) {$\,1\,$};
    \node[gauge] (g1) at (1,1) {$\,3\,$};
    \node[gauge] (g2) at (2,2) {$\,1\,$};
    \node[gauge] (g3) at (3,1) {$\,3\,$};
    \node[gauge] (g4) at (4,0) {$\,2\,$};
    \node[gauge] (g5) at (5,-1) {$\,1\,$};
    \draw (g0)--(g1)--(g2)--(g3)--(g4)--(g5);
    \draw (g1)--(g3);
    \node[above left] at (1.5,1.5) {$2$};
\begin{scope}[shift={(6,0)}]
    \node[] at (0,0) {\large$\longrightarrow$};
\end{scope}
\begin{scope}[shift={(7.2,0)}]
    \node[gauge] (g0) at (0,0) {$\,1\,$};
    \node[gauge] (g1) at (1,0) {$\,3\,$};
    \node[gauge] (g2) at (2,0) {$\,3\,$};
    \node[gauge] (g3) at (3,0) {$\,2\,$};
    \node[gauge] (g4) at (4,0) {$\,1\,$};
    \node[flavor] (f0) at (1,1) {$\,2\,$};
    \node[flavor] (f1) at (2,1) {$\,1\,$};
    \draw (g0)--(g1)--(g2)--(g3)--(g4);
    \draw (g1)--(f0);
    \draw (g2)--(f1);
\end{scope}
\end{tikzpicture}
\end{equation}
where on the right we have removed the overall decoupled $\mathfrak{u}(1)$ gauge symmetry from the top node. Note that this can be identified as the $T_\sigma^\rho(\mathfrak{su}(7))$ theory with $\sigma=[2,1^5]$ and $\rho=[3,2^2]$, which is the 3d mirror of the $T_\rho^\sigma(\mathfrak{su}(7))$ theory with the same $\rho$ and $\sigma$ that instead describes the 3d direct reduction \cite{Closset:2020afy,Giacomelli:2020ryy}. The $\mathfrak{su}(5)$ flavour symmetry can be understood from the quiver by noticing that all the gauge nodes are balanced except for the leftmost one, which instead provides the $\mathfrak{u}(1)$ symmetry.

Our proposal is that this theory is equivalent to the following type I Argyres--Douglas theory:
\begin{equation}\label{eq:Genex5}
    \ADAF{3}{4}{5} \cong \ADAI{3}{7}{[2,1^5]}~.
\end{equation}
This type I theory can be easily confirmed to have the same Coulomb branch spectrum as \eqref{eq:CBSpecEx5}. Since the type I theory has a reduced regular puncture, it can obtained from $\ADAIF{3}{7}$ via nilpotent Higgsing of the form $[2,1^5]$. The $c$ and $a$ central charges computed this way can be shown to be in agreement with \eqref{eq:acex5} as well.

The 3d mirror of the type I theory $\ADAI{3}{7}{[2,1^5]}$ can be obtained from that of $\ADAIF{3}{7}$ as follows. The quiver for the case with the maximal regular puncture, $[Y]=[1^7]$, is given by,
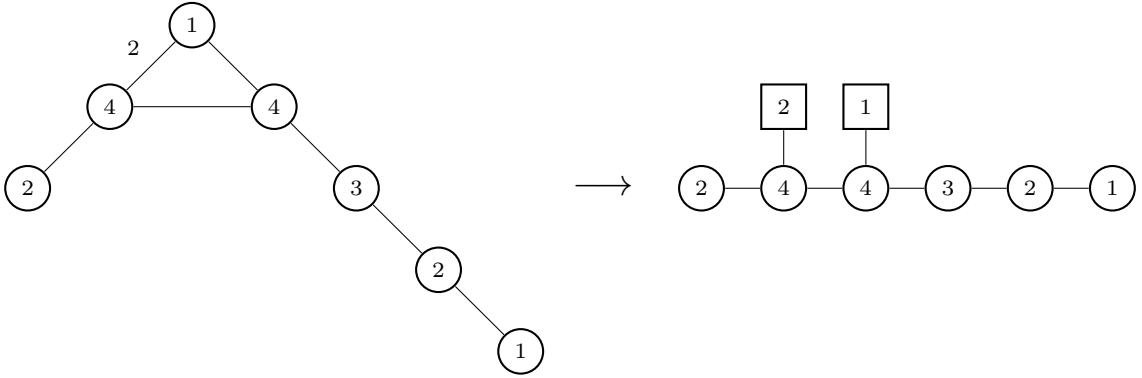
\begin{equation}\label{eq:ex5quiv2}
\resizebox{\textwidth}{!}{
    \begin{tikzpicture}[scale=1.1,every node/.style={scale=1.2},font=\scriptsize]
    \node[gauge] (g0) at (0,0) {$\,2\,$};
    \node[gauge] (g1) at (1,1) {$\,4\,$};
    \node[gauge] (g2) at (2,2) {$\,1\,$};
    \node[gauge] (g3) at (3,1) {$\,4\,$};
    \node[gauge] (g4) at (4,0) {$\,3\,$};
    \node[gauge] (g5) at (5,-1) {$\,2\,$};
    \node[gauge] (g6) at (6,-2) {$\,1\,$};
    \draw (g0)--(g1)--(g2)--(g3)--(g4)--(g5)--(g6);
    \draw (g1)--(g3);
    \node[above left] at (1.5,1.5) {$2$};
\begin{scope}[shift={(7,0)}]
    \node[] at (0,0) {\large$\longrightarrow$};
\end{scope}
\begin{scope}[shift={(8.2,0)}]
    \node[gauge] (g0) at (0,0) {$\,2\,$};
    \node[gauge] (g1) at (1,0) {$\,4\,$};
    \node[gauge] (g2) at (2,0) {$\,4\,$};
    \node[gauge] (g3) at (3,0) {$\,3\,$};
    \node[gauge] (g4) at (4,0) {$\,2\,$};
    \node[gauge] (g5) at (5,0) {$\,1\,$};
    \node[flavor] (f0) at (1,1) {$\,2\,$};
    \node[flavor] (f1) at (2,1) {$\,1\,$};
    \draw (g0)--(g1)--(g2)--(g3)--(g4)--(g5);
    \draw (g1)--(f0);
    \draw (g2)--(f1);
\end{scope}
\end{tikzpicture}
}
\end{equation}
Once again, this quiver can be identified with the $T_\sigma^\rho(\mathfrak{su}(7))$ theory with $\rho=[3,2^2]$ and $\sigma=[1^7]$. The full $\mathfrak{su}(7)$ symmetry is manifest in the quiver on the right since all the gauge nodes are balanced. Obtaining the quiver that describes the 3d mirror of the $\ADAI{3}{7}{[2,1^5]}$ Argyres--Douglas theory just corresponds to changing $\sigma=[1^7] \to [2,1^5]$. Note that this Higgs branch nilpotent vev in the 4d theory corresponds to changing the $\sigma$ describing the 3d Coulomb branch of the $T_\sigma^\rho(\mathfrak{su}(7))$ theory as we are describing the 3d mirror of the original Argyres--Douglas theory. Thus the 3d mirror of the $\ADAI{3}{7}{[2,1^5]}$ Argyres--Douglas theory is the $T_{[2,1^5]}^{[3,2^2]}(\mathfrak{su}(7))$ which is the same as the 3d mirror of the original theory $\ADAF{3}{4}{5}$ in \eqref{eq:ex5quiv1}.

We will now rephrase this agreement of the 3d mirrors in the language of the decay algorithm implemented via quiver subtraction \cite{Bourget:2023dkj,Bourget:2024mgn,Bourget:2021siw}, which will be useful for the general proof we will give in Subsection \ref{subsec:iso3dmatch}. In particular, this is important because generically the 3d mirrors are not $T_\sigma^\rho(\mathfrak{su}(N))$ theories. 

The brane setup for the quiver theory in \eqref{eq:ex5quiv2} can be expressed as follows,
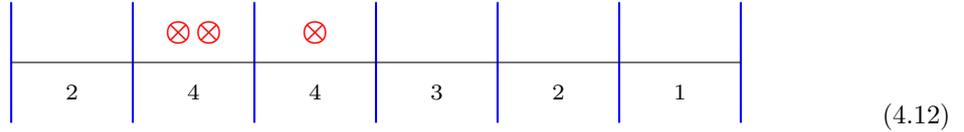
\begin{equation}
        \begin{tikzpicture}[scale=0.8,every node/.style={scale=1.2},font=\scriptsize]
        \draw[thick,blue] (0,2)--(0,0);
        \draw[thick,blue] (2,2)--(2,0);
        \draw[thick,blue] (4,2)--(4,0);
        \draw[thick,blue] (6,2)--(6,0);
        \draw[thick,blue] (8,2)--(8,0);
        \draw[thick,blue] (10,2)--(10,0);
        \draw[thick,blue] (12,2)--(12,0);
        \draw (12,1)--(0,1);
        \node at (2.75,1.5) {\color{red}\small $\bm\otimes$};
        \node at (3.25,1.5) {\color{red}\small $\bm\otimes$};
        \node at (5,1.5) {\color{red}\small $\bm\otimes$};
        \node at (1,0.5) {$2$};
        \node at (3,0.5) {$4$};
        \node at (5,0.5) {$4$};
        \node at (7,0.5) {$3$};
        \node at (9,0.5) {$2$};
        \node at (11,0.5) {$1$};
    \end{tikzpicture}
\end{equation}
By performing a sequence of Hanany--Witten moves to pull all the D5--branes to the left, we obtain
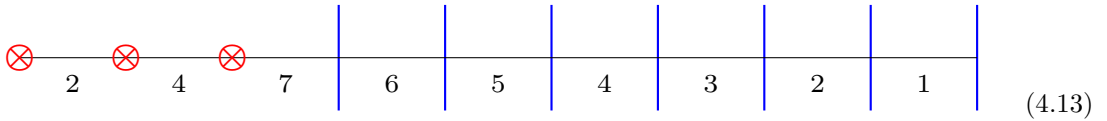
\begin{equation}
        \begin{tikzpicture}[scale=0.7,every node/.style={scale=1.4},font=\scriptsize]
        \draw[thick,blue] (0,2)--(0,0);
        \draw[thick,blue] (2,2)--(2,0);
        \draw[thick,blue] (4,2)--(4,0);
        \draw[thick,blue] (6,2)--(6,0);
        \draw[thick,blue] (8,2)--(8,0);
        \draw[thick,blue] (10,2)--(10,0);
        \draw[thick,blue] (12,2)--(12,0);
        \draw (12,1)--(-6,1);
        \node at (-2,1) {\color{red}\small $\bm\otimes$};
        \node at (-4,1) {\color{red}\small $\bm\otimes$};
        \node at (-6,1) {\color{red}\small $\bm\otimes$};
        \node at (-5,0.5) {$2$};
        \node at (-3,0.5) {$4$};
        \node at (-1,0.5) {$7$};
        \node at (1,0.5) {$6$};
        \node at (3,0.5) {$5$};
        \node at (5,0.5) {$4$};
        \node at (7,0.5) {$3$};
        \node at (9,0.5) {$2$};
        \node at (11,0.5) {$1$};
    \end{tikzpicture}
\end{equation}
which enables us to read off the partitions $\sigma=[1^7]$ and $\rho=[3,2^2]$ of the $T^\rho_\sigma(\mathfrak{su}(7))$ theory by counting the linking numbers of the five-branes. However, the configuration we want to obtain is the one corresponding to $\sigma=[2,1^5]$ and $\rho=[3,2^2]$. As explained before, this amounts to a CB vev on the 3d mirror or a Higgs branch nilpotent vev of the 4d theory. This can be achieved in the brane setup by moving a single D3--brane between every couple of adjacent NS5--branes to infinity to obtain
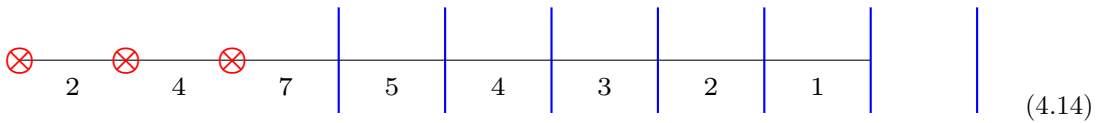
\begin{equation}
        \begin{tikzpicture}[scale=0.7,every node/.style={scale=1.4},font=\scriptsize]
        \draw[thick,blue] (0,2)--(0,0);
        \draw[thick,blue] (2,2)--(2,0);
        \draw[thick,blue] (4,2)--(4,0);
        \draw[thick,blue] (6,2)--(6,0);
        \draw[thick,blue] (8,2)--(8,0);
        \draw[thick,blue] (10,2)--(10,0);
        \draw[thick,blue] (12,2)--(12,0);
        \draw (10,1)--(-6,1);
        \node at (-2,1) {\color{red}\small $\bm\otimes$};
        \node at (-4,1) {\color{red}\small $\bm\otimes$};
        \node at (-6,1) {\color{red}\small $\bm\otimes$};
        \node at (-5,0.5) {$2$};
        \node at (-3,0.5) {$4$};
        \node at (-1,0.5) {$7$};
        \node at (1,0.5) {$5$};
        \node at (3,0.5) {$4$};
        \node at (5,0.5) {$3$};
        \node at (7,0.5) {$2$};
        \node at (9,0.5) {$1$};
    \end{tikzpicture}
\end{equation}
Notice indeed that the linking numbers of the NS5--branes after this operations reproduce exactly the new partition $\sigma=[2,1^5]$ we are interested in. The removal of a D3--brane that stretches between two NS5--branes corresponds to reducing the rank of the corresponding gauge node in the quiver by 1. Therefore changing $\sigma=[1^7] \to [2,1^5]$ in the 3d mirror can be expressed as subtracting the following quiver from the 3d mirror with $\sigma=[1^7]$ in \eqref{eq:ex5quiv2},
\begin{equation}
    \begin{tikzpicture}[scale=1.2,every node/.style={scale=1.3},font=\scriptsize]
    \node[gauge] (g0) at (0,0) {$\,1\,$};
    \node[gauge] (g1) at (1,0) {$\,1\,$};
    \node[gauge] (g2) at (2,0) {$\,1\,$};
    \node[gauge] (g3) at (3,0) {$\,1\,$};
    \node[gauge] (g4) at (4,0) {$\,1\,$};
    \node[gauge] (g5) at (5,0) {$\,1\,$};
    \draw (g0)--(g1)--(g2)--(g3)--(g4)--(g5);
\end{tikzpicture}
\end{equation}
which exactly reproduces the quiver in \eqref{eq:ex5quiv1}.

\subsubsection*{Example 6: $\ADAF{4}{2}{4}$}

This is an Argyres--Douglas theory that is neither a type I nor type II theory and has one exactly marginal deformation. The Coulomb branch spectrum of this theory is
\begin{equation}
    \left\{\frac32,\frac32,2,\frac52,\frac52,3,\frac72\right\}~.
\end{equation}
Although the regular puncture here is maximal, this theory has no obvious known construction in the geometric engineering picture and therefore we do not know the $c$ and $a$ central charges. However from the Shapere--Tachikawa relation \eqref{eq:ShapereTachikawa} we know that
\begin{equation}\label{eq:ex6ST}
    2a-c = \frac{13}{2}~.
\end{equation}
The flavour symmetry algebra of this theory is $\mathfrak{u}(2)\oplus\mathfrak{u}(1)\oplus\mathfrak{su}(4)$, where the $\mathfrak{su}(4)$ is the contribution from the regular puncture. The quiver gauge theory that describes the 3d mirror of this Argyres--Douglas theory is given by
\begin{equation}\label{eq:ex6quiv1}
    \begin{tikzpicture}[scale=1.2,every node/.style={scale=1.3},font=\scriptsize]
    \node[gauge] (g0) at (0,0) {$\,1\,$};
    \node[gauge] (g1) at (1,0) {$\,2\,$};
    \node[gauge] (g2) at (2,0) {$\,3\,$};
    \node[gauge] (g3) at (3,0) {$\,2\,$};
    \node[gauge] (g4) at (4,0) {$\,1\,$};
    \node[gauge] (g5) at (1,1) {$\,1\,$};
    \node[gauge] (g6) at (2,1) {$\,1\,$};
    \draw (g0)--(g1)--(g2)--(g3)--(g4);
    \draw (g1)--(g5)--(g6)--(g2)--(g5);
    \draw (g1)--(g6);
\end{tikzpicture}
\end{equation}
The $\mathfrak{su}(4)$ flavour symmetry can be understood from this quiver by noticing that the three gauge nodes on the right are balanced while the $\mathfrak{su}(2)$ can be seen in the balanced node on the left. Remembering that one overall $\mathfrak{u}(1)$ gauge node should be decoupled, this leaves us with two more $\mathfrak{u}(1)$ topological symmetries.

Our proposal is that this is equivalent to the following type I Argyres--Douglas theory,
\begin{equation}\label{eq:Genex6}
    \ADAF{4}{2}{4} \cong \ADAI{4}{10}{[3^2,1^4]}~.
\end{equation}
One can check that the two theories have the same Coulomb branch spectrum. The central charges of the type I theory can be computed by using the fact that this theory can be obtained via nilpotent Higgsing of the form $[3^2,1^4]$ from the Argyres--Douglas theory $\ADAIF{4}{10}$, and they turn out to be
\begin{equation}
    c = \frac{22}{3} \;,\quad a = \frac{83}{12}~.
\end{equation}
These can be seen to satisfy the same Shapere--Tachikawa relation as \eqref{eq:ex6ST}, which is of course a consequence of the two theories having the same Coulomb branch spectrum. The regular puncture now contributes $\mathfrak{u}(2)\oplus\mathfrak{su}(4)$ to the flavour symmetry whereas the irregular singularity only contributes a $\mathfrak{u}(1)$. Thus, even though the flavour symmetry algebras agree on both sides of \eqref{eq:Genex6}, they arise in different ways.

Checking that the 3d mirror of this type I theory is the same as in \eqref{eq:ex6quiv1} is trickier than the previous example. This is because of the complete graph of two $\mathfrak{u}(1)$ nodes, since $m=2$ here. For $m=1$, the overall decoupled $\mathfrak{u}(1)$ can be chosen to be the one associated to the complete graph and one ends up with a linear quiver $T_\rho^\sigma$ theory. Once we have identified the 3d mirror as a $T_\rho^\sigma$ theory, nilpotent vev on the Higgs branch simply corresponds to changing the partition corresponding to the Coulomb branch of the 3d mirror as demonstrated in the previous example. Instead in this example, and more generally for $m\neq1$, there are $m=2$ $\mathfrak{u}(1)$ nodes in the complete graph and one cannot express this as a $T_\rho^\sigma$ theory. However, we can proceed as follows. 

We choose to decouple one of the $\mathfrak{u}(1)$ nodes in the complete graph and ungauge the other one by gauging its $\mathfrak{u}(1)$ topological symmetry, as explained in Subsubsection \ref{subsec:typeIEMCex}. We end up with the following quiver,
\begin{equation}\label{eq:ex6quiv2}
    \begin{tikzpicture}[scale=1.2,every node/.style={scale=1.3},font=\scriptsize]
    \node[gauge] (g0) at (0,0) {$\,1\,$};
    \node[gauge] (g1) at (1,0) {$\,2\,$};
    \node[gauge] (g2) at (2,0) {$\,3\,$};
    \node[gauge] (g3) at (3,0) {$\,2\,$};
    \node[gauge] (g4) at (4,0) {$\,1\,$};
    \node[flavor] (g5) at (1,1) {$\,2\,$};
    \node[flavor] (g6) at (2,1) {$\,2\,$};
    \draw (g0)--(g1)--(g2)--(g3)--(g4);
    \draw (g1)--(g5);
    \draw (g2)--(g6);
\end{tikzpicture}
\end{equation}
and a decoupled hypermultiplet. This quiver can now be identified with that of the $T_\sigma^\rho(\mathfrak{su}(10))$ theory with $\sigma=[3^2,1^4]$ and $\rho=[3^2,2^2]$. The original quiver \eqref{eq:ex6quiv1} can be recovered by gauging the $\mathfrak{u}(1)$ flavour symmetry that we gained.

Let us now consider the 3d mirror of the type I theory with the full puncture, $\ADAIF{4}{10}$,
\begin{equation}\label{eq:ex6quiv3}
    \begin{tikzpicture}[scale=1.2,every node/.style={scale=1.3},font=\scriptsize]
    \node[gauge] (g0) at (0,0) {$\,3\,$};
    \node[gauge] (g1) at (1,0) {$\,6\,$};
    \node[gauge] (g2) at (2,0) {$\,7\,$};
    \node[gauge] (g3) at (3,0) {$\,6\,$};
    \node[gauge] (g4) at (4,0) {$\,5\,$};
    \node[gauge] (g7) at (5,0) {$\,4\,$};
    \node[gauge] (g8) at (6,0) {$\,3\,$};
    \node[gauge] (g9) at (7,0) {$\,2\,$};
    \node[gauge] (g10) at (8,0) {$\,1\,$};
    \node[gauge] (g5) at (1,1) {$\,1\,$};
    \node[gauge] (g6) at (2,1) {$\,1\,$};
    \draw (g0)--(g1)--(g2)--(g3)--(g4)--(g7)--(g8)--(g9)--(g10);
    \draw (g1)--(g5)--(g6)--(g2)--(g5);
    \draw (g1)--(g6);
\end{tikzpicture}
\end{equation}
Once again, we decouple one of the $\mathfrak{u}(1)$ gauge nodes in the complete graph and ungauge the other by gauging the $\mathfrak{u}(1)$ topological symmetry to obtain
\begin{equation}\label{eq:ex6quiv4}
    \begin{tikzpicture}[scale=1.2,every node/.style={scale=1.3},font=\scriptsize]
    \node[gauge] (g0) at (0,0) {$\,3\,$};
    \node[gauge] (g1) at (1,0) {$\,6\,$};
    \node[gauge] (g2) at (2,0) {$\,7\,$};
    \node[gauge] (g3) at (3,0) {$\,6\,$};
    \node[gauge] (g4) at (4,0) {$\,5\,$};
    \node[gauge] (g7) at (5,0) {$\,4\,$};
    \node[gauge] (g8) at (6,0) {$\,3\,$};
    \node[gauge] (g9) at (7,0) {$\,2\,$};
    \node[gauge] (g10) at (8,0) {$\,1\,$};
    \node[flavor] (g5) at (1,1) {$\,2\,$};
    \node[flavor] (g6) at (2,1) {$\,2\,$};
    \draw (g0)--(g1)--(g2)--(g3)--(g4)--(g7)--(g8)--(g9)--(g10);
    \draw (g1)--(g5);
    \draw (g2)--(g6);
\end{tikzpicture}
\end{equation}
plus a decoupled hypermultiplet. This can directly be identified as the $T_\sigma^\rho(\mathfrak{su}(10))$ theory with $\sigma=[1^{10}]$ and $\rho=[3^2,2^2]$. If we now perform the CB nilpotent vev $\sigma=[1^{10}] \longrightarrow [3^2,1^4]$, we exactly obtain the theory in \eqref{eq:ex6quiv2}, including the decoupled hypermultiplet. 

The crucial observation now is that this CB nilpotent vev commutes with the gauging of the flavour $\mathfrak{u}(1)$ that we need to do to go back to the actual 3d mirror of the AD theory. Hence, we have shown that the 3d mirrors of $\ADAI{4}{10}{[3^2,1^4]}$ and of $\ADAF{4}{2}{4}$ are identical. Notice that this is the 3d mirror analogue of the strategy that we used in Subsection \ref{subsec:typeIEMC} to match the 3d reductions of the theories in \eqref{eq:typeIwEMD}.

Once again, we can re-express this manipulation to derive the quiver after the nilpotent Higgsing in the 3d mirror of the type I theory via quiver subtraction according to the decay algorithm. The brane setup describing the 3d mirror \eqref{eq:ex6quiv4} of $\ADAIF{4}{10}$ after ungauging the $\mathfrak{u}(1)$ nodes in the complete graph is
\begin{equation}
        \begin{tikzpicture}[scale=0.55,every node/.style={scale=1.1},font=\scriptsize]
        \draw[thick,blue] (0,2)--(0,0);
        \draw[thick,blue] (2,2)--(2,0);
        \draw[thick,blue] (4,2)--(4,0);
        \draw[thick,blue] (6,2)--(6,0);
        \draw[thick,blue] (8,2)--(8,0);
        \draw[thick,blue] (10,2)--(10,0);
        \draw[thick,blue] (12,2)--(12,0);
        \draw[thick,blue] (14,2)--(14,0);
        \draw[thick,blue] (16,2)--(16,0);
        \draw[thick,blue] (18,2)--(18,0);
        \draw (18,1)--(-8,1);
        \node at (-2,1) {\color{red}\small $\bm\otimes$};
        \node at (-4,1) {\color{red}\small $\bm\otimes$};
        \node at (-6,1) {\color{red}\small $\bm\otimes$};
        \node at (-8,1) {\color{red}\small $\bm\otimes$};
        \node at (-7,0.5) {$2$};
        \node at (-5,0.5) {$4$};
        \node at (-3,0.5) {$7$};
        \node at (-1,0.5) {$10$};
        \node at (1,0.5) {$9$};
        \node at (3,0.5) {$8$};
        \node at (5,0.5) {$7$};
        \node at (7,0.5) {$6$};
        \node at (9,0.5) {$5$};
        \node at (11,0.5) {$4$};
        \node at (13,0.5) {$3$};
        \node at (15,0.5) {$2$};
        \node at (17,0.5) {$1$};
    \end{tikzpicture}
\end{equation}

The CB nilpotent vev $[1^{10}]\to[3^2,1^4]$ is implemented by removing suitable D3-branes stretched between the NS5--branes
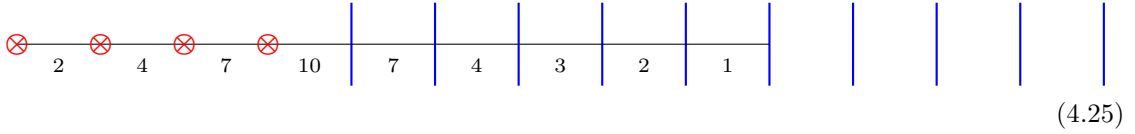
\begin{equation}\label{eq:ex5bs2}
        \begin{tikzpicture}[scale=0.55,every node/.style={scale=1.1},font=\scriptsize]
        \draw[thick,blue] (0,2)--(0,0);
        \draw[thick,blue] (2,2)--(2,0);
        \draw[thick,blue] (4,2)--(4,0);
        \draw[thick,blue] (6,2)--(6,0);
        \draw[thick,blue] (8,2)--(8,0);
        \draw[thick,blue] (10,2)--(10,0);
        \draw[thick,blue] (12,2)--(12,0);
        \draw[thick,blue] (14,2)--(14,0);
        \draw[thick,blue] (16,2)--(16,0);
        \draw[thick,blue] (18,2)--(18,0);
        \draw (10,1)--(-8,1);
        \node at (-2,1) {\color{red}\small $\bm\otimes$};
        \node at (-4,1) {\color{red}\small $\bm\otimes$};
        \node at (-6,1) {\color{red}\small $\bm\otimes$};
        \node at (-8,1) {\color{red}\small $\bm\otimes$};
        \node at (-7,0.5) {$2$};
        \node at (-5,0.5) {$4$};
        \node at (-3,0.5) {$7$};
        \node at (-1,0.5) {$10$};
        \node at (1,0.5) {$7$};
        \node at (3,0.5) {$4$};
        \node at (5,0.5) {$3$};
        \node at (7,0.5) {$2$};
        \node at (9,0.5) {$1$};
    \end{tikzpicture}
\end{equation}
Therefore changing $\sigma=[1^{10}]\rightarrow[3^2,1^4]$ in the 3d mirror can be expressed as subtracting the following quiver from \eqref{eq:ex6quiv4}:
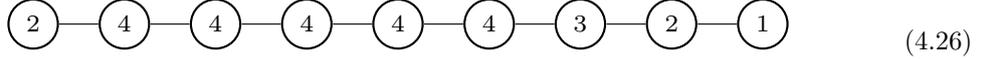
\begin{equation}
    \begin{tikzpicture}[scale=1.2,every node/.style={scale=1.3},font=\scriptsize]
    \node[gauge] (g0) at (0,0) {$\,2\,$};
    \node[gauge] (g1) at (1,0) {$\,4\,$};
    \node[gauge] (g2) at (2,0) {$\,4\,$};
    \node[gauge] (g3) at (3,0) {$\,4\,$};
    \node[gauge] (g4) at (4,0) {$\,4\,$};
    \node[gauge] (g5) at (5,0) {$\,4\,$};
    \node[gauge] (g6) at (6,0) {$\,3\,$};
    \node[gauge] (g7) at (7,0) {$\,2\,$};
    \node[gauge] (g8) at (8,0) {$\,1\,$};
    \draw (g0)--(g1)--(g2)--(g3)--(g4)--(g5)--(g6)--(g7)--(g8);
\end{tikzpicture}
\end{equation}
which exactly reproduces the quiver in \eqref{eq:ex6quiv2}. Since, as explained above, this CB nilpotent Higgsing commutes with the procedure of gauging the flavour symmetry, we can instead directly subtract this quiver from \eqref{eq:ex6quiv3}, the actual 3d mirror of $\ADAIF{4}{10}$, which gives the 3d mirror of the theory $\ADAI{4}{10}{[3^2,1^4]}$ we are interested in. This direct ``quiver subtraction" exactly reproduces the 3d mirror \eqref{eq:ex6quiv1} of the $\ADAF{4}{2}{4}$ theory. From now on, we will only use the quiver subtraction method in accordance with the decay algorithm applied to the quivers with all the nodes gauged describing the actual 3d mirrors of the AD theories in consideration.

\subsection{\label{subsec:iso3dmatch}Matching the 3d mirrors}

In this section, we show how to match the 3d mirrors of the Argyres--Douglas theories on the two sides of \eqref{eq:GentypeIiso} for $[Y]=[1^N]$ and generic $N$, $p$, and $b$ using the same technique described in the previous examples. 

We consider the quiver describing the 3d mirror of the general Argyres--Douglas theory $\ADAF{p}{b}{N}$, where we recall that $m=\gcd(p,b)$, $n=\frac{b}{m}$, and $p=m(n+k)$. We split the discussion into two parts, one each for $p>b$ and $p<b$. In both cases our conjecture states that this must be identical to the 3d mirror of a type I Argyres--Douglas theory as follows:
\begin{equation}\label{eq:bigiso}
    \ADAF{p}{b}{N} \cong \ADAI{p}{(N-b)p+b}{[(p-1)^{N-b},1^N]}~.
\end{equation}
The 3d mirror of the l.h.s.\ of \eqref{eq:bigiso} has already been discussed in full generality in Subsection \ref{subsubsec:AD3dmirrgen}, see Figures \ref{fig:3dmirrnkp} and \ref{fig:3dmirrnkn}. We will show how to match this with the quiver for the 3d mirror of the type I theory on the r.h.s.\ of \eqref{eq:bigiso}, which we will obtain starting from the one for the theory with the maximal regular puncture $[1^{(N-nm)p+nm}]$ and implementing the 4d HB vev by subtracting the right quiver as dictated by the decay and fission algorithm.

The quiver that needs to be subtracted turns out to be the same for both $p>b$ and $p<b$. Recall from the discussion in the examples that the procedure consists of first ungauging the $\mathfrak{u}(1)$ gauge nodes in the complete graph to obtain a $T_\sigma^\rho(\mathfrak{su}((N-nm)p+nm))$ theory. The $\sigma$ will always be the maximal one $\sigma=[1^{(N-nm)p+nm}]$ since this corresponds to the maximal puncture of the 4d theory we begin with, while the form of the $\rho$ will be different for $p>b$ and $p<b$. From the brane setup of this $T_\sigma^\rho(\mathfrak{su}((N-nm)p+nm))$ theory, one removes D3--branes that stretched between adjacent NS5--branes to achieve the 4d Higgsing corresponding to changing the $\sigma$, which can be implemented in the 3d mirror by the inverted quiver subtraction algorithm. Finally, one gauges back the $\mathfrak{u}(1)$ symmetries that we ungauged after the subtraction. Thus the difference in the form of $\rho$ for $p>b$ and $p<b$ does not play a role in determining the quiver that needs to be subtracted to achieve the transition $\sigma=[1^{(N-nm)p+nm}] \rightarrow [(p-1)^{N-nm},1^N]$, since $\rho$ only dictates how the D3--branes end on the D5--branes which are instead left untouched.

Let us now determine the quiver that needs to be subtracted by proceeding in a similar way as in the previous examples. We consider the $T_\sigma^\rho(\mathfrak{su}((N-nm)p+nm))$ theory with the maximal $\sigma=[1^{(N-nm)p+nm}]$ and the transition $\sigma=[1^{(N-nm)p+nm}]\to[(p-1)^{N-nm},1^N]$. Comparing the two partitions, one can easily figure out the number of D3--branes that need to be removed in each interval between NS5--branes to achieve this, which gives us the ranks of the gauge nodes that need to be subtracted. This way we obtain the following quiver that needs to be subtracted to implement the 4d Higgsing:
\begin{equation}\label{eq:subquiver}
        \begin{tikzpicture}[scale=1.7,every node/.style={scale=1.3},font=\scriptsize]
    \node[rotate=-45] at (0.8,1.3) {\fontsize{42pt}{42pt}\selectfont$\thinbraceleft{\begin{array}{l}\color{white}
    A \\ \color{white}B
    \end{array}}$};
    \node at (-0.2,1.65) {$N-1$ times};
    \node[gauge] (glm3) at (-1.75,-1.75) {$1$};
    \node[gauge] (glm2) at (-1.25,-1.25) {$2$};
    \node (glm1) at (-0.75,-0.75) {$\cdots$};
    \node[gauge] (gl0) at (-0.25,-0.25) {\tiny$(N-nm)(p-2)-1$};
    \node[gauge] (gl1) at (0.5,0.5) {\tiny$(N-nm)(p-2)$};
    \node[] (gl2) at (1,1) {$\cdots$};
    \node[gauge] (gl3) at (1.5,1.5) {\tiny$(N-nm)(p-2)$};
    \node[gauge] (gl4) at (2.25,2.25) {\tiny$(N-nm)(p-2)$};
    \node[gauge] (gr4) at (4.25+0.5,2.25) {\tiny$(N-nm)(p-2)$};
    \node[gauge] (gr3) at (5+0.5,1.5) {\tiny$(N-nm-1)(p-2)$};
    \node[] (gr2) at (5.5+0.5,1) {$\cdots$};
    \node[gauge] (gr1) at (6+0.5,0.5) {\tiny$2(p-2)$};
    \node[gauge] (gr0) at (6.5+0.5,0) {\tiny$p-2$};
    \draw (glm3)--(glm2)--(glm1)--(gl0)--(gl1)--(gl2)--(gl3)--(gl4)--(gr4)--(gr3)--(gr2)--(gr1)--(gr0);
\end{tikzpicture}
\end{equation}

Now we want to consider the 3d mirror of the analogue of the theory on the r.h.s.\ of \eqref{eq:bigiso} when the puncture is maximal, $\ADAI{p}{(N-b)p+b}{[1^{(N-nm)p+nm}]}$.

\subsubsection{The case $p>b$}

For $p>b$, we have
\begin{eqnarray}
    &N' = (N-nm)p+nm \;,\quad p' = p =mq=m(n+k)~,\nn\\
    &\nu' = \frac{N'}{m} \mod\left(\frac{p'}{m}\right) = n \;,\quad \mu' = \frac{N'-\nu'm}{p'} = N-nm~.
\end{eqnarray}
since obviously $p'<N'$. Using the quiver explicitly drawn in Figure \ref{fig:3dmirrIkn} and substituting the above values, we have that the 3d mirror of the theory $\ADAI{p}{(N-b)p+b}{[1^{(N-nm)p+nm}]}$ for $p>b$ is
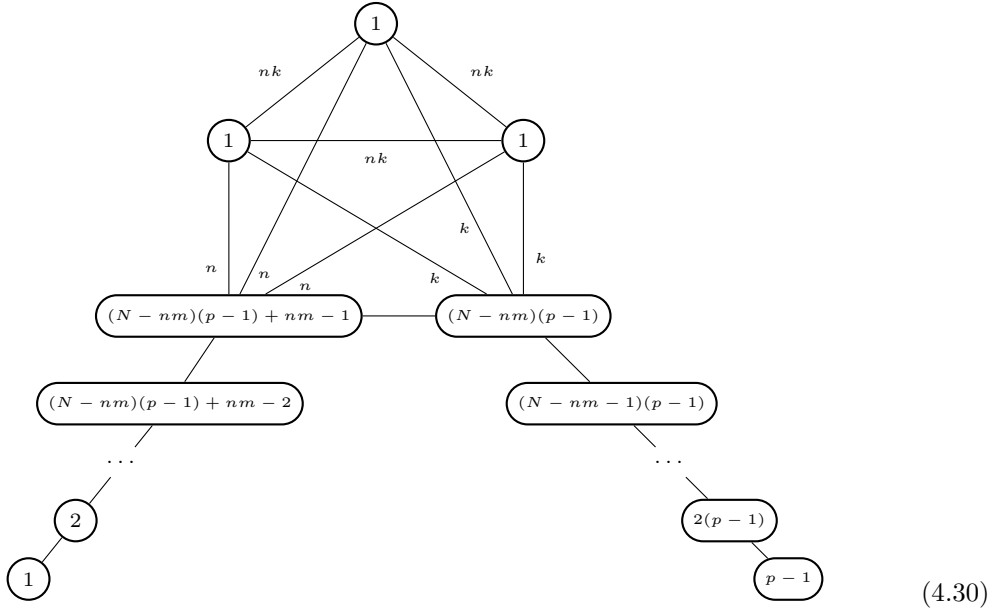
\begin{equation}\label{eq:befsub}
        \begin{tikzpicture}[scale=1.55,every node/.style={scale=1.1},font=\scriptsize]
    \node[gauge] (gl0) at (0.55,0) {$1$};
    \node[gauge] (gl1) at (0.95,0.5) {$2$};
    \node[] (gl2) at (1.35,1) {$\cdots$};
    \node[gauge] (gl3) at (1.75,1.5) {\tiny$(N-nm)(p-1)+nm-2$};
    \node[gauge] (gl4) at (2.25,2.25) {\tiny$(N-nm)(p-1)+nm-1$};
    \node[gauge] (gr4) at (4.25+0.5,2.25) {\tiny$(N-nm)(p-1)$};
    \node[gauge] (gr3) at (5+0.5,1.5) {\tiny$(N-nm-1)(p-1)$};
    \node[] (gr2) at (5.5+0.5,1) {$\cdots$};
    \node[gauge] (gr1) at (6+0.5,0.5) {\tiny$2(p-1)$};
    \node[gauge] (gr0) at (6.5+0.5,0) {\tiny$p-1$};
    \node[gauge] (gu0) at (2.25,3.25+0.5) {$1$};
    \node[gauge] (gu1) at (4.25+0.5,3.25+0.5) {$1$};
    \node[gauge] (gu2) at (3.25+0.25,4.25+0.5) {$1$};
    \draw (gl0)--(gl1)--(gl2)--(gl3)--(gl4)--(gr4)--(gr3)--(gr2)--(gr1)--(gr0);
    \draw (gu0)--(gu1);
    \draw (gu0)--(gu2);
    \draw (gu2)--(gu1);
    \draw (gl4)--(gu0);
    \draw (gl4)--(gu1);
    \draw (gl4)--(gu2);
    \draw (gr4)--(gu0);
    \draw (gr4)--(gu1);
    \draw (gr4)--(gu2);
    \node[] at (2.1,2.65) {\tiny$n$};
    \node[] at (2.55,2.6) {\tiny$n$};
    \node[] at (2.9,2.5) {\tiny$n$};
    \node[] at (4.7+0.2,2.65+0.1) {\tiny$k$};
    \node[] at (3.9+0.35,2.6+0.4) {\tiny$k$};
    \node[] at (3.4+0.6,2.475+0.1) {\tiny$k$};
    \node[] at (2.6,3.85+0.5) {\tiny$nk$};
    \node[] at (3.9+0.5,3.85+0.5) {\tiny$nk$};
    \node[] at (3.25+0.25,3.1+0.5) {\tiny$nk$};
\end{tikzpicture}
\end{equation}
where we have used that $\nu'=n$ and $\frac{N'+k'}{m}-\nu'=k$. Moreover, we simplify the drawing to represent the case with only three $\mathfrak{u}(1)$ gauge nodes in the complete graph as in Figure \ref{fig:3dmirrIkn}. The complete graph will not be relevant in the following discussion, since we will only focus on the linear sub-quiver consisting of the two tails that manifests the $\mathfrak{su}((N-nm)p+nm)$ flavour symmetry.

We now want to subtract the quiver \eqref{eq:subquiver} from \eqref{eq:befsub} to recover the quiver in Figure \ref{fig:3dmirrnkp} that is the 3d mirror of $\ADAN{p}{nm}{N}{[1^N]}$ for $p>b$. Upon doing this, we observe that the right tail immediately matches with that in Figure \ref{fig:3dmirrnkp}. Meanwhile for the left tail, we have $(N-nm)(p-1)+nm-1-(N-nm)(p-2)=N-1$ and so on until the gauge node with rank $(N-nm)(p-1)+nm-N+1$ whose rank is reduced to one. The rest of the left tail in the quiver \eqref{eq:befsub} has gauge nodes whose rank is reduced by one starting from $(N-nm)(p-2)$ all the way to one, which exactly matches the corresponding part in the quiver \eqref{eq:subquiver} that has to be subtracted, so this part of the tail is completely removed. This exactly reproduces the left tail in Figure \ref{fig:3dmirrnkp}. The complete graphs of the quiver \eqref{eq:befsub} and the one in Figure \ref{fig:3dmirrnkp} instead match independently of the quiver subtracted, since we recall that $q=n+k$. Finally, the number of free hypermultiplets also matches as
\begin{equation}
    \frac12m(\nu'-1)\left(\frac{p'}{m}-\nu'-1\right) = \frac12m(n-1)(k-1)~.
\end{equation}
This shows the matching of the 3d mirrors of the theories in \eqref{eq:bigiso} for $p>b$.

\subsubsection{The case $p<b$}

Now for $p<b$, we have $n=\mu\frac{p}{m}+\nu$, and so
\begin{eqnarray}
    &N' = (N-nm)p+nm \;,\quad p'=p=mq=m(n+k)\,, \nonumber\\
    &\nu' = \frac{N'}{m} \mod\left(\frac{p'}{m}\right) = \nu \;,\quad \mu' = \frac{N'-\nu'm}{p'} = N-nm+\mu~.
\end{eqnarray}
Note that again $p'<N'$ here, therefore we have defined $\mu'$ and $\nu'$ above. Using the quiver explicitly drawn in Figure \ref{fig:3dmirrIkn} and substituting the above values, we have that the 3d mirror of the theory $\ADAI{p}{(N-b)p+b}{[1^{(N-nm)p+nm}]}$ for $p<b$ is
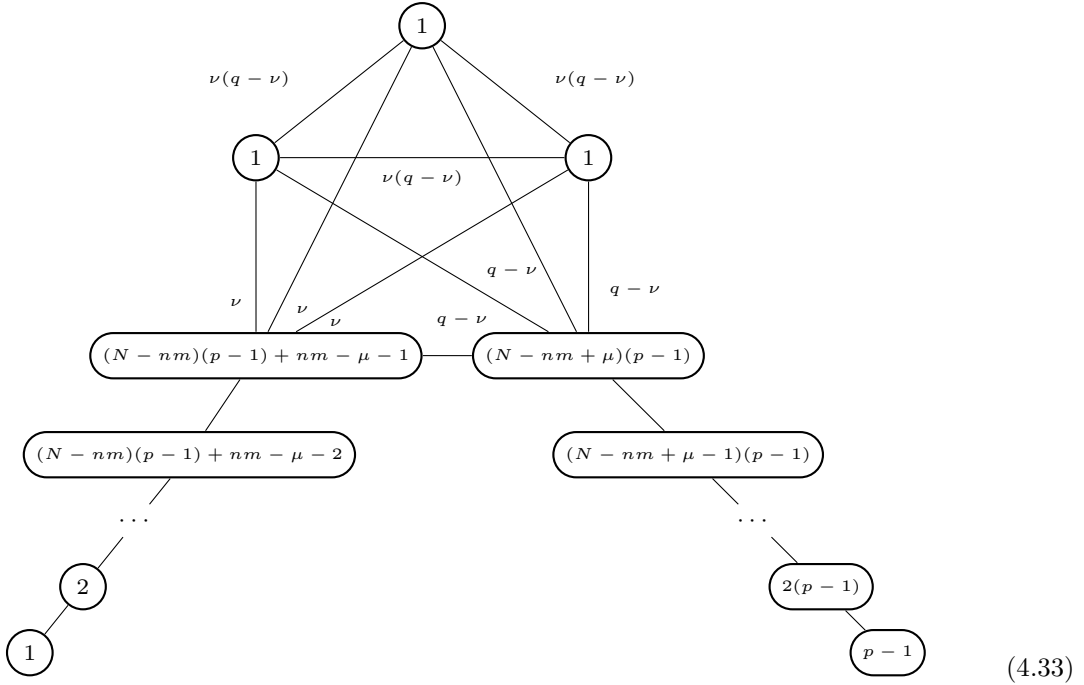
\begin{equation}\label{eq:befsub2}
    \begin{tikzpicture}[scale=1.75,every node/.style={scale=1.2},font=\scriptsize]
    \node[gauge] (gl0) at (0.55,0) {$1$};
    \node[gauge] (gl1) at (0.95,0.5) {$2$};
    \node[] (gl2) at (1.35,1) {$\cdots$};
    \node[gauge] (gl3) at (1.75,1.5) {\tiny$(N-nm)(p-1)+nm-\mu-2$};
    \node[gauge] (gl4) at (2.25,2.25) {\tiny$(N-nm)(p-1)+nm-\mu-1$};
    \node[gauge] (gr4) at (4.25+0.5,2.25) {\tiny$(N-nm+\mu)(p-1)$};
    \node[gauge] (gr3) at (5+0.5,1.5) {\tiny$(N-nm+\mu-1)(p-1)$};
    \node[] (gr2) at (5.5+0.5,1) {$\cdots$};
    \node[gauge] (gr1) at (6+0.5,0.5) {\tiny$2(p-1)$};
    \node[gauge] (gr0) at (6.5+0.5,0) {\tiny$p-1$};
    \node[gauge] (gu0) at (2.25,3.25+0.5) {$1$};
    \node[gauge] (gu1) at (4.25+0.5,3.25+0.5) {$1$};
    \node[gauge] (gu2) at (3.25+0.25,4.25+0.5) {$1$};
    \draw (gl0)--(gl1)--(gl2)--(gl3)--(gl4)--(gr4)--(gr3)--(gr2)--(gr1)--(gr0);
    \draw (gu0)--(gu1);
    \draw (gu0)--(gu2);
    \draw (gu2)--(gu1);
    \draw (gl4)--(gu0);
    \draw (gl4)--(gu1);
    \draw (gl4)--(gu2);
    \draw (gr4)--(gu0);
    \draw (gr4)--(gu1);
    \draw (gr4)--(gu2);
    \node[] at (2.1,2.65) {\tiny$\nu$};
    \node[] at (2.6,2.6) {\tiny$\nu$};
    \node[] at (2.85,2.5) {\tiny$\nu$};
    \node[] at (4.6+0.5,2.65+0.1) {\tiny$q-\nu$};
    \node[] at (4.1+0.075,2.6+0.3) {\tiny$q-\nu$};
    \node[] at (3.4+0.4,2.475+0.05) {\tiny$q-\nu$};
    \node[] at (2.2,3.85+0.5) {\tiny$\nu(q-\nu)$};
    \node[] at (4.3+0.5,3.85+0.5) {\tiny$\nu(q-\nu)$};
    \node[] at (3.25+0.25,3.1+0.5) {\tiny$\nu(q-\nu)$};
\end{tikzpicture}
\end{equation}
where we have used that $\nu'=n$, $\frac{N'+k'}{m}-\nu'=k$, and that $q=n+k$. Once again, one can easily check that subtracting the quiver in Figure \eqref{eq:subquiver} from \eqref{eq:befsub2} precisely reproduces the quiver in Figure \ref{fig:3dmirrnkn} and that the number of free hypermultiplets also agrees.

Thus in this subsection we have provided very strong evidence for our conjecture \eqref{eq:GentypeIiso} by explicitly matching the quivers describing the 3d mirrors of the Argyres--Douglas theories on the two sides of the equivalence with $[Y]=[1^N]$. The result immediately holds for general $[Y]$ as these cases can be seen as the nilpotent Higgsings of the manifest $\mathfrak{su}(N)$ symmetries in the theories with $[Y]=[1^N]$. Further, the 3d mirrors for generic $[Y]$ can be figured out easily in a similar way by the decay algorithm.

We will now justify and derive the number of extra free hypermultiplets in the equivalence \eqref{eq:GentypeIisofree} in the next subsection. These did not show up in this section since we only considered the case with $[Y]=[1^N]$ for which $H_{{\rm free}}=0$ in \eqref{eq:GentypeIisofree}. 

\subsection{\label{subsec:isofree}Counting the free hypermultiplets}

In this subsection we will derive the number of extra free hypermultiplets \eqref{eq:GentypeIisofree} that appear on the right hand side of \eqref{eq:GentypeIiso}. From a practical perspective these free hypermultiplets must be added so that the anomalies on the two sides of the equivalence agree. However, there is a direct way of understanding the origin of these extra hypermultiplets.

We focus on the case with the maximal regular puncture \eqref{eq:bigiso}, where in particular $H_{\text{free}}=0$. The $\mathfrak{su}(N)$ subalgebra of the flavour symmetry algebra, which is the entire contribution of the regular puncture on the l.h.s.\ of \eqref{eq:bigiso}, is identified with the $\mathfrak{su}(N)$ subalgebra on the r.h.s.\ of \eqref{eq:bigiso} which is only a part of the contribution from the regular puncture. One can now give a nilpotent vev of the form $[Y]$ to the moment map of this $\mathfrak{su}(N)$ symmetry and consider the IR limit of the triggered RG flow. This na\"ively leads to the following, incorrect for $p-1<N$, equivalence,
\begin{equation}\label{eq:Genisowrong}
    \ADAN{p}{n}{N}{[Y]} \cong \ADAI{(p}{(N-b)p+b}{[(p-1)^{N-b},Y]}~.
\end{equation}
This is incorrect because for $p-1<N$ the nilpotent Higgsings on the two sides look different. On the l.h.s.\ of \eqref{eq:Genisowrong}, the nilpotent vev has the form $[N^{l_N},\dots,(p-1)^{l_{p-1}},\dots,1^{l_1}]$, whereas the nilpotent vev on the r.h.s.\ of \eqref{eq:Genisowrong} has the form $[N^{l_N},\dots,(p-1)^{N-b+l_{p-1}},\dots,1^{l_1}]$. This can alternatively be viewed as the consequence of the na\"ive $\mathfrak{su}(N-b)\oplus\mathfrak{su}(l_{p-1})$ subalgebra of the flavour symmetry in the IR SCFT being actually enhanced to $\mathfrak{su}(N-b+l_{p-1})$. Therefore the number of massless Nambu--Goldstone modes in the IR is different on the two sides and thus the equivalence is not between just the interacting parts of the SCFTs.

We will focus on the case $p-1<N$ for the remainder of this subsection. We denote the nilpotent Higgsing on the l.h.s.\ of \eqref{eq:bigiso} of the form $[1^N]\rightarrow[Y]$ by \textbf{L}. The associated number of massless Nambu--Goldstone modes in the IR can be expressed as
\begin{equation}
    H_{\text{free}}(\textbf{L}) = \frac12\left(N^2-\sum_{i=1}^N s_{\textbf{L},i}^2\right)~,
\end{equation}
where $s_{\textbf{L},i}$ labels the entries of the conjugate partition $[{}^t Y]$. On the other hand, we denote the nilpotent Higgsing of the form $[(p-1)^{N-b},1^N]\to[(p-1)^{N-b},Y]$ on the r.h.s.\ of \eqref{eq:bigiso} by $\textbf{R}$ and the number of massless Nambu--Goldstone modes in the IR as $H_{\text{free}}(\textbf{R})$. Therefore the correct form of the equivalence is
\begin{eqnarray}
    \ADAN{p}{b}{N}{[Y]} \cong &&\ADAI{p}{(N-b)p+b}{[(p-1)^{N-b},Y]} \nn\\
    &\otimes& (H_{\text{free}}(\textbf{R})-H_{\text{free}}(\textbf{L}))\text{ free hypermultiplets}~.
\end{eqnarray}

\begin{figure}
\centering
    \begin{tikzpicture}[scale=1.5,every node/.style={scale=1.2},font=\normalsize]
        \draw[->,thick,black] (3,3.5)--(3,2.5);
        \draw[->,thick,black] (3,1.5)--(3,0.5);
        \draw[->,thick,black] (5.5,3.5)--(5.5,0.5);
        \draw[->,thick,black] (-1,1.5)--(-1,0.5);
        \node at (-0.5,1) {\textbf{L}};
        \node at (2.5,3) {\textbf{R}$_2$};
        \node at (2.5,1) {\textbf{R}};
        \node at (6,2) {\textbf{R}$_1$};
        \node at (0.35,2) {$\cong$};
        \node at (0.9,0) {$\cong$};
        \node at (4,4) {$\ADAI{p}{(N-b)p+b}{\left[1^{(N-b)p+b}\right]}$};
        \node at (3,2) {$\ADAI{p}{(N-b)p+b}{\left[(p-1)^{N-b},1^N\right]}$};
        \node at (4,0) {$\ADAI{p}{(N-b)p+b}{\left[(p-1)^{N-b},Y\right]}$};
        \node at (-1,2) {$\ADAN{p}{b}{N}{[1^N]}$};
        \node at (-1,0) {$\ADAN{p}{b}{N}{[Y]}$};
    \end{tikzpicture}
    \caption{Schematic representation of the various RG flows triggered by nilpotent vevs that relate the theories in \eqref{eq:GentypeIiso} for different choices of the partition $[Y]$.}
\label{fig:VariousHiggs}
\end{figure}

As schematically represented in Figure \ref{fig:VariousHiggs}, $H_{\text{free}}(\textbf{R})$ can be obtained by considering the difference of the Nambu--Goldstone modes produced in nilpotent Higgsing of the form $[1^{(N-b)p+b}]\rightarrow[(p-1)^{N-b},Y]$, denoted by $\textbf{R}_1$, and the one in the nilpotent Higgsing of the form $[1^{(N-b)p+b}]\rightarrow[(p-1)^{N-b},1^N]$, denoted by $\textbf{R}_2$,
\begin{equation}
    H_{\text{free}}(\textbf{R}) = H_{\text{free}}(\textbf{R}_1) - H_{\text{free}}(\textbf{R}_2)~.
\end{equation}
The corresponding number of massless Nambu--Goldstone modes in the IR is given by,
\begin{eqnarray}
    H_{\text{free}}(\textbf{R}_1) &= \frac12\left(\Big((N-b)p+b\Big)^2-\sum_{i=1}^{(N-b)p+b}s_{\textbf{R}_1,i}^2\right) \nn \\ H_{\text{free}}(\textbf{R}_2) &= \frac12\left(\Big((N-b)p+b\Big)^2-\sum_{i=1}^{(N-b)p+b}s_{\textbf{R}_2,i}^2\right)~,
\end{eqnarray}
where $s_{\textbf{R}_1,i}$ labels the transpose of the partition $[(p-1)^{N-b},Y]$ and $s_{\textbf{R}_2,i}$ labels the transpose of the partition $[(p-1)^{N-b},1^N]$.

Thus the correct number of free hypermultiplets that need to be added to the r.h.s.\ of the incorrect equivalence \eqref{eq:Genisowrong} is
\begin{equation}
    H_{\text{free}} = H_{\text{free}}(\textbf{R}_1) - H_{\text{free}}(\textbf{R}_2) - H_{\text{free}}(\textbf{L})~.
\end{equation}
We will now show that this computation gives \eqref{eq:GentypeIisofree} by considering the representation theory involved in these various nilpotent Higgsings in the general case.

Following the discussion in Subsection \ref{subsec:ADflavourcc}, we investigate the decomposition of the adjoint representation of the appropriate symmetry. This allows us to keep track of the number of massless Nambu--Goldstone modes as well as their representation under the corresponding global symmetry. Let us consider the decomposition of the adjoint representation in the three cases \textbf{L}, $\textbf{R}_1$, $\textbf{R}_2$ independently.

\paragraph{L: \boldmath$[1^N] \to [Y]=[N^{l_N},\dots,(p-1)^{l_{p-1}},\dots,1^{l_1}]$.}
The adjoint representation of $\mathfrak{su}(N)$ decomposes in the following way:
\begin{equation}
    \begin{split}
        \text{adj}_{\mathfrak{su}(N)} \longrightarrow \left(V_{\frac{p-2}{2}}\otimes V_{\frac{p-2}{2}},l_{p-1}\otimes\overline{l_{p-1}}\right) &\bigoplus_{p-1\neq i=1}^N \left(V_{\frac{p-2}{2}}\otimes V_{\frac{i-1}{2}},l_{p-1}\otimes\overline{l_i}\oplus l_i\otimes\overline{l_{p-1}}\right) \\
        &\bigoplus_{p-1\neq i,j=1}^N \left(V_{\frac{i-1}{2}}\otimes V_{\frac{j-1}{2}},l_i\otimes\overline{l_j}\oplus l_j\otimes\overline{l_i}\right)~.
    \end{split}
\end{equation}
This gives rise to the following set of massless Nambu--Goldstone modes that transform in the specified representations of the IR global symmetry
\begin{eqnarray}\label{eq:GBinL}
    &n_{p-1,p-1}=\frac{(p-1)(p-2)}{2} &\text{ in } l_{p-1}\otimes\overline{l_{p-1}}~,\nn\\
    &n_{p-1,i}=\frac{2ip-p-3i+1+|i-p+1|}{4} &\text{ in } l_{p-1}\otimes\overline{l_i}\oplus l_i\otimes\overline{l_{p-1}}~,\nn\\
    &n_{ij}=\frac{(i+j-|i-j|)(i+j-2+|i-j|)}{8} &\text{ in } l_i\otimes\overline{l_j}\oplus l_j\otimes\overline{l_i}~.
\end{eqnarray}

\paragraph{\boldmath$\text{R}_2$: $[1^{(N-b)p+b}] \to [Y]=[(p-1)^{N-b},1^N]$.}

The adjoint representation of $\mathfrak{su}((N-b)p+b)$ decomposes in the following way:
\begin{equation}
    \begin{split}
        \text{adj}_{\mathfrak{su}((N-b)p+b)} \longrightarrow &\left(V_{\frac{p-2}{2}}\otimes V_{\frac{p-2}{2}},N-n\otimes\overline{N-b}\right) \\
        &\oplus \left(V_{\frac{p-2}{2}},N-b\otimes\overline{N}\oplus N\otimes\overline{N-b}\right) \oplus \left(V_0,N-b\otimes\overline{N}\right)~.
    \end{split}
\end{equation}
This gives rise to the following set of massless Nambu--Goldstone modes that transform in the specified representations of the IR global symmetry
\begin{eqnarray}\label{eq:GBinR2}
    &n_{p-1,p-1}=\frac{(p-1)(p-2)}{2} &\text{ in } N-b\otimes\overline{N-b}~,\nn\\
    &n_{p-1,1}\frac{p-2}{2} &\text{ in } N-b\otimes\overline{N}\oplus N\otimes\overline{N-b}~.
\end{eqnarray}

\paragraph{\boldmath$\text{R}_1$: $[1^{(N-b)p+b}] \to [N^{l_N},\dots,(p-1)^{N-
b+l_{p-1}},\dots,1^{l_1}]$.}
The adjoint of $\mathfrak{su}((N-b)p+b)$ decomposes in the following way
\begin{equation}
    \begin{split}
        \text{adj}_{\mathfrak{su}_N} \longrightarrow &\left(V_{\frac{p-2}{2}}\otimes V_{\frac{p-2}{2}},N-b+l_{p-1}\otimes\overline{N-b+l_{p-1}}\right) \\
        &\bigoplus_{p-1\neq i=1}^N \left(V_{\frac{p-2}{2}}\otimes V_{\frac{i-1}{2}},N-b+l_{p-1}\otimes\overline{l_i}\oplus l_i\otimes\overline{N-b+l_{p-1}}\right) \\
        &\bigoplus_{p-1\neq i,j=1}^N \left(V_{\frac{i-1}{2}}\otimes V_{\frac{j-1}{2}},l_i\otimes\overline{l_j}\oplus l_j\otimes\overline{l_i}\right)~.
    \end{split}
\end{equation}
This gives rise to the following set of massless Nambu--Goldstone modes that transform in the specified representations of the IR global symmetry
\begin{alignat}{3}\label{eq:GBinR1}
    &n_{p-1,p-1}=\tfrac{(p-1)(p-2)}{2} && \text{ in } && {\color{blue} N-b\otimes\overline{N-b}} \oplus N-b\otimes\overline{l_{p-1}}\oplus l_{p-1}\otimes\overline{N-b}\oplus {\color{red} l_{p-1}\otimes\overline{l_{p-1}}}~,\nn\\
    &n_{p-1,i}=\tfrac{2ip-p-3i+1+|i-p+1|}{4} &&\text{ in } &&N-b\otimes\overline{l_i}\oplus l_i\otimes\overline{N-b}\oplus {\color{red} l_{p-1}\otimes\overline{l_i}\oplus l_i\otimes\overline{l_{p-1}}}~,\nn\\
    &n_{ij}=\tfrac{(i+j-|i-j|)(i+j-2+|i-j|)}{8} &&\text{ in } &&{\color{red} l_i\otimes\overline{l_j}\oplus l_j\otimes\overline{l_i}}~.
\end{alignat}
The Nambu--Goldstone modes that appear in \eqref{eq:GBinL} are denoted in {\color{red} red} whereas those that appear in \eqref{eq:GBinR2} are denoted in {\color{blue} blue}. Note that all the Nambu--Goldstone modes appearing in \textbf{L} appear in the same representations in $\textbf{R}_2$, whereas $\frac{p-2}{2}$ Nambu--Goldstone modes transforming in $N-b\otimes\overline{N}\oplus N\otimes\overline{N-b}$ representation of $\mathfrak{su}(N-b)\oplus\mathfrak{su}_N$ do not appear in $\textbf{R}_1$. 

Focusing just on the net number of Nambu--Goldstone bosons and not on their representations under the flavour symmetry algebra, we find
\begin{equation}
    \begin{split}
        H_{\text{free}} &= H_{\text{free}}(\textbf{R}_1) - H_{\text{free}}(\textbf{R}_2) - H_{\text{free}}(\textbf{L})\\
        &= \sum_{i=1}^{N} \frac{2pi-3i-p+1}{4}\times2(N-b)\,l_i - \frac{p-2}{2}\times2(N-b)N~, \\
        &= (N-b)\left[(p-2)\sum_{i\leqslant p-1}il_i+(p-1)\sum_{i>p-1}(i-1)l_i-(p-2)N\right]~,
    \end{split}
\end{equation}
where in the second line we only write the terms that are not cancelled due to the identifications we made in the previous paragraph. At this stage we recall that $N=\sum_{i=1}^N il_i$, which lets us simplify this expression to
\begin{equation}
    \begin{split}
        H_{\text{free}} &= (N-b)\left[(p-1)\sum_{i>p-1}il_i-(p-1)\sum_{i>p-1}l_i-(p-2)\sum_{i>p-1}il_i\right] \\
        &= (N-b)\sum_{i>p-1}(i-p+1)l_i~,
    \end{split}
\end{equation}
which reproduces \eqref{eq:GentypeIisofree}. This means that for each $\mathfrak{su}(l_i)$ simple factor in the flavour symmetry algebra for $i> p-1$ there are $(N-b)(i-p+1)$ hypermultiplets in its fundamental representation.

\section{\label{sec:conclusions}Conclusions and outlook}

We have seen in this paper that the landscape of (non-degenerate, \emph{i.e.}, not-of-\emph{type-III}) type $A$ Argyres--Douglas theories (without outer automorphism twists) can be simplified substantially, and for theories without exactly marginal deformations one essentially has only the coprime type I theories. Further equivalences between type I theories simplify the picture more; this also leads to a variety of conjectures/expectations for the associated vertex operator algebras, Higgs branches, and Hitchin moduli spaces. The main consistency checks that we have performed for our proposals are the matching of the Coulomb branch spectra and of the 3d theories arising from circle compactification or their mirror duals, as well as the matching of the conformal and flavour central charges whenever possible. For the most interesting equivalence between coprime theories, an elegant interpretation at the level of associated vertex algebras is provide by the matrix extended $\mathcal{W}_{1+\infty}$ algebras of \cite{Prochazka:2017qum}.

These various isomorphisms are not special to type $A$ Argyres--Douglas theories, and the analysis pursued here can be generalised to, \emph{e.g.}, type $D$ as well as to constructions using outer automorphism twists. In particular, there are partial results for Argyres--Douglas theories without exactly marginal deformations of more general types in \cite{Xie:2019yds} and the auxiliary Riemann sphere algorithm for treating theories with exactly marginal deformations and a maximal regular puncture is described in \cite{Xie:2017aqx}. Our methods can straightforwardly be extended to those theories as well, since one has access to Coulomb branch spectra, 3d mirrors, and (in some cases) 3d reductions \cite{Carta:2021whq,Xie:2021ewm,Carta:2021dyx,Carta:2022spy,Carta:2022fxc}. Even though the relevant 3d theories are orthosymplectic quivers rather than the unitary/special unitary quivers we have encountered in this work, we expect an analogous approach to be applicable to them. An additional test that could be performed would be a matching of one-form symmetries \cite{Gaiotto:2014kfa} across the equivalences. This is trivially true for the equivalences discussed in the present work, since all the theories considered have trivial one-form symmetry groups \cite{DelZotto:2020esg,Hosseini:2021ged,Buican:2021xhs,Bhardwaj:2021mzl}. Furthermore, the landscape of degenerate/type III Argyres--Douglas theories should enjoy similar simplifications. While most central charges are not known in general for this class of theories, the Coulomb branch spectra \cite{Xie:2017vaf} and the quivers describing the 3d mirror of their circle compactification \cite{Xie:2021ewm} are known and can be used to perform this analysis. We hope to present a detailed account of some of these simplifications in the near future.

An interesting consequence of the equivalence \eqref{eq:typeItypeIIeq} between type I and type II Argyres--Douglas theories arises in relation to their geometric engineering construction in type IIB \cite{Xie:2012hs,Cecotti:2012jx,Cecotti:2013lda,Wang:2015mra}. The only case where both sides of this isomorphism admit a known geometric engineering realisation is for full punctures and $p=2$, which gives the equivalence
\begin{equation}
    \ADAF{2}{N-1}{N} \cong \ADAIF{2}{N+1}~.
\end{equation}
It would be interesting to explore possible relationships between the corresponding Calabi--Yau singularities.

Finally, we note that holographic duals have been conjectured for the large $N$ limits of type I Argyres--Douglas theories of type $A$. For regular punctures labelled by a rectangular Young diagram, proposals were made in \cite{Bah:2021mzw,Bah:2021hei,Bomans:2023ouw}. This was generalised to type I theories with an arbitrary regular puncture in \cite{Couzens:2022yjl, Bah:2022yjf}. The two primary equivalences between type I Argyres--Douglas theories \eqref{eq:isopl} and \eqref{eq:isoconj} can in fact be realised in the holographic setting \cite{Couzens:2023kyf}.

\acknowledgments{The authors would like to thank Pieter Bomans, Dylan Butson, Christopher Couzens, Thomas Creutzig, Tudor Dimofte, Jacques Distler, Laura Fredrickson, Neil Lambert, Sujay Nair, Andy Neitzke, Tom\'{a}\v{s} Proch\'{a}zka, G\'{e}rard Watts, and Zhenghao Zhong for useful discussions related to this work. We would also like to thank the Pollica Physics Centre for hospitality while some of this work was completed. The work of CB and MS is supported in part by ERC Consolidator Grant 864828 ``Algebraic Foundations of Supersymmetric Quantum Field Theory (SCFTAlg)'' and by the Simons Collaboration for the Nonperturbative Bootstrap under grant 494786 from the Simons Foundation. CB is also supported in part by the STFC through grant number ST/T000864/1. PS would like to acknowledge support from the Clarendon Fund and the Mathematical Institute, University of Oxford. JS would like to thank the STFC for their support under the studentship grant ST/T506187/1. MM is supported in part by the STFC through grant number ST/X000753/1.}

\appendix

\section{\label{app:Trhosigma}Review of \texorpdfstring{$T_\rho^\sigma(\mathfrak{su}(N))$}{T[rho][sigma](\mathfrak{su}(N))} theories}

An important class of three-dimensional $\mathcal{N}=4$ gauge theories that arise throughout this work are those of unitary linear quiver type:
\medskip
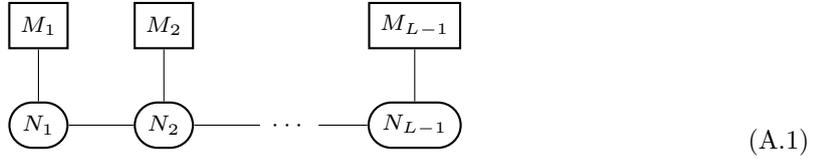
\begin{equation}\label{eq:quiverTrhosigmaSUN}
\begin{tikzpicture}[scale=1.1,every node/.style={scale=1.2},font=\scriptsize]
    \node[gauge] (g0) at (0,0) {$N_1$};
    \node[gauge] (g1) at (1.5,0) {$N_2$};
    \node (g2) at (3,0) {$\cdots$};
    \node[gauge] (g3) at (4.5,0) {$N_{L-1}$};
    \node[flavor] (f0) at (0,1.2) {$M_1$};
    \node[flavor] (f1) at (1.5,1.2) {$M_2$};
    \node[flavor] (f3) at (4.5,1.2) {$M_{L-1}$};
    \draw (g0)--(g1)--(g2)--(g3);
    \draw (g0)--(f0);
    \draw (g1)--(f1);
    \draw (g3)--(f3);
\end{tikzpicture}
\end{equation}
We adopt the standard notation for quivers in which square nodes denote flavour symmetries and circular/rounded nodes denote gauge symmetries, all of unitary type, while lines connecting nodes represent hypermultiplets transforming in the bi-fundamental representation of the symmetries associated with the nodes they connect. In this appendix we review some standard aspects of these theories that are utilised in the body of this paper.

Theories of this type are denoted by $T_\rho^\sigma(\mathfrak{su}(N))$ \cite{Gaiotto:2008ak}, where the ranks of the flavour and gauge symmetries are encoded in two partitions of $N$, $\rho$ and $\sigma$. For partitions given as
\begin{equation}\label{Partrhosigma}
\begin{split}
    \rho&=[\rho_1,\cdots,\rho_L]=[N^{l_N},\cdots,1^{l_1}]~,\\
    \sigma&=[\sigma_1,\cdots,\sigma_K]=[N^{k_N},\cdots,1^{k_1}]~,
\end{split}
\end{equation}
satisfying
\begin{equation}
    \sum_{n=1}^Nn\,l_n=\sum_{m=1}^Nm\,k_m=N\,,\quad L=\sum_{n=1}^Nl_n\,,\quad K=\sum_{m=1}^Nk_m~,
\end{equation}
the flavour and gauge symmetry ranks are determined as follows:
\begin{equation}
    M_{L-i}=k_i~,\quad N_{L-i}=\sum_{j=i+1}^L\rho_j-\sum_{j=i+1}^N(j-i)k_j~,\qquad i=1,\cdots,L-1~.
\end{equation}
When the partitions satisfy the condition $\sigma^T<\rho$ then the theory is \emph{good} in the sense of \cite{Gaiotto:2008ak}. This means in particular that for each gauge node the number of fundamental hypermultiplets is greater or equal than twice its rank, which is equivalently the condition for all monopole operators to have scaling dimensions computed in terms of the UV $R$-symmetry that are above the unitarity bound $\Delta>\frac{1}{2}$.

The partitions encode various properties of the theory. First of all, the global symmetries are given by
\begin{equation}\label{eq:symmTrhosigmaSUN}
    \mathfrak{g}=\mathfrak{g}_{\text{Higgs}}\oplus \mathfrak{g}_{\text{Coulomb}}=\frac{\bigoplus_{m=1}^N\mathfrak{u}(k_m)}{\mathfrak{u}(1)}\oplus \frac{\bigoplus_{n=1}^N\mathfrak{u}(l_n)}{\mathfrak{u}(1)}~.
\end{equation}
The first factor is the flavour symmetry acting on the Higgs branch and is manifest in the Lagrangian description of \eqref{eq:quiverTrhosigmaSUN}. The second factor is the symmetry acting on the Coulomb branch and is enhanced at low energies from the $\mathfrak{u}(1)^{L-1}$ topological symmetry that is manifest in the quiver description \eqref{eq:quiverTrhosigmaSUN}. As explained in \cite{Gaiotto:2008ak}, such an enhancement occurs whenever a gauge node is \emph{balanced}, \emph{i.e.}, the number of fundamental hypermultiplets is equal to twice the rank. If there is a sequence of $p$ consecutive balanced gauge nodes, then the associated $\mathfrak{u}(1)^p$ topological symmetry is enhanced to $\mathfrak{su}(p+1)$.

Moreover, the Higgs and the Coulomb branch, as algebraic varieties, are given by \cite{Nakajima:1994nid,Gaiotto:2008ak}
\begin{equation}\label{eq:MTsigmarho}
    \mathcal{M}_{\text{HB}}=\overline{\mathbb{O}}_{\rho^T}\cap\mathcal{S}_{\sigma}~, \qquad\mathcal{M}_{\text{CB}}=\overline{\mathbb{O}}_{\sigma^T}\cap\mathcal{S}_{\rho}~,
\end{equation}
where $\overline{\mathbb{O}}_{\rho^T}$ and $\overline{\mathbb{O}}_{\sigma^T}$ are the closures of the nilpotent orbits of types $\rho^T$ and $\sigma^T$, while $\mathcal{S}_{\sigma}$ and $\mathcal{S}_{\rho}$ are the Slodowy slices of type $\sigma$ and $\rho$.

The partitions $\rho$ and $\sigma$ are also related to the Type IIB brane setup that engineers $T_\rho^\sigma(\mathfrak{su}(N))$ \cite{Hanany:1996ie}. The brane configuration consists of $N$ D3-branes stretched between $L$ NS5--branes and $K$ D5--branes. The entries of the two partitions $\rho_i$ and $\sigma_i$ represent the net number of D3-branes ending on the $i$-th NS5--brane and D5--brane, respectively, where we count from the interior to the exterior of the configuration if the partitions are put in decreasing order.

The $T_\rho^\sigma(\mathfrak{su}(N))$ theories are related in pairs by mirror symmetry \cite{Intriligator:1996ex}. Specifically, $T_\rho^\sigma(\mathfrak{su}(N))$ is mirror dual to $T^\rho_\sigma(\mathfrak{su}(N))$. The swap of the two partitions is compatible with the exchange of the Higgs and Coulomb branches under mirror symmetry. Mirror symmetry for these theories follows from the action of $S$-duality of the Type IIB string theory, which converts NS5--branes into D5--branes and \emph{vice versa}.

The $T_\rho^\sigma(\mathfrak{su}(N))$ theories are additionally related by moduli space renormalisation group (RG) flows that are triggered by turning on certain expectation values for Higgs or Coulomb branch operators. In fact, any $T_\rho^\sigma(\mathfrak{su}(N))$ theory can be obtained via Higgs and Coulomb branch deformations of the $T(\mathfrak{su}(N))$ theory, which is the member of the family corresponding to two trivial partitions $\rho=\sigma=[1^N]$:
\begin{equation}\label{eq:quiverTSUN}
\begin{tikzpicture}[scale=1.2,every node/.style={scale=1.2},font=\scriptsize]
    \node[gauge] (g0) at (0,0) {$N-1$};
    \node (g1) at (1.5,0) {$\cdots$};
    \node[gauge] (g2) at (3,0) {$\,2\,$};
    \node[gauge] (g3) at (4.5,0) {$\,1\,$};
    \node[flavor] (f0) at (-1.5,0) {$\,N\,$};
    \draw (g0)--(g1)--(g2)--(g3);
    \draw (g0)--(f0);
\end{tikzpicture}
\end{equation}

This theory is self-dual under mirror symmetry. Accordingly, it has the same Higgs and Coulomb branch symmetries which are the maximal possible ones $\mathfrak{su}(N)_{\text{Higgs}}\times \mathfrak{su}(N)_{\text{Coulomb}}$, and both its Higgs and Coulomb branch are given by the closure of the principal nilpotent orbit of $\mathfrak{sl}_N^*$, a.k.a.~the nilpotent cone of $\mathfrak{sl}_N$,
\begin{align}
    \mathcal{M}_{\text{HB}}=\mathcal{M}_{\text{CB}} = \overline{\mathbb{O}}_{\text{prin}}=\mathcal{N}_{\mathfrak{sl}_N}=\left\{M\in SL(N,\mathbb{C})\,|\,M^N=0\right\}~.
\end{align}
Turning on an expectation value for the $\mathfrak{sl}_N^\ast$-valued $\mathfrak{su}(N)_{\text{Coulomb}}$ moment map lying in the nilpotent co-adjoint orbit labelled by $\rho$ and for the $\mathfrak{su}(N)_{\text{Higgs}}$ moment map lying in the nilpotent co-adjoint orbit labelled by $\sigma$ leads to a flow from $T(\mathfrak{su}(N))$ to $T_\rho^\sigma(\mathfrak{su}(N))$ (see for example \cite{Cremonesi:2014uva,Hwang:2020wpd} for more details). This corresponds to moving to the leaves of the nilpotent cone of $\mathfrak{sl}_N$ whose transverse slices are described by \eqref{eq:MTsigmarho}.

At the level of the type IIB brane realisation, these expectation values (and the subsequent RG flow) are realised by moving certain D3-brane segments to infinity. For Higgs branch expectation values, these are D3-branes suspended between pairs of D5--branes, while for the Coulomb branch they are D3-branes suspended between pairs of NS5--branes. It is then clear that a Higgs branch expectation value labelled by $\sigma$ does not affect the Coulomb branch symmetry, while it breaks the Higgs branch symmetry as
\begin{equation}
    \mathfrak{su}(N)_{\text{Higgs}}\to \frac{\prod_{m=1}^N\mathfrak{u}(k_m)}{\mathfrak{u}(1)}~.
\end{equation}
Similarly, a Coulomb branch expectation value does not affect the Higgs branch symmetry, but it breaks the Coulomb branch symmetry according to
\begin{equation}
    \mathfrak{su}(N)_{\text{Coulomb}}\to \frac{\prod_{n=1}^N\mathfrak{u}(l_n)}{\mathfrak{u}(1)}~.
\end{equation}
We then obtain from the $\mathfrak{su}(N)_{\text{Higgs}}\times \mathfrak{su}(N)_{\text{Coulomb}}$ symmetry of $T(\mathfrak{su}(N))$ the symmetry \eqref{eq:symmTrhosigmaSUN} of $T_\rho^\sigma(\mathfrak{su}(N))$.

From $T_\rho^\sigma(\mathfrak{su}(N))$, one can further deform the theory to other $T_{\rho'}^{\sigma'}(\mathfrak{su}(N))$ theories by activating appropriate expectation values that modify the partitions $\rho\to\rho'$ and $\sigma\to\sigma'$. In the main text we considered several such flows.

\section{\label{app:notations}A survey of naming conventions}

For the benefit of the reader, we provide in this appendix the dictionary relating various notations used for Argyres--Douglas theories throughout the literature, including the one introduced in the present paper.

The popular notation $D_p^b(SU(N))$ introduced in \cite{Cecotti:2012jx,Cecotti:2013lda} designates Argyres--Douglas theories with a full regular puncture. Additionally, this notation is normally reserved for type I and type II theories, which are characterised by $b=N,N-1$, respectively. Our suggested notation is the natural generalisation of this to include $A$-type Argyres--Douglas theories of more general types $b$ as well as a generic regular puncture, the exact relation being
\begin{equation}
    D_p^b(SU(N)) = \ADAN{p}{b}{N}{[1^N]}~.
\end{equation}
An alternative notation for Argyres--Douglas theories is of the form $\ADA{N-1}{b}{k}{[Y]}$ \cite{Wang:2015mra}. This is related to the notation of this paper according to,
\begin{equation}
    \ADA{N-1}{b}{k}{[Y]} = \ADAN{b+k}{b}{N}{[Y]}~.
\end{equation}
A further labelling for these untwisted Argyres--Douglas theories is by the set $<\mathfrak{sl}_N,b,k,\mathfrak f_{[Y]}>$ \cite{Xie:2021ewm,Li:2022njl} or equivalently $\mathcal T_{\mathfrak{sl}_N,b,k,\mathfrak f_{[Y]}}$ \cite{Shan:2023xtw}, where $\mathfrak f_{[Y]}$ denotes the nilpotent element specified by the partition $[Y]$. In our notation, these are described as,
\begin{equation}
    \mathcal T_{\mathfrak{sl}_N,b,k,\mathfrak f_{[Y]}} =\; <\mathfrak{sl}_N,b,k,\mathfrak f_{[Y]}> \;= \ADAN{b+k}{b}{N}{[Y]}
\end{equation}
In the case of type I Argyres--Douglas theories without regular punctures, additional popular designations are of the form $(A_{N-1},A_{k-1})$ \cite{Cecotti:2010fi} or $I_{N,k}$ \cite{Xie:2013jc}. The relation to our notation is as follows,
\begin{equation}
    (A_{N-1},A_{k-1}) = I_{N,k} = \ADAI{N+k}{N}{[N]}~.
\end{equation}
Finally, the so-called $(A_1,D_N)$ Argyres--Douglas theories of \cite{Cecotti:2010fi} correspond in our notation to
\begin{equation}
    (A_1,D_N) = \ADAIF{N}{2}~.
\end{equation}

\bibliographystyle{JHEP}
\bibliography{refs.bib}

\end{document}